\documentclass{aastex63}
\usepackage{amsmath, amsthm, amsfonts}
\pdfoutput=1

\newcommand\mml{mm-$\lambda$}

\received{October 1, 2024}
\revised{December 3, 2024}
\accepted{December 31, 2024}
\submitjournal{ApJ}

\shorttitle{Solar Spicules at Millimeter Wavelengths}
\shortauthors{Bastian et al.}

\begin{document}

\title{ALMA Observations of Solar Spicules \\ in a Polar Coronal Hole}

\correspondingauthor{T. S. Bastian}
\email{tbastian@nrao.edu}

\author[0000-0002-0713-0604]{T. S. Bastian}
\affiliation{National Radio Astronomy Observatory, 520 Edgemont Road, 
Charlottesville, VA 22903USA}

\author{C. Alissandrakis}
\affiliation{Section of Astro-Geophysics, Department of Physics, University of Ioannina,
GR-45110 Ioannina, Greece}

\author{A. Nindos}
\affiliation{Section of Astro-Geophysics, Department of Physics, University of Ioannina,
GR-45110 Ioannina, Greece}

\author{M. Shimojo}
\affiliation{National Astronomical Observatory of Japan, Mitaka, Japan}
\affiliation{Graduate University for Advanced Studies, SOKENDAI, Mitaka, Japan}

\author{S. M. White}
\affiliation{Space Vehicles Directorate, Air Force Research Laboratory, Kirtland AFB, NM, USA}

\begin{abstract}
We report observations of solar spicules at millimeter wavelengths (\mml) using the {\sl Atacama Large Millimeter/submillimeter Array} (ALMA). These are supplemented by observations in optical (O), ultraviolet (UV) and extreme ultraviolet (EUV) wavelengths. The observations were made on 2018 December 25 of the northern polar coronal hole. ALMA obtained time-resolved imaging observations at wavelengths of 3 mm (100 GHz; 2~s cadence) and 1.25~mm (239 GHz; $\approx 2$~min cadence) with an angular resolution of $2.2\times 1.3"$ and $1.5"\times0.7"$, respectively. Spicules observed at \mml\ are easily seen low in the chromosphere whereas spicules in UV bands are seen to extend higher. The spicules observed at \mml\ are seen in absorption against coronal EUV  emission, allowing us to estimate the column depth of neutral hydrogen. Spicular emission at \mml, due to thermal free-free radiation, allows us to estimate the electron number density as a function of height. We find that spicule densities, inferred from the \mml\ data are uniquely insensitive to assumptions regarding the temperature of plasma in spicules. We suggest that the upward mass flux carried by spicules is unlikely to play a significant role in the mass budget of the solar corona and solar wind, and the transport of hot material into the corona by spicules may not play a significant role in coronal heating.  However, the possibility that electric currents, fast kink and torsional waves, or other wave modes carried by spicules  may play a role in transporting energy into the solar corona cannot be excluded. 
\end{abstract}


\bigskip\bigskip

\section{Introduction} \label{sec:intro}

Solar spicules are a ubiquitous phenomenon in which multitudes of dynamic, filamentary jets with temperatures of order $10^4$ K extend thousands of kilometers from the chromosphere up into the low solar corona. They occur predominantly in the chromospheric network,  the boundaries of supergranular cells (see reviews by \citealt{Beckers1968, Beckers1972, Sterling2000, Tsiropoula2012}). Although known for more than a century spicules remained at the limits of available angular resolution at optical wavelengths for many years. This changed during the past two decades thanks to the high-angular-resolution observations at optical (O), ultraviolet (UV), and extreme ultraviolet (EUV) wavelengths that became possible with the {\sl Hinode} \citep{Kosugi2007} Solar Optical Telescope (SOT; \citealt{Tsuneta2008}), the Solar Dynamics Observatory (SDO; \citealt{Pesnell2011}) Atmospheric Imaging Assembly (AIA; \citealt{Lemen2011}), and the Interface Region Imaging Spectrometer (IRIS; \citealt{DePontieu_2014}), leading to renewed interest in spicules and their role in the mass and energy budget of the solar atmosphere. In 2007, based on high resolution Ca~II~H line observations of the limb made with the SOT, \citet{DePontieu_2007} proposed that there were in fact two types of spicules classified, naturally, as type I and type II spicules\footnote{This designation should not be confused with the two spicule types described by \citet{Beckers1968}, who based the distinction on spicules with wide (type I) and narrow (type II) Ca II line widths.}. Type I spicules are characterized by time scales of 3-7 min, upward speeds of $\sim 25$ km/s, and often display parabolic trajectories in height-time diagrams. In contrast, type II spicules are much more dynamic than type I spicules. They are very thin (few $\times 100$ km), have lifetimes of 30-110 s, display speeds of 50-150 km/s, and appear to fade with height in Ca~II and, as a consequence, are not observed to fall back toward the Sun. However, this may be primarily the result of a decrease of opacity in Ca~II due to temperature and/or density changes \citep{Skogsrud2015, Carlsson2019}. Spicules observed in lines of Mg~II, which is 18 times more abundant than Ca~II, but form at similar temperatures, show parabolic trajectories in height-time diagrams. \citet{Pereira2014}, again from SOT limb observations, reported that Type II spicules are the most common type seen in the quiet Sun (QS) and coronal holes (CH), whereas Type I spicules are seen mostly in active regions. 

On the disk, spicules were first identified with dark mottles \citep{Macris1957}, which are elongated absorbing features observed above network structures. They are most visible in the blue wing of the H$\alpha$ line, suggesting ascending motions. An example of modern observations of spicules on the disk in the H$\alpha$ blue wing are those of \citet{Samanta2019}. \citet{DePontieu_2007} conclude, based on a detailed assessment of type I spicule kinematics that they may be the limb manifestation of shock-wave-driven fibrils seen against the disk in active regions. \citet{Langangen2008} report ``rapid blueshifted excursions" (RBEs) in the blue wing wing of the Ca~II line at 8542\AA\  near chromospheric network. These were suggested to be the on-disk counterparts to spicules, an identification that was subsequently strengthened in studies by \citet{Sekse2013}, \citet{RouppevanderVoort2009} and \citet{Bose2019}.  Neverthless, they closely resemble classic dark mottles seen at high resolution \citep{Alissandrakis2022}. Disk spicules near the limb appear bright in the Mg II k line \citep{Pereira2014}.
\citet{Tian2014} suggest that high speed ``network jets" seen by IRIS on the solar disk may be counterparts to type-II spicules. \citet{RouppevanderVoort2015} report examples of network jets connected with RBEs and rapid red-shifted excursions (RREs), downflows observed in H$\alpha$. A study by \citet{Bose2021} in both Ca~II and the Mg~II~k line considered both RBEs and RREs and their relation to spicular heating and dynamics.  \citet{DePontieu2017} and \citet{Chintzoglou2018} suggest that the high apparent speeds of at least some network jets may be the result of heat fronts, not mass motions.

The distinction between the two types of spicules described above has not been without controversy  \citep{Carlsson2019, Sterling2021}. Using the same dataset as \citet{DePontieu2007}, \citet{Zhang2012} contested the type~I/type~II spicule classification. An analysis by \citet{Pereira2012}, however, confirmed the two spicules types. Furthermore, \citet{Pereira2013} showed that, by degrading modern observations of spicules in Ca~II~H and H$\alpha$ to an angular resolution and sampling cadence commensurate with older, ground-based observations, the ascending speeds inferred for type~II spicules were similar to those reported by \citet{Beckers1972}. \citet{Sterling2021} has suggested that spicules observed prior to 2000 -- solely from the ground, largely in H$\alpha$, mainly in quiet Sun (QS) and coronal hole (CH) regions --  be referred to as ``classical" spicules and that they correspond to type~II spicules. While the terminology persists in the literature, it may be a distinction without a difference - at least for QS and CH observations. Here, we will generally refer simply to ``spicules".

Past estimates suggest that spicules carry $\sim\!100$ times the mass flux of the solar wind upward \citep{Beckers1972, Athay1982} and may therefore play a significant role in the mass and energy budget of the chromosphere and corona \citep{Pneuman1978, Athay1982, Withbroe1983}, an idea that has persisted until the present time. \citet{Beckers1972} expressed doubt that the kinetic energy carried by spicules was energetically important to coronal heating but believed that spicules may play an important role in supplying mass to the solar wind. \citet{Pneuman1978} suggested spicular material was heated to coronal temperatures, with most of it returning to the chromosphere as gentle downflows. \citet{Withbroe1983} concluded, based on a study of EUV emission, that spicules may contribute to heating the upper chromosphere but are unlikely to play a significant role in the chromosphere-corona energy balance unless they are significantly heated below $\lesssim 15$ Mm. {Modern observations show that spicular plasma is indeed heated.} \citet{Pereira2014} analyzed spicule observations made by {\sl Hinode}, SDO, and IRIS covering lines of Ca II, Mg II, Si IV, and He II, sensitive to temperature ranging from $\sim 10^4$ K to transition region (TR) temperatures of $\sim 10^5$~K finding that, while spicules fade from view in Ca II as noted above, they become visible in hotter spectral lines as they rise; see also \citet{Tian2014}. \citet{Skogsrud2015}, who also included C~II in their analysis, confirmed that heating occurs in spicules to at least TR temperatures, possibly toward the top of the spicules. Interestingly, with the exception of Ca~II, parabolic trajectories are seen in the height-time diagrams of the majority of spicules they analyzed, even in He~II, and Si~IV, implying the presence of TR temperatures along their length \citep{Carlsson2019}.

These modern observations of spicules have provoked intense interest, renewing proposals that spicules play an important role as a source of hot plasma in the solar corona (see, e.g., \citealt{DePontieu2009, DePontieu2011, Tian2014, RouppevanderVoort2015, DePontieu2017}) which have, in turn, stimulated new models of spicules \citep{MartinezSykora2013, Iijima2017, MartinezSykora2017, MartinezSykora2018, Antolin2018}. The importance of spicules in coronal heating has been questioned, however, by \citet{Klimchuk2012, Judge2012, Zhang2012, Patsourakos2014}, and \citet{SowMondal2022}. However, these results are based on simplifying assumptions that do not necessarily address the full complexity of the environment in which spicules occur. In summary, while a great deal of progress has been made over the past fifteen years in refining the observational characteristics of spicules and how they may be heated, questions remain as to whether or to what degree spicules contribute to the mass-energy budget of the solar atmosphere. 

Observations in the O/UV/EUV wavelength bands have been mainstays in characterizing spicule temperatures, densities, and velocities.  Here, we offer a complementary perspective on solar spicules: high-angular- and high-time-resolution observations of spicules at \mml\ with the Atacama Large Millimeter-submillimeter Array (ALMA; \citealt{Wootten2009}). Observations of solar phenomena at \mml\ complement those at O/UV/EUV wavelengths. Emission from the Sun at \mml\ is in local thermodynamic equilibrium (LTE; e.g., \citealt{Vernazza1976}). The source function is therefore Planckian and since $h\nu/k_B T\ll1$ the Rayleigh-Jeans approximation is valid. It has been known for some time that the ionization/recombination time scales of hydrogen are longer than the dynamical time scales \citep{Carlsson2002} in the chromosphere. The degree of ionization may therefore depart significantly from equilibrium conditions. Hence, while the source function is in LTE at \mml\ the opacity can depart from its LTE value \citep{Loukitcheva2004}, a point we discuss further in \S4.2.


Observations of the solar chromosphere at mm/submm wavelengths have been performed for decades but most studies were performed with single dishes and, therefore, with low angular resolution. For example, \citet{Kundu1971} used an 11~m telescope to map the Sun at 1.2~mm; in the 1980s and 1990s, the {\sl Infrared Telescope Facility} \citep{Lindsey1981}, {\sl James Clerk Maxwell Telescope}  \citep{Lindsey1992}, and the {\sl Kuiper Astronomical Observatory} \citep{Roellig1991} became available for occasional use for solar observing at mm/submm wavelengths. While these were still relatively low in angular resolution they were sufficient to establish the gross properties of the solar chromospheric emission at mm/submm wavelengths across the solar disk. In 1991, a number of groups exploited a total solar eclipse of the Sun, using the knife edge of the lunar limb to provide extremely high angular resolution in one dimension. Using the {\sl Caltech Submillimeter Observatory} (CSO) at 850$\mu$ \citet{Ewell1993} measured the position of the solar limb extended well above the optical limb. A similar measurement made at 3~mm of the same eclipse at the {\sl Owens Valley Radio Observatory} (OVRO) also found a significant limb extension \citep{Belkora1992}. Combining these with other measurements made in previous eclipses, spanning 200$\mu$ to 3 mm, it is quite clear that the mean radius of the solar limb increases systematically with wavelength -- far above the radius expected from semi-empirical models for which hydrostatic equilibrium is assumed such as those of \citet{Vernazza1981} and \citet{Fontenla1990}. The limb extension has been attributed, at least in part, to the presence of spicules. In particular, the ensemble of spicules along the line of sight have sufficient optical depth to free-free absorption to render the chromosphere optically thick at significantly greater heights than the optical limb. \citet{Rutten2017} suggested that the chromospheric canopy also likely contributes to limb extension. In any case,  single dish limb extension measurements, even using an eclipse, necessarily represent a mean height since azimuthal structure remained unresolved.

ALMA provides dramatic improvements in the angular resolution and time resolution with which spicules may be studied at \mml. The first observations of solar spicules using ALMA were performed at a wavelength of 3~mm \citep{Yokoyama2018, Nindos2018, Shimojo2020}; see also the review by \citet{Alissandrakis2022a} for additional details. In this paper, we report observations of spicules in a polar coronal hole made by ALMA on 2018 December 25 in two \mml\ bands: 3~mm and 1.25~mm. Although our focus is on \mml\ emission we qualitatively compare our \mml\ observations with those made in O/UV/EUV wavelength bands.  More quantitative comparisons are deferred to other publications. In \S2 we describe the observations obtained by each instrument and then describe the approach we took to ALMA imaging in \S3.  We discuss our results in \S4. Specifically, we compare \mml\ images with those in O/UV/EUV wavelengths. We then discuss sources of opacity at \mml, the mean height of the chromosphere at \mml, spicule temperatures, electron number densities, the impact of spicule filling factors, and briefly touch on spicule lifetimes and kinematics at \mml. We summarize our observational results in \S5 and discuss their implications for the mass-energy budget of the low solar atmosphere. We conclude in \S6 with prospects for future work. 

\section{Observations} \label{sec:OA}

The observations reported here were made during ALMA observing Cycle~6 on 2018 December 25 -- Christmas Day. The Sun was extremely quiet, with no numbered NOAA active regions visible on the disk. The target was the limb of the north polar coronal hole (Fig.~1). Observations were performed in ALMA band 3 (100 GHz; 3~mm) and in band 6 (239 GHz; 1.25~mm). The helioprojective pointing of ALMA in both bands was $(0", 984")$. In addition to the ALMA observations at \mml, we obtained observations in O, UV and EUV wavelengths with GONG, IRIS, and SDO/AIA, respectively. Table~I lists the times for which ALMA and IRIS observations were made. GONG observations are generally available from the Cerro Tololo site in Chile, which is proximate to ALMA. SDO observations are available at all times. 

\begin{deluxetable}{lc}
\tablecaption{Observing Log: 2018 Dec 25}
\tablehead{\colhead{Instrument} & \colhead{UT Time} }
\startdata 
ALMA (3~mm)&\\
\tableline
Scan 1&14:02:48.4 - 14:12:55.6 \\
Scan 2&14:15:28.0 - 14:25:35.2\\
Scan 3&14:28:09.0 - 14:38:16.2\\
Scan 4&14:40:52.0 - 14:47:57.7 \\
\tableline
ALMA (1.25~mm)& \\
\tableline
Scan 1&17:17:29.1 - 17:24:12.2 \\
Scan 2&17:26:32.7 - 17:33:15.8 \\
Scan 3&17:36:04.2 - 17:42:47.3 \\
Scan 4&17:45:06.2 - 17:51:49.3 \\
Scan 5&17:54:36.7 - 18:01:19.8 \\
Scan 6&18:03:37.8 - 18:10:20.9 \\
\tableline
IRIS&\\
\hline
Interval 1&12:00:00 - 12:58:00 \\
Interval 2&13:29:00 - 14:34:00 \\
interval 3&16:44:00 - 17:48:00 \\
Interval 4&18:21:00 - 19:26:00 \\
\enddata
\end{deluxetable}

\bigskip\bigskip

\subsection{Optical, Ultraviolet, and Extreme Ultraviolet Observations}

Observations in H$\alpha$ were obtained from GONG \citep{Harvey1996}, operated under the National Solar Observatory Integrated Synoptic Program (NISP). GONG comprises a network of six observing stations that provide high cadence imaging at selected wavelengths for the purposes of helioseismology and space weather monitoring. We make use of 860$\times 860$ pixel H$\alpha$ images with a pixel size of $2.5"$ obtained at 1 min cadence by the GONG site at Cerro Tololo in Chile. Cutouts of the north polar coronal hole in H$\alpha$ were obtained and the mean brightness as a function of radius was subtracted. 

Observations by IRIS were coordinated with ALMA observations. We later compare ALMA images with those obtained by the IRIS Slit-Jaw Imager. The SJI observes in 4 spectral windows in the UV, including the 2796 \AA\ window with the Mg~II~k lines formed at temperatures from $\sim\!5-15\times 10^3$~K, and the 1400 \AA\ window with the Si IV doublet formed at temperatures between $\sim\!5-80\times 10^3$~K. Only the second IRIS observing interval overlaps with ALMA 3~mm observations whereas the third interval overlaps with ALMA 1.25~mm observations. A total of 208 images is available for intervals 2 and 3 in each bandpass with a cadence of 18.7~s except for those times between intervals when the spacecraft was eclipsed.  Each SJI image has a pixel size of $0.33"$ and Nyquist resolution of $0.66",$ and a size of approximately $102"\times 113"$. The spacecraft roll angle was fixed at $30^\circ$ relative to the north polar axis during the course of the observations, as illustrated in Fig.~1.

SDO/AIA observations are available in seven EUV bandpasses sensitive to plasma temperatures ranging from $6\times 10^4$~K to $2\times 10^7$~K. In addition, the AIA provides UV observations in the 1600 and 1700 \AA\ bandpasses. As SDO is in a geosynchronous orbit, full disk images are available on a continuous basis in each of these bandpasses with an angular resolution of $1.5"$ on a cadence of 12s. Cutouts of the AIA images in each of the EUV bands centered on the ALMA coronal hole target were obtained. The time series of images in each bandpass are $396\times 383$ pixels, and each pixel is $0.6"\times0.6"$. The image size in each bandpass is therefore approximately $238"\times 230"$. 
For the present work, we compare AIA observations in the 171 \AA\ and 304 \AA\ EUV bandpasses with the \mml\ observations. The former corresponds to plasma temperatures of $\sim 8\times 10^5$~K \citep{Dere1997, DelZanna2021} whereas the latter, dominated by the He II line, corresponds to temperatures of $\sim\!75-100\times 10^3$~K \citep{Golding2014}. Note that these temperatures are somewhat higher than those given by \citet{Lemen2011}. O/UV/EUV images were processed to enhance the visibility of spicules above the limb, as described in \S4.1

\begin{figure}[ht]
\begin{center}
\includegraphics[angle=0,clip,width=6.5in]{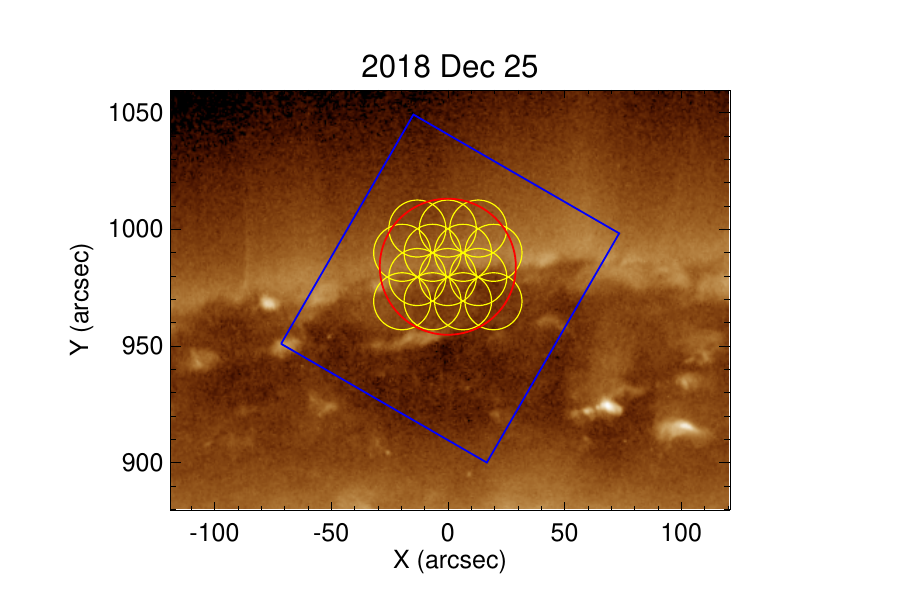}
\caption{Detail of the northern coronal hole on 2018 Dec 25. The background image shows the solar limb as observed by the SDO/AIA in the 171 \AA\ filter. The circle outlined in red shows the ALMA field of view -- the FWHM of the primary beam -- at 3~mm (100 GHz). The yellow circles show the 1.25~mm field of view for each of the 14 mosaic pointings (see text). The box outlined in blue shows the IRIS field of view. }
\end{center}
\end{figure}

\subsection{ALMA Observations}

An overview of solar observing with ALMA is given by \citet{Bastian2022}. ALMA solar observing programs typically include two complementary types of observations: high resolution interferometric (INT) maps with a limited field of view \citep{Shimojo2017} and low resolution full disk total power (TP) maps \citep{White2017}. Both types of data were used in the present study. 

\subsubsection{Interferometric Observations}

A total of fifty antennas was available in the interferometric array on 2018 Dec 25, forty-one 12~m antennas (the ``12-m array") and nine 7~m antennas (the Atacama Compact Array -- ACA). The 12-m array is reconfigurable while the ACA is fixed. The 12-m array was in the C43-3 antenna configuration, the highest resolution configuration available for solar observations in ALMA Cycle 6. 



ALMA observations at 3~mm (100~GHz) were performed from 13:44-14:49~UT using a single pointing (Fig.~1). Initial calibrations typically require $\approx\!20$ min; these include pointing, sideband ratio, and spectral bandpass calibration. Source observations were carried out from 14:03-14:48 UT in three source scans, each 10~min in duration, plus a partial scan (Table 1). The time resolution of the 3~mm observations is 2~sec. Each source scan was interleaved with scans to observe a gain calibrator for 2~min and to measure quantities used to calibrate the flux scale (see \citealt{Shimojo2017}).  In the present case, J1256-0547 (3C279) was used as the bandpass calibrator and J1733-1304 was used as the gain calibrator.  Observations were made in dual-polarization mode (linearly polarized correlations XX and YY) in four spectral windows: 92-94, 94-96, 104-106, and 106-108~GHz. The data were acquired in the Time Domain Mode where each spectral window is subdivided into 128 spectral channels. Note that true spectral line observations of the Sun with ALMA have not yet undergone science verification and commissioning and are therefore not yet supported. In practice, it is convenient to average over channels in each spectral window to form pseudo-continuum databases for imaging and analysis. Amplitude loss due to bandwidth smearing is small for ALMA frequency bands \citep{Thompson2017}. As we discuss in the Appendix, only the 12~m array was used to image the 3~mm emission; i.e., correlations between 7~m antennas and between 7~m and 12~m antennas were discarded. The field of view (FOV) was therefore determined by the response function of a 12~m antenna, referred to as the primary beam. The primary beam is well approximated by a Gaussian with a full width at half maximum (FWHM) of $58.3"$ (average frequency of 100 GHz). The angular resolution is determined by the array configuration. For the C43-3 observations reported here the synthesized beam of the 3~mm observations is characterized by an elliptical Gaussian with FWHM dimensions of $2.2" \times 1.3"$ and a position angle measured East from (celestial) North of approximately $73^\circ$. 

The 1.25~mm (239 GHz) observations followed the 3~mm observations from 16:58-18:19~UT. The observations began with the same calibrations as those performed for 3~mm but J1924-2914 was used as the bandpass calibrator; the gain calibrator was again J1733-1304. A total of six source scans, each approximately 6.5~min in duration, was observed between 17:17-18:18 UT, again interleaved with observations of the gain calibrator and measurements used for flux calibration. The four spectral windows observed were 229-231, 231-233, 245-247, and 247-249~GHz and each spectral window was again subdivided into 128 spectral channels. The FOV of a single pointing with a 12~m antenna  at 1.25~mm is  $24.3"$ (239~GHz).  To form a larger FOV in the 1.25~mm wavelength band, comparable to that of the 3~mm band, mosaicking techniques were used as shown in Fig.~1 and discussed in the Appendix. We have used both 7~m and 12~m antennas in the heterogeneous array to image the 1.25~mm emission, which has implications discussed in the Appendix. The angular resolution of the 1.25~mm observations is characterized by an elliptical Gaussian with FWHM dimensions of $1.5" \times 0.7"$ and a position angle of approximately $84^\circ$.


\subsubsection{Full Disk Mapping Observations}

In addition to the 12-m array and the ACA, ALMA has four total power antennas, each 12~m in diameter. One or more of these are used in solar observing programs to make full disk low-resolution TP maps of the Sun to supplement the interferometric data, as discussed in greater detail in the Appendix. The resolution of the TP maps is determined by the FOV of a 12~m ALMA: $58.3"$ and $24.3"$ at 3~mm and 1.25~mm, respectively. TP mapping employs fast-scan mapping techniques developed by \citet{Phillips2015} where a ``double-circle" scanning pattern is used to map a region centered on the Sun that is 2400" in diameter \citep{White2017}. The 3~mm TP map on 2018 Dec 25 was scanned in 13~min, centered on 13:57~UT.  We used the recommended brightness temperature scaling of \citet{White2017} for the 3~mm map, a central quiet Sun brightness temperature of 7300~K. We note that \citet{Linsky1973}, who recalibrated extant measurements of the quiet Sun brightness against the Moon, reports 3~mm values of 7425~K and 7148~K for linear and quadratic fits to the data, respectively.  The 1.25~mm TP map was scanned in 17~min, centered on 12:58~UT. A multi-band analysis of the center-to-limb brightness distribution of the quiet Sun by \citet{Alissandrakis2022} has shown that ALMA TP maps in band 6 (1.25~mm) and 7 (0.86~mm) can be cross-calibrated against band 3. The brightness temperature of the center of the solar disk at 1.25~mm is 6347~K, a value that is 7.6\% larger than the 5900~K value recommended by \citet{White2017}. The linear and quadratic fits of \citet{Linsky1973} yield values of 5990~K and 6311~K, respectively.  



\section{ALMA Data Reduction and Analysis}

The ALMA 3~mm and 1.25~mm data were calibrated by analysts at the North American ALMA Science Center using the Common Astronomy Software Applications package (CASA; \citealt{Mcmullin2007, Team2022}) and special utilities for solar data calibration \citep{Shimojo2017}. The calibrated data were then averaged over spectral channels in each spectral window as described in \S2 and measurement sets containing only solar target data were exported. Two analysis paths were used to image the ALMA data. For the 3~mm data, the Astronomical Image Processing System (AIPS; \citealt{Greisen2003}) was used as the more expedient approach for mapping the continuous time series of snapshot images made by the 12-m array, similar to the approach of \citet{Nindos2018}. CASA was used as the more convenient approach for handling the 1.25~mm mosaic data made with a heterogeneous array. Only the first three scans of 3~mm data were analyzed. The fourth scan was a partial scan and had no overlap with IRIS observations. 

We note that ALMA data are calibrated in spectral flux density units of Jansky (1~Jy $\equiv 10^{-23}$~W~m$^{-2}$~Hz$^{-1}$) and images are typically expressed in terms of flux density per resolution element (the ``synthesized beam"). When the synthesized beam $\Omega$ is converted to steradians the images have the same units as specific intensity. It is  convenient to convert Jy/beam to brightness temperature $T_b$, where the flux density per beam $S$ is related to $T_b$ through $S=2k_B T_b \Omega/\lambda^2$, where $k_B$ is Boltzmann's constant and $\lambda$ is the wavelength of the radiation in question. We have converted all images to brightness temperature with Kelvin units. It is important to distinguish between $T_b$, a proxy for specific intensity, and the effective temperature $T_e$ of the emitting material. Consider thermal emission from a uniform slab of material with a temperature $T$.  In this case, the effective temperature of the plasma $T_e=T$ (in general, $k_BT_{e}=\langle E\rangle$, the average energy of the emitting particles). The radiative transfer equation can be then be expressed simply as $T_b=T_e[1-\exp(-\tau)]$ where $\tau$ is the optical depth along the line of sight. When $\tau >> 1$ the slab is optically thick and $T_b=T_e=T$; the observed brightness is a direct measure of the temperature of the emitting material. When $\tau<<1$ the slab is optically thin and we have $T_b=\tau T_e$. For a given plasma temperature, the brightness temperature is directly related to the optical depth.

Solar observations with ALMA present a number of data analysis challenges. First, the sky is variable as a result of water vapor over the array which, in turn, can introduce significant phase variability, degrading the ``seeing". Hence, self-calibration techniques are essential. Second, as an interferometer, ALMA acts as a high-pass filter. It does not measure largest angular scales corresponding to spatial frequencies less than the minimum antenna separation. Since most of the flux density is contained in the largest angular scales, their recovery is important for those programs that require accurate brightness temperatures. Third, as previously noted,  the small field of view of the array at 1.25~mm necessitates the use of mosaicking, where the instrument samples a user-specified grid of antenna pointings on the sky that are combined to form a larger image in post-processing (Fig.~1). Finally, the use of a heterogeneous array introduces subtleties that users should be aware of. We discuss each of these challenges in greater detail in the Appendix and now describe our approach to imaging the ALMA observations of spicules on the solar limb.


\subsection{Imaging and Self-calibration}

The 12-m array INT data were imaged and self-calibrated using the AIPS tasks {\sl imagr} and {\sl calib}, respectively, to deconvolve the sidelobe response of ALMA's point spread function (PSF or the ``dirty beam") from the raw snapshot images. Atmospheric seeing largely manifested as image wander as well as a small loss of coherence on the longest baselines. We adopted a hierarchical approach to self-calibration, initially referring all data to an average image of the limb per scan. The procedure solved for antenna-based phase corrections; solving for both amplitude and phase gain corrections did not lead to significant improvement in the images. The solution interval was reduced over the course of successive iterations down to the integration time of  2~s.  We refer to images made at the integration time of 2~s as ``snapshot" images. Fig.~2 shows the angular displacement of each snapshot image relative to the scan average for each of the three 10~min scans (top row) and as time series of the radial displacement relative to the mean (bottom row). The displacements were determined by cross-correlating a final self-calibrated snapshot image with the corresponding un-self-calibrated  image. We characterize image wander as the standard deviation $\sigma_d$ of the angular displacement from the mean. We find that $\sigma_d=0.83"$, $1.38"$ and $0.70"$ for scans 1, 2, and 3, respectively.  The time scale for image wander is typically several 10s of seconds, comparable to the time $t_D=D/v_W$ for turbulent eddies comparable in size to the maximum baselines ($D\sim500$~m), to drift over the array for a wind speed $v_W\sim 5-10$ m/s. The net effect of self-calibration for the 3~mm data is to stabilize image wander and to sharpen the images somewhat. While the relative positions of snapshot images within a given scan are quite stable after self-calibration, their absolute position is subject to a systematic error due to uncorrected scan-to-scan variation. We address this in the next section. The magnitude of this error relative to the nominal pointing is of order $\sigma_d$, or $1"$. 

\begin{figure}[ht]
\begin{center}
\includegraphics[angle=90, width=6in]{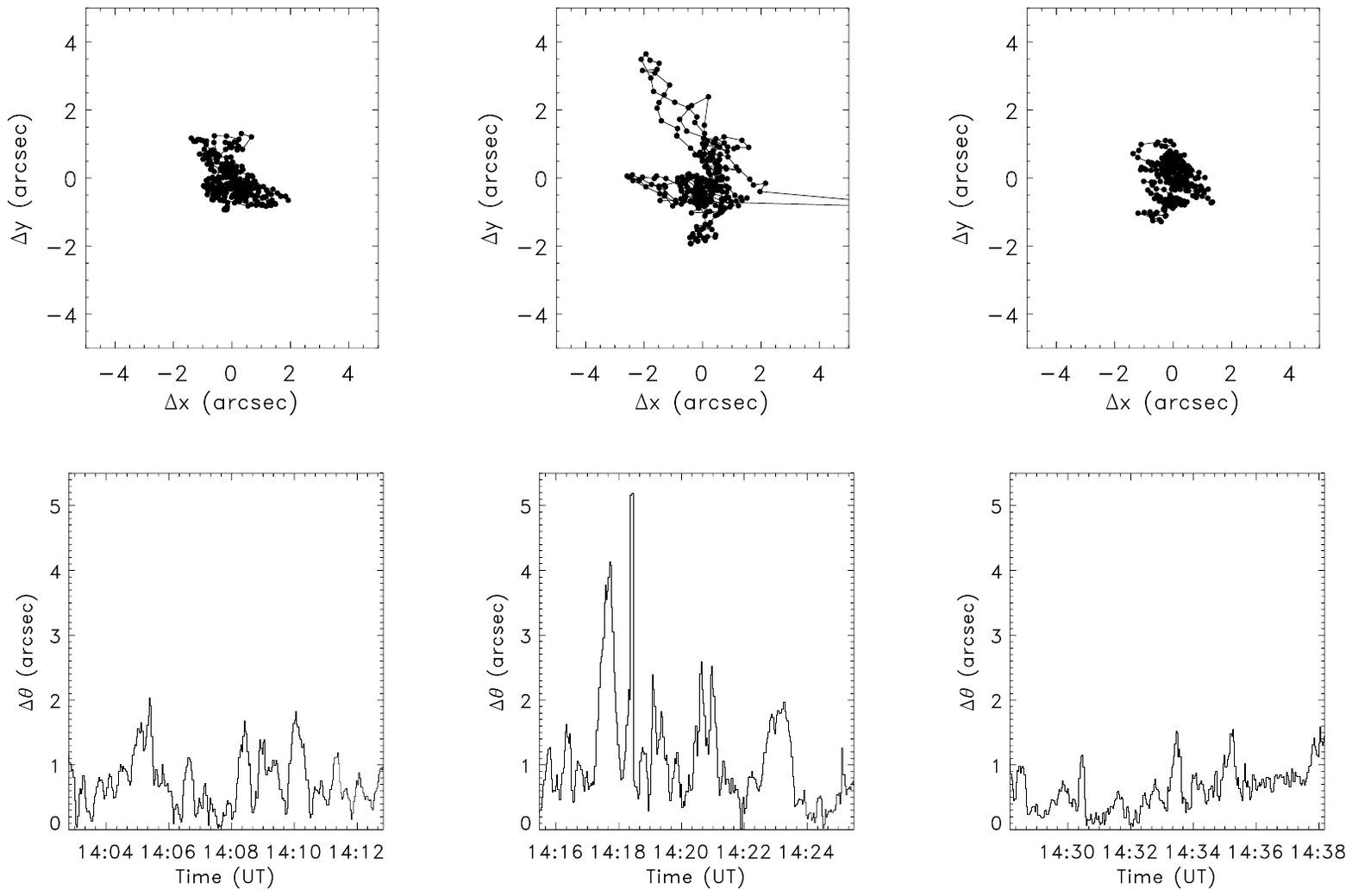}
\caption{Image wander as deduced from self-calibration of the band 3 data for each of the three successive 10~min scans. The top row shows the shift of the image relative to the mean position for each 2~s snapshot. The bottom row shows the radial deflection from the mean as a function of time. }
\end{center}
\end{figure}

In the case of the 1.25~mm observations, a fourteen-point mosaic pattern was used to image spicules, as illustrated in Fig.~1. Each mosaic required nearly 2~min to execute, limiting the time resolution to no better than that. Moreover, there was a limitation imposed by the number of subscans per scan (discrete mosaic pointings per scan) that was non-commensurate with completing an integer number of mosaics per scan.  As a result roughly every fourth mosaic was interrupted by a calibration scan.  These limitations will be addressed in future observations of this kind. The implication for these 1.25~mm observations is that the dynamics of the spicules were poorly resolved at 1.25~mm. A total of 21 mosaic images was completed but only 15 of these were formed from uninterrupted pointing sequences.

Mosaic images were produced in CASA using the task {\sl tclean} with the {\sl mosaicft} gridding function to properly handle mosaicking.  Self-calibration was somewhat less straightforward for the 1.25~mm data than it was for the 3~mm data as a result of the mosaic imaging strategy. For the 1.25~mm data, antenna-based phase solutions were first calculated jointly for the 14 mosaic pointings within a scan; that is, a common phase solution was applied to each antenna for all pointings. Antenna-based phase solutions were then calculated jointly using each pointing within a scan. Finally, solutions were calculated for each pointing in each mosaic within a scan. The CASA task {\sl gaincal} was used to self-calibrate the data in phase each iteration and the task {\sl applycal} transferred the solutions to the visibility data. A given antenna pointing was revisited every $\approx2$ min, at best, which is too long to track image wander on time scales less than the mosaic completion time. While self-calibration can improve mosaic alignment relative to an average model, there is a systematic error in the position of the mosaic images that is difficult to determine and correct in the absence of a reference. As we discuss in the next subsection, we adopt the solar limb as our reference and, based on the shift required to reconcile the solar limb at 1.25~mm with the mosaics, the systematic error is of order $2"$. 

\subsection{Solar Limb Reference}

To establish an absolute limb position, we refer the self-calibrated interferometric snapshot maps and mosaics to the limb determined from the corresponding full disk TP maps. Note that we find no significant azimuthal dependence of the limb brightening in either wavelength band; the coronal polar hole has the same degree of limb brightening as quiet regions on the limb. Following \cite{Alissandrakis2020}, we proceeded as follows: first, the map was passed through a radial gradient filter, which results in an annulus, the maximum of which corresponds to the location of the limb (Fig.~15, Appendix). The 3~mm TP map observed on 2018 Dec 25 is shown in Fig.~3a.  Inspection of the limb revealed faint, regular azimuthal artifacts that we attribute to small residual timing errors in the double-circle scanning pattern used, an effect first noted by \citet{Alissandrakis2017}. We fit a circle to the annulus, excising points affected by scanning artifacts. A radius of $982.2"$ was found for the 3~mm TP map. A similar procedure was performed for the 1.25~mm data (Fig.~3b) for which a radius of $978.3"$ was found for the 1.25~mm TP map. We estimate that the uncertainty for both fits is less than 10\% of the primary beam width for each band; i.e, $\lesssim 5.8"$ and $\lesssim 2.4"$ for the fitted heights of the 3~mm and 1.25~mm limb, respectively. We acknowledge that the uncertainty in the limb position is relatively large and the use of TP maps as a position reference is therefore not entirely satisfactory. As we shall see, small adjustments to the reference radii are in fact needed to bring the two bands into relative agreement (\S\S4.2,4.5). We also note that the procedure described is helpful in establishing reference radii but it cannot provide a reference for correcting positional uncertainties in the east-west direction. While this is unimportant for analysis of the \mml\ data, we found it necessary to introduce small EW shifts in comparing ALMA maps with images from other instruments (\S4.1).


\begin{figure}
\begin{center}
\includegraphics[width=3.5in]{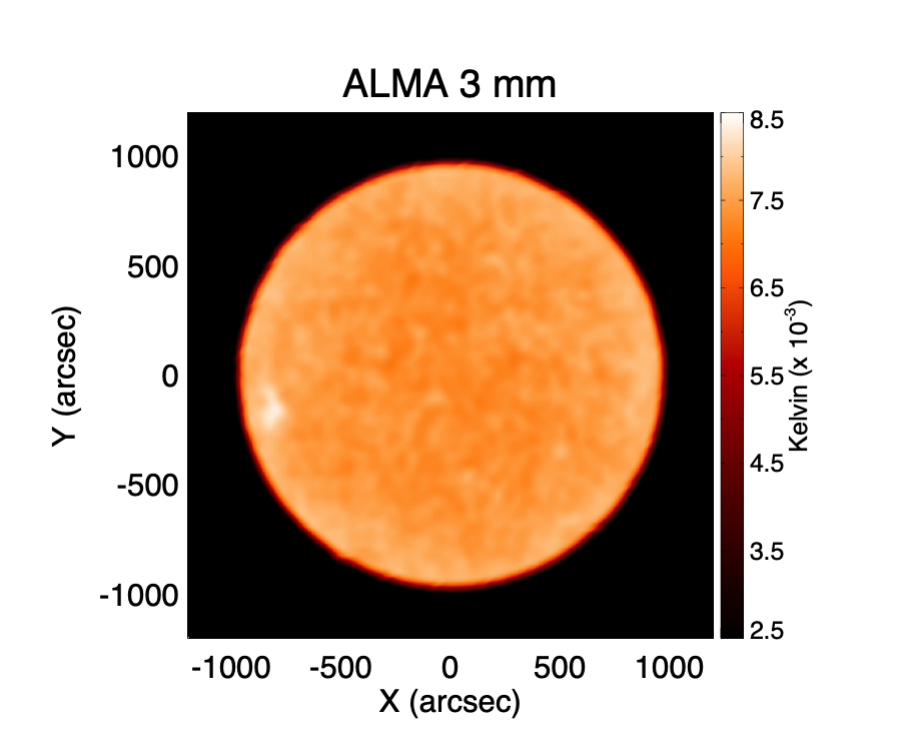}
\includegraphics[width=3.5in]{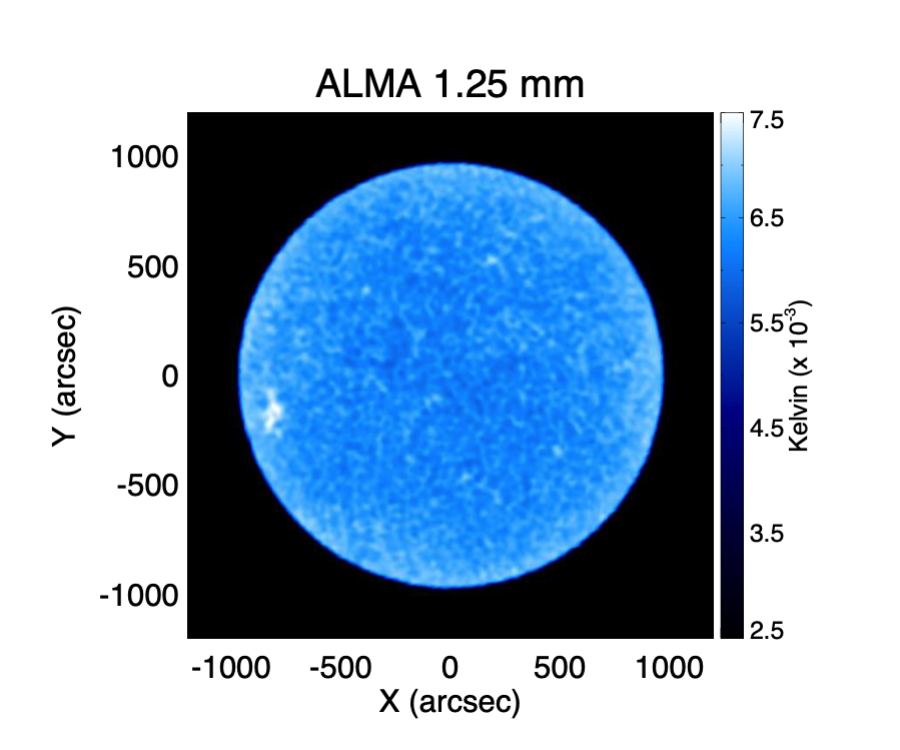}
\includegraphics[width=3.25in]{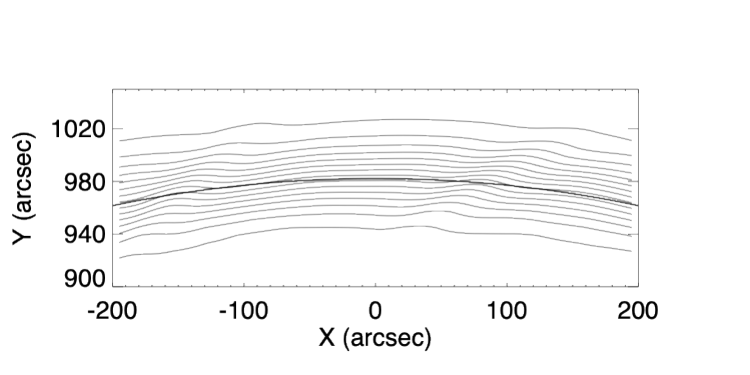}
\includegraphics[width=3.25in]{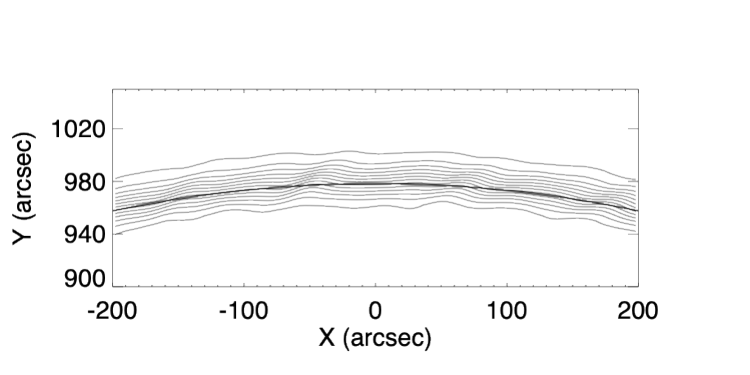}
\caption{Top: Full disk maps of the Sun on 2018 December 25 using fast-scan total power mapping techniques at 3~mm (left) and 1.25~mm (right). Bottom: Artifacts resulting from residual timing errors in the fast-scan TP maps are visible as faint regular ridges at the north pole in both the 3~mm TP map (left) and the 1.25~mm TP map (right). The contour interval is 500~K and the lowest contour level is 500~K (top of each panel). The heavy solid line in each panel represents the mean limb radius.}
\end{center}
\end{figure}


The limb determined from the TP map represents an average over any structure present in the (single-pointing) field of view. The radius of an un-restored INT map at a given azimuth is the location where the apparent brightness crosses zero (Fig.~15). Therefore, for each INT map, we found the zero-level contour and calculated its mean radius. Using the fitted radius of the 3~mm TP map, we shifted the 3~mm INT snapshot maps to ensure that their mean radius matched the nominal radius of $982.2"$.  In the case of the 1.25~mm maps, the zero-contour was similarly determined for each mosaic, its mean calculated, and the map was shifted to match the nominal mean limb radius of $978.3"$.

\subsection{Restoration of Large Angular Scales}

We used two different approaches to recover the largest angular scales in each band, both involving models of the background solar disk instead of the TP maps as scanned. Models were used: i) to avoid introducing the faint scanning artifacts into the restored maps, and ii) to allow increased flexibility in the restoration process. For the 3~mm band we formed the model disk using the quiet Sun brightness temperature and center-to-limb brightness profile determined by \citet{Alissandrakis2022}, with a radius as determined by fitting the TP data.  The model TP map was then multiplied by the 12~m beam response function at the telescope pointing coordinate $(0,984")$. A model of the 1.25~mm limb-brightened disk of the Sun was similarly constructed, but multiplied by effective beam response resulting from mosaic imaging. 

The first approach to recovering the largest angular scales was to subtract a model of the ALMA response to the solar disk, multiplied by the primary beam response, from the self-calibrated visibility data. This was achieved by Fourier transforming the model of the background solar disk at the limb, sampling it with the relevant snapshot {\sl uv} coverage of the 12-m array, and subtracting the result from the INT data. This step was performed using AIPS task {\sl uvsub} (3~mm data) and the pair of CASA tasks {\sl ft} and {\sl uvsub} (1.25~mm data). Maps of the residual were then formed and the PSF was deconvolved. The model disk was then added back to the (clean) residual maps in the image domain. Finally, each snapshot map or mosaic was corrected for the 12~m beam response in the case of the 3~mm data or the mosaic beam response for the 1.25~mm data. 

The second approach to this problem was to use a modified version of ``feathering " \citep{Cotton2015} to combine model TP maps with snapshots or mosaics.  Briefly, feathering a TP map and interferometric data involves the following steps: i) ensure that both the TP and interferometric images are the same size and resolution and scaled to identical physical units; ii) Fourier transform (FT) each map; iii) form a weighting mask $m=1-g_{TP}$ where $g_{TP}$ is the real part of the FT of the TP antenna power pattern normalized to unity; iv) multiply the interferometric data by the weighting mask $m$; v) sum the TP and masked interferometric data in the Fourier domain; and vi) perform an inverse FT to recover the result. We used a modified feathering scheme in which INT snapshot maps (3~mm) or mosaics (1.25~mm), uncorrected for the telescope beam response, were feathered with the model TP data, also uncorrected for the beam response. Moreover, we ``tuned" the resolution of the model TP map, and hence the mask applied to the INT data, to ensure sufficient overlap between the two in the {\sl uv} domain. For the 3~mm data we used an effective resolution of $35"$ for the TP data; for the 1.25~mm data we used $15"$. 

We found that the first approach (disk-subtraction) approach worked well for the 3~mm data, eliminating the overshoot response. It was not successful for the 1.25~mm data, resulting in mosaics with large residual errors. We attribute the failure of the approach for the 1.25~mm data to the sparse snapshot sampling of the inner {\sl uv} plane, and to time variability of the source during the nearly 2~min required to complete a given mosaic. In contrast, we found that the second approach (feathering) was less effective in removing overshoot from the 3~mm snapshot maps but was superior to the disk-subtraction method for the 1.25~mm mosaics. 

Fig.~4a-c shows summarizes the various reduction steps for a representative 3~mm snapshot. The results of imaging, self-calibration, and restoration are three data cubes each containing time series of 300 snapshot images. The relative alignment of snapshots within a 10~min data cube is of order 0.5 pixels, or $\approx 0.1"$ but the uncertainty in the absolute radial position, which we take to be the same as the uncertainty in the position of the limb, $\approx 5.5"$. The reduction steps for a representative 1.25~mm mosaic are shown in Fig.~4d-f. We have far fewer maps of spicules at 1.25~mm due the need to mosaic, a total of only 21 maps. Here, too, the uncertainty of the mean radius of the limb at 1.25~mm is taken to be that given by the full disk fit, $\approx2.5"$. Examples of restored maps at 3~mm are shown in Fig.~5 for three different times in the left-hand column. The right-hand column shows the corresponding images for which the minimum brightness at each radius has been subtracted to better reveal spicules above the limb. Fig.~6 shows the same for 1.25~mm mosaic images. For both figures, the dotted line represents the average limb whereas the solid line represents the top of the chromosphere (see \S4.2).

We conclude this section with a brief discussion of the residual uncertainties in the brightness temperature $T_b$ in the reconstructed snapshot maps (3~mm) and mosaics (1.25~mm). While maps are ultimately limited by thermal noise due to antenna electronics, ground spillover, the atmosphere, and noise from the Sun itself \citep{Shimojo2017}, in practice variations in $T_b$ are dominated by residual sidelobe clutter due to imperfect calibration and deconvolution. We note, too, that after correcting for the primary beam taper, noise toward the edges of the FOV is amplified by a factor of $\sim\!2$. Observers typically establish the {\sl rms} brightness temperature variation, $\sigma_{T_b}$, empirically by measuring it in a region free of signal. We find $\sigma_{T_b}\sim 100-200$~K in both wavelength bands, the upper end of the range being at the edge of our FOV. In our analysis in later sections, we conservatively limit our measurements to features with a signal-to-noise ratio $5-10\sigma_{T_b}$, or 1000~K.


\begin{figure}[ht]
\begin{center}
\includegraphics[width=7.25in]{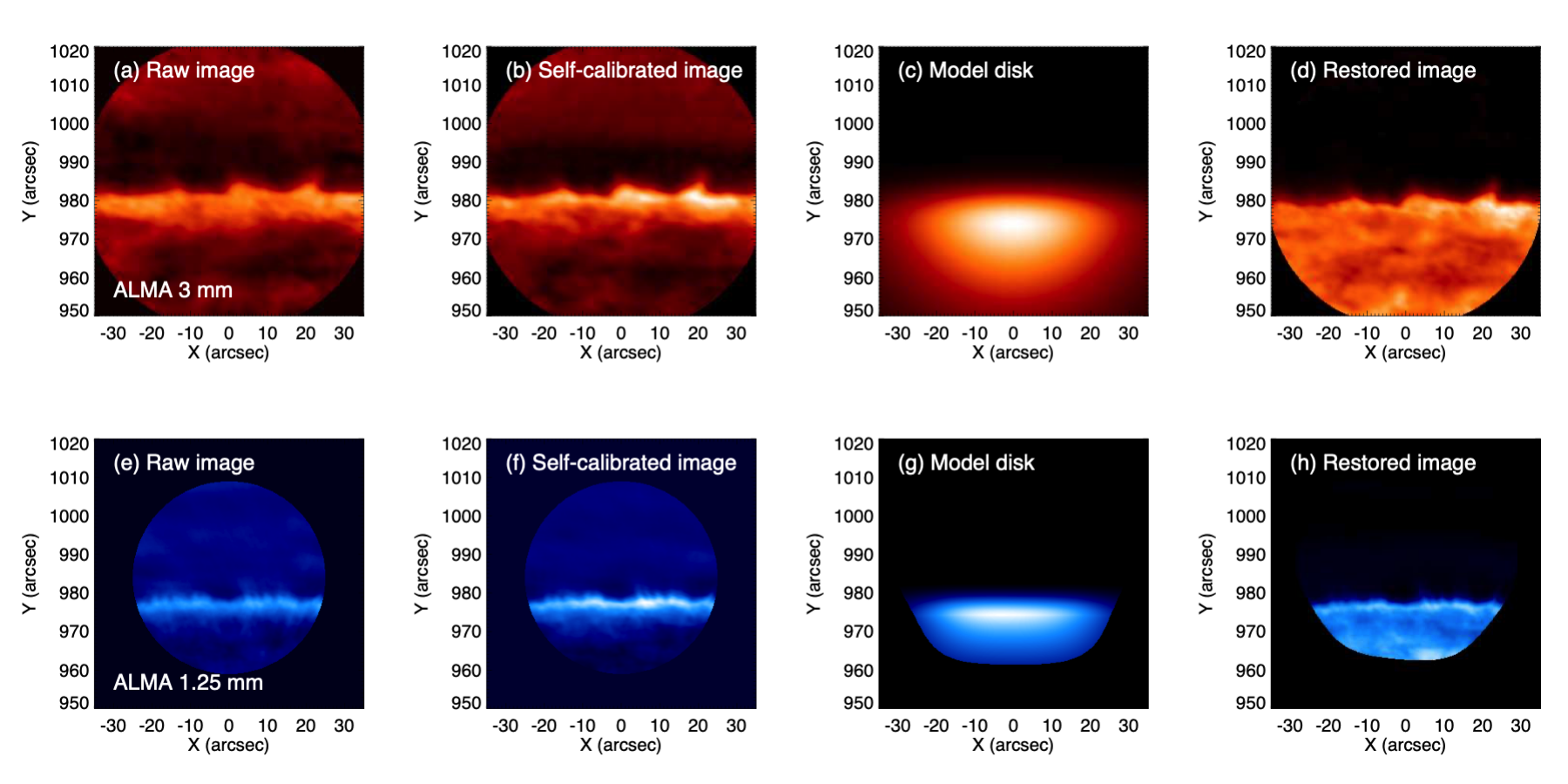}
\caption{Examples of images at various stages of processing. Top row: a 2-s snapshot image (14:02:49.4 UT) at a wavelength of 3~mm. a) an image made from interferometric data that has not been self-calibrated; b) an image made from self-calibrated data; c) the model brightness distribution used to restore large angular scales; d) a final image in which the large angular scales have been restored. Bottom row: a mosaic at a wavelength of 1.25~mm constructed from interferometric data between 17:19:52.8-17:21:48.6~UT. e) using data that have not been self-calibrated; f) using self-calibrated data; g) the model brightness distribution used to restore large angular scales; h) a final image for which the large angular scales have been restored.}
\end{center}
\end{figure}

\newpage

\begin{figure}[t]
\begin{center}
\includegraphics[width=6in]{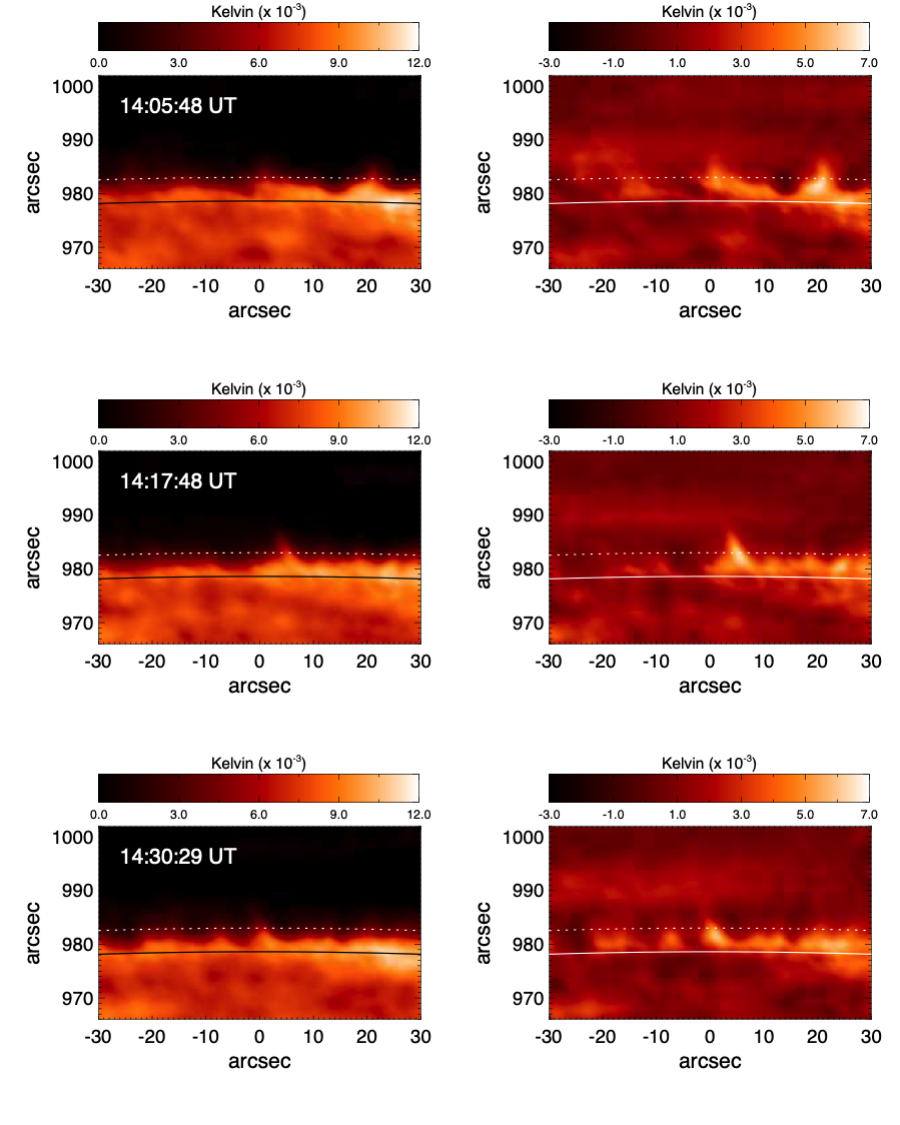}
\caption{Selection of ALMA 3~mm snapshot images (left) and their background-subtracted counterparts (right). The solid line indicates the top of the chromosphere at 3~mm whereas the dotted line indicates the mean limb. Emission from structures above the limb is clearly visible. }
\end{center}
\end{figure}

\newpage

\begin{figure}[t]
\begin{center}
\includegraphics[width=6in]{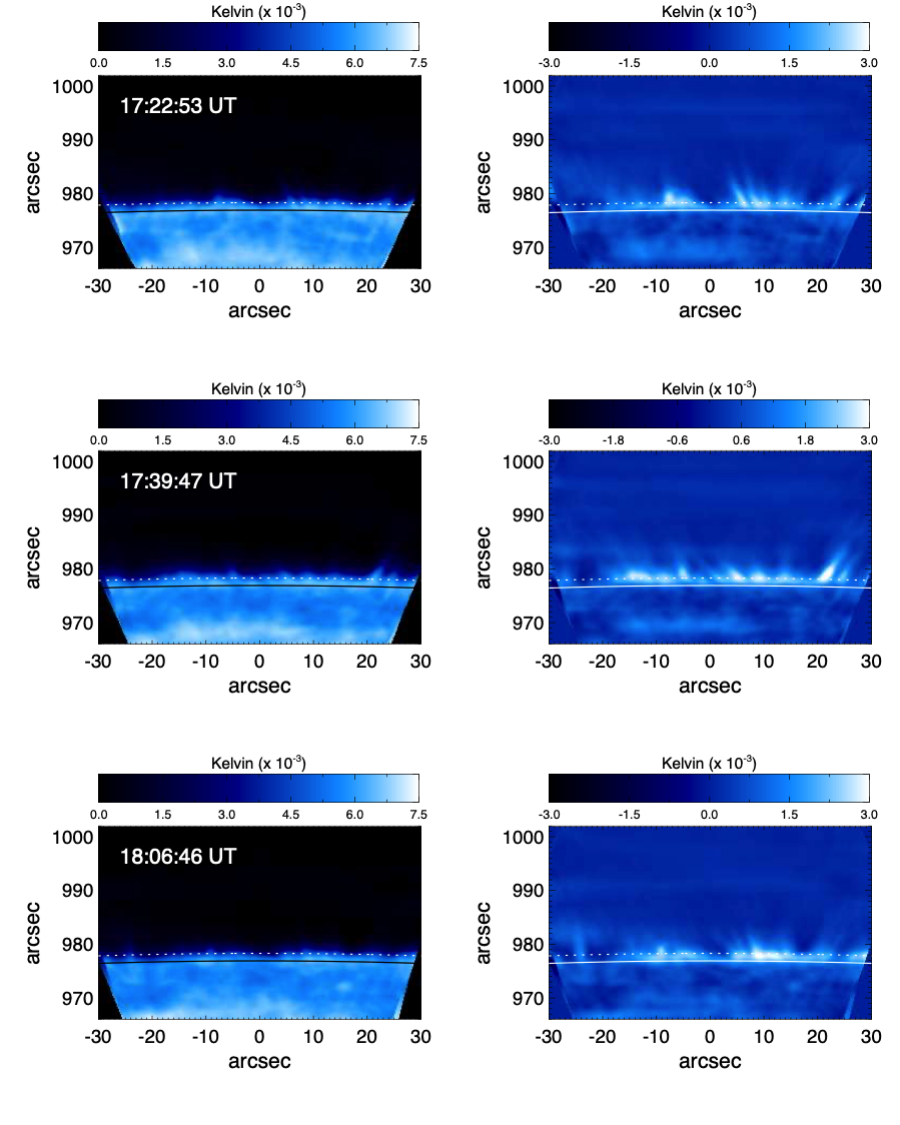}
\caption{Selection of ALMA 1.25~mm mosaic images (left) and their background-subtracted counterparts (right). The solid line indicates the top of the chromosphere at 3~mm whereas the dotted line indicates the mean limb. The examples shown are cases for which complete 14-point mosaics were observed without interruption. }
\end{center}
\end{figure}

\newpage

\section{Results for Coronal Hole Spicules}

In this section we present basic properties of spicules observed in a polar coronal hole at wavelengths of  3~mm and 1.25~mm, recognizing that the both wavelength bands have specific limitations. We begin by making a qualitative comparison between spicules observed in the two ALMA millimeter bands with those in UV and EUV passbands. All references to height above the photosphere assume a photospheric radius at the limb of $975.5"$  corresponding to the height where $\tau_{5000}=1$.

 \subsection{Comparison of mm-$\lambda$ and O/UV/EUV Spicule Observations}
 
Fig.~7 compares snapshot maps of the ALMA 3\,mm and 1.25\,mm emission with the corresponding optical, EUV and UV images obtained by GONG, SDO and IRIS. Although the GONG resolution is inferior to that of ALMA the two image sets are similar, as noted by \cite{Nindos2018}. Next are IRIS SJI images in the 2796 \AA\ band, dominated by the Mg~II~k line ($\sim10^4$~K), again corrected for center-to-limb variation; the spectrograph slit has been removed from the SJI images by dividing each image row by the average slit profile and a rotation correction of -0.7 degrees was applied to match features with those of AIA images, as in \citet{Alissandrakis2023}. These images show the spicules quite well, extending above the ALMA spicules - typically to heights of 10~Mm but in excess of 15~Mm in two instances. 


\begin{figure}[hb]
\begin{center}
\includegraphics[width=\textwidth]{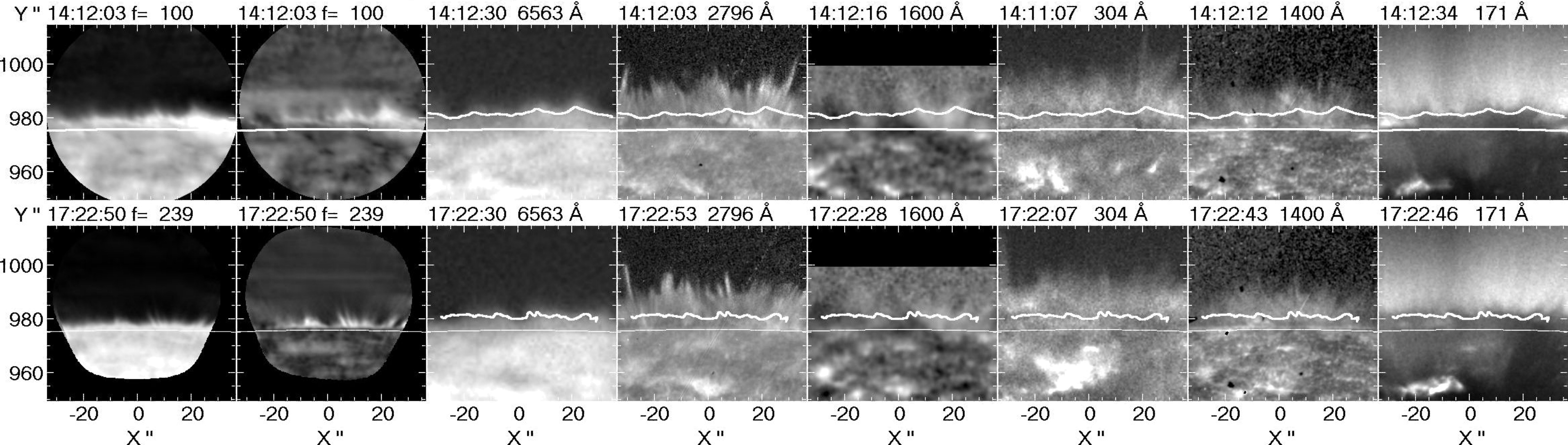}
\caption{A comparison between 3~mm (top) and 1.25\,mm (bottom) images and O/UV/EUV images. The first column shows the ALMA images; the second shows the corresponding background-subtracted images; the third shows H$\alpha$ images from GONG; the fourth shows IRIS slit-jaw images in the 2976 \AA\ band,  the fifth and the sixth show AIA images in the 1600 \AA\ and 304 \AA\ bands, the seventh shows IRIS slit-jaw images in the 1400 \AA\ band and the last column shows AIA images in the 171 \AA\ band. The contour in the top row images corresponds to a 3\,mm brightness temperature of 4500\,K and the one in the bottom row images is at a 1.25~mm brightness temperature of 800~K. The white arc marks the photospheric limb. }
\end{center}
\end{figure}

\begin{figure}[ht]
\begin{center}
\includegraphics[width=6in]{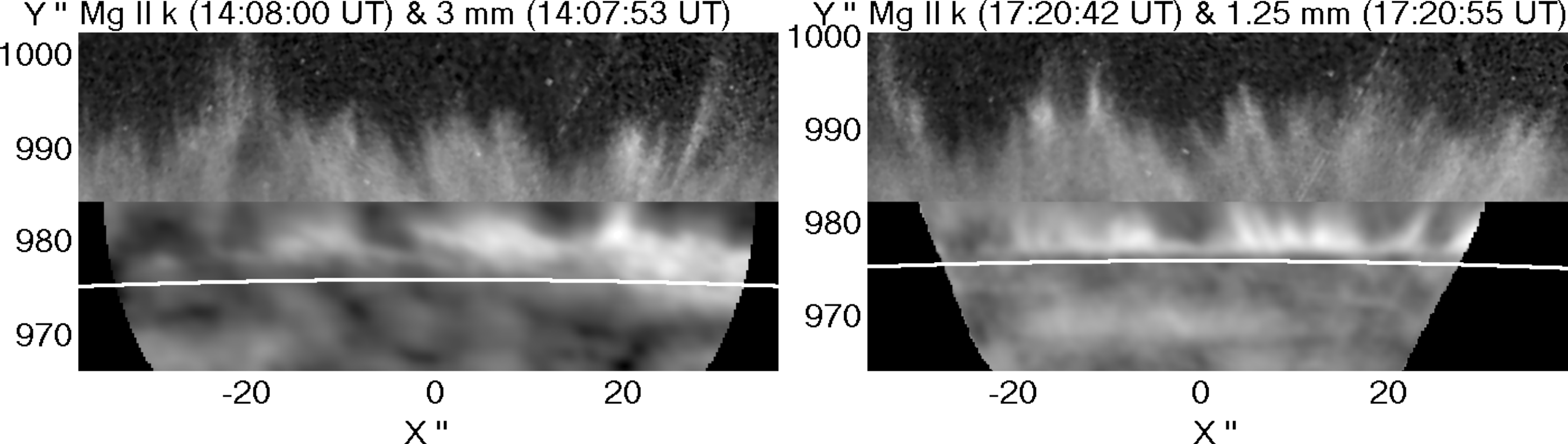}
\caption{Composite ALMA filtered images (3\,mm in the left column and 1.25\,mm in the right column) and IRIS slit-jaw images in the 2796\,\AA\ band. The ALMA images are shown in the lower half of each panel and the IRIS images are shown in the upper half. The  white arc indicates the photospheric limb. }
\end{center}
\end{figure}

A better comparison between ALMA and the Mg~II~k line is provided in Fig.~8, where the lower half of the FOV shows a filtered ALMA image and the upper half shows the nearly simultaneous Mg~II~k line SJI image. 
We note that at 1.25\,mm (right column) the ALMA resolution is more comparable to that of IRIS ($1.5"\times 0.7"$ versus $0.66\times 0.66$, respectively) and there is a close correspondence between the spicules seen at 1.25~mm and in the Mg~II~k line; for example, in the spicule group near $X=8$\arcsec. At 3\,mm the ALMA resolution is inferior, but still the association is quite evident. 

The fifth column in Fig.~7 shows images in the AIA 1600\,\AA\ band (formed at low chromospheric temperatures), after correction for center-to-limb variation; due to the low signal to noise ratio we had to integrate over 5\,min to see structures beyond the limb. 
The sixth column shows images in the AIA 304\,\AA\ band, dominated by the He\,{\sc ii} line, averaged over 2\,min to reduce noise. In spite of the residual noise it is clear that here spicules extend much higher, still there is a good association with ALMA limb structures.The seventh column shows IRIS slit jaw images in the 1400\,\AA\ band, dominated by the Si\,{\sc iv} lines ($\sim 8\times 10^4$ K). The emission is more diffuse here but, again, it extends above the 3~mm contour.

The last column in Fig.~7 shows AIA 171 \AA\ images, formed around  $10^6$\,K. As has been pointed out previously \citep{Daw1995, Anzer2005, DePontieu2011, Alissandrakis2019}, plasma at chromospheric temperatures, including spicules, appears in absorption against the coronal background in spectral regions where the emission is formed at temperatures above $\sim10^6$\,K. The ALMA 3~mm contour is correlated with the outer limit of the 171 \AA\ absorption, a correlation also noted by \citet{Yokoyama2018}. The absorption of the 171 \AA\ emission is presumed to be due to neutral H, He, and singly ionized He, which allows us to place limits on the column depth of H\,{\sc i} at a given height above the limb as we discuss in the next section.

Overall, spicular structures are similar in all spectral ranges that we examined, differences being due to the fact that the physical conditions affect the emission of radiation in different ways and due to instrumental issues such as noise, resolution and dynamic range. 
Before continuing, it is useful to identify the primary sources of opacity at \mml, their dependencies on physical parameters, and their relationship to the observed quantity, the brightness temperature $T_b$. 

\subsection{Sources of Opacity at mm-$\lambda$}

The dominant sources of opacity at mm- and submm-$\lambda$ in the chromosphere are thermal H and H$^-$ free-free absorption \citep{Vernazza1976}, resulting from collisions between electrons and ions, and between electrons and neutral hydrogen (HI) atoms, respectively. The absorption of coronal EUV emission noted in \S4.1 is well-correlated with \mml\ emission in the sense that the variable height of the 3~mm emission is comparable to the height below which the background EUV emission from coronal plasma is absorbed. Is the amount of H~I sufficiently large to play a role in the opacity at 3~mm and 1.25~mm? To answer this we must estimate the column depth of H~I with height. The absorption of EUV photons is largely due to photoionization of H, and neutral and singly ionized He. The total optical depth to neutrals is 

\begin{equation}
\tau_{tot}=N_{HI}\sigma_{HI}+N_{HeI}\sigma_{HeI}+N_{HeII}\sigma_{HeII}
\end{equation}

\noindent where $N_{HI}$ is the neutral hydrogen column density, $\sigma_{HI}$ is the H~I photoionization cross section, and so on. Interestingly, using the fitted parameters of \citet{Verner1996}, the cross sections of He~I and He~II are nearly the same for wavelengths $\lesssim 250$\AA\ and so $\tau_{tot}\approx N_{HI}\sigma_{HI}+N_{He}\sigma_{He}$ where now $N_{He}=N_{HeI}+N_{HeII}$ and $\sigma_{He}\approx \sigma_{He~I}\approx \sigma_{He~II}$. Taking the fractional ionization of hydrogen to be $f=n_p/(n_{HI}+n_p)$ where $n_p$ is the proton number density, we then have 

\begin{equation}
\tau_{tot}\approx N_{HI}\sigma_{tot}=N_{HI}\Bigl[\sigma_{HI}+{{A_{He}\sigma_{He}}\over{1-f}}\Bigr]
\end{equation}

\noindent We take the helium abundance to be $A_{He}=n_{He}/n_H=0.085$ \citep{Grevesse2007}.
Using observations of the limb in multiple EUV passbands from the TRACE and SDO missions, \citet{Alissandrakis2019} (see also \citealt{Alissandrakis2019a}) measured the variation of $N_{HI}$ with height by exploiting the fact that when $\tau_{tot}=1$ for a given EUV wavelength, $N_{HI}=\sigma_{tot}^{-1}$. The data are well fit by a simple exponential variation with height above the limb. Assuming the H~I number density is $n_{HI}=n_o\exp(-h/L_{HI})$, the column density then varies (see \citealt{Zhang1998, Alissandrakis2019}) as 

\begin{equation}
N_{HI}\approx n_o\sqrt{2\pi L_{HI} R_\odot}\exp(-h/L_{HI}). 
\end{equation}

\noindent With a fractional ionization $f=0.5$,  for which $\sigma_{tot}\approx 1.88 \times 10^{-19}$ cm$^2$, Alissandrakis \& Valentino find that $L_{HI}\approx 0.97$ Mm and $n_o\approx 6\times 10^{10}$ which then leads to $N_{HI}\approx 3.9\times 10^{20}\exp(-h/0.97)$ where $h$ is measured in Mm. This, in turn, implies that $N_{HI}$ decreases from $5\times 10^{19}$ to $2.3\times 10^{18}$ cm$^{-2}$, and that $n_{HI}$ decreases from $7.6\times 10^9$ to $3.5\times 10^8$ cm$^{-3}$, respectively, between heights $h=2-5$ Mm. For a higher fractional ionization these values must be adjusted downward -- by a factor $\approx\!2$ when $f=0.8$, for example. We note that \citet{Daw1995}, analyzing the absorption of soft X-rays at the limb, suggest a much higher value of $n_H\sim 10^{10}$ cm$^{-3}$ at a height of 5~Mm than was inferred by Alissandrakis \& Valentino. The difference may lie in the fact that Daw {\sl et al.} assume a filling factor of just a few percent while Alissandrakis \& Valentino assume a uniform atmosphere. 

We have computed the free-free and H$^-$ absorption coefficients \citep{Vernazza1976} for a wide range of $n_e$ and $n_{HI}$ and find that free-free absorption dominates by far except for cases when $n_{HI}>5\times 10^{10}$ cm$^{-3}$ and $n_e\lesssim10^8$ cm$^{-3}$. Hence, regardless of assumptions about the HI filling factor, we conclude that while neutrals can account for the absorption of background coronal EUV radiation they play no significant role in emission at 3~mm and 1.25~mm. 

The free-free absorption coefficient is (e.g., \citealt{Lang1999})


\begin{equation}
\kappa_{ff}\approx 9.786\times 10^{-3} n_e n_i \nu^{-2} T_e^{-3/2} \xi(T_e,\nu) 
\end{equation}

\begin{figure}
\begin{center}
\includegraphics[angle=-90,width=5in]{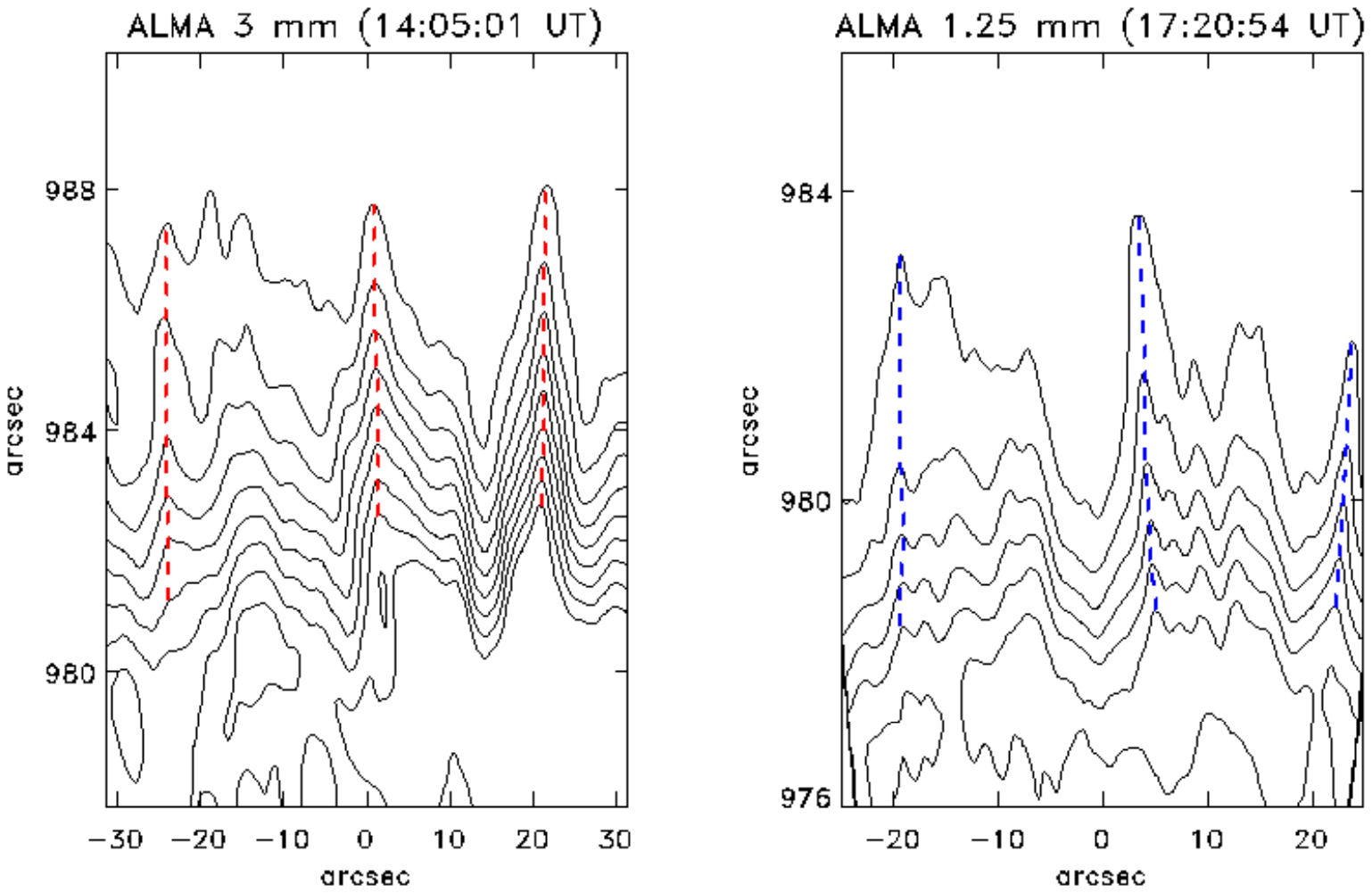}
\caption{Left: Example of a 3~mm snapshot used to measure the variation of brightness temperature $T_b$ with height in discrete structures (red dashed lines); Right: the same for a 1.25~mm mosaic (blue dashed lines). Note the difference in scale between to two wavelengths, with the 3~mm data offset to greater heights than the 1.25~mm data. Contours decrease with height and are shown at 1000~K intervals, the outermost contour being 1000~K.}
\end{center}
\end{figure}

\noindent where $n_e$ is the electron number density, $n_i$ is the ion number density, and $\xi(T_e,\nu) \approx 17.72 +  \ln(T_e^{3/2}/\nu)$ for a plasma temperature $T_e<3.2\times10^5$~K. In the following, we assume $n_e=n_i$ where $n_i$ mostly comprises protons and singly and/or doubly ionized He. The optical depth is $\tau_{ff}(h)=\int \kappa_{ff}(h) dl$ and we then have 

\begin{equation}
\tau_{ff}(h)\approx 9.786 \times 10^{-3} \Phi(h) \nu^{-2} T_e(h)^{-3/2} \xi(T_e,\nu)
\end{equation}

\noindent where $\Phi(h)=\int n_e(h)^2 dl$ is the column emission measure. 

As discussed in \S1, the degree of ionization is not in equilibrium owing to the fact that the ionization and recombination time scales are much longer than dynamical time scales in the chromosphere. Conditions of nonequilibrium ionization result in greater numbers of free electrons than would be the case for equilibrium conditions. While this can have profound impacts on the interpretation and modeling of UV/EUV line emissions (e.g., \citealt{MartinezSykora2017}) it has little impact on our analysis below. 
Opacity measurements derived from the \mml\ observations are a convenient way to measure the column emission measure, regardless of whether the emitting plasma is in equilibrium conditions or not. 



\subsection{Height of the Chromosphere}

The mean radius of the Sun at 3~mm and 1.25~mm was referenced to fits of the limb of the full disk TP maps (\S3.2. The mean value of $982.2"$ for the 3~mm TP map represents an azimuthal average on the angular scale of $\sim 1'$. From inspection of Figs.~5 and 6 it is clear that the 3~mm and 1.25~mm limbs are corrugated due to dynamic structures extending above and below the mean. The range of the radius measured at 3~mm is $\approx$ 979.4"-985". Converting these values to heights above the limb we find the mean height of the limb at 3~mm is $\langle r_3\rangle=4.8$~Mm and the range is $2.8-6.8$~Mm. For the 1.25~mm emission, we have a mean radius of $978.3"$, averaged on a azimuthal scale of $\approx 24"$, corresponding to a mean height $\langle r_{1.25}\rangle=2$~Mm above the limb. The range is $\approx $976.9"-980.5" corresponding to a height range of $1-3.6$~Mm. However, as we show in \S4.5, the mean limb heights deduced for each wavelength or more precisely, the height difference, are inconsistent with the implied variation of spicule optical depth with height. A correction to the height difference between the 1.25 and 3~mm maps of $\approx\!1"$ is implied. While it is unknown whether either, or both, height determinations are in error, we distribute the height difference between both wavelengths by adjusting the  3~mm limb upward by $\approx\!0.7"$  to 982.9" and the 1.25~mm limb downward by $0.3"$ to 978.0" -- significant, but well within the uncertainty of the fits to the TP data. We therefore adopt a value of 5.3~Mm for the mean 3~mm limb, with a range of 2.6-6.6~Mm; and a value of 1.8~Mm for the mean 1.25~mm limb with a range of 0.8-3.4~Mm. Hence, while a height difference of $\approx\!3.5$~Mm is consistent with the data presented in \S4.5, there is a systematic uncertainty in the absolute height of the chromosphere in both bands. 
 
  
There are few high-resolution measurements of the solar limb with which to compare these results. \citet{Wannier1983} used the OVRO interferometer at 2.6~mm with a resolution of $6"$ to infer a limb extension of $5700-8560$~km, somewhat above our 3~mm measurement. \citet{Belkora1992} used the same interferometer at 3~mm during the 1991 July 11 solar eclipse to infer a 3~mm limb extension of $5.53\pm 0.59$~Mm in a quiet Sun region, consistent with the revised value 5.3~Mm we adopted above. \citet{White1994}, however, observing the same eclipse with the BIMA interferometer at 3~mm report a limb height of $8\pm0.13$~Mm. The difference in limb extension heights reported by \citet{Belkora1992} and \citet{White1994} is not understood. At shorter wavelengths, \citet{Horne1981} used a single 10~m antenna at OVRO with a beam width of $29"$ to make radial scans across the solar disk at 1.3~mm. They report a limb extension of $8.6"\pm 1"$, corresponding to a height of nearly 6.3~Mm, well above our measurement of the mean radius at 1.25~mm. \citet{Ewell1993} used the CSO during the eclipse on 1991 July 11 to infer the limb extension at 0.85~mm, finding a value of $3.38\pm 0.14$~Mm, again larger than the mean value inferred for the 1.25~mm radius here. \citet{Alissandrakis2020}, however, measured the limb at 3~mm and 1.25~cm using ALMA TP maps and a procedure similar to our own; they report values of $4.2\pm 2.5$~Mm and $2.4\pm 1.7$~Mm, respectively, consistent with the values we have reported here. We emphasize that our values for the chromospheric height, as well as those reported by others, average over dynamic structure on the limb. More appropriate values for the height of the chromosphere are near the minima of the range of heights measured in the restored maps: 2.6~Mm for 3~mm emission and 0.8~Mm for 1.25~mm emission. We note that the filtered images in both bands clearly show the top of the chromosphere; i.e., the level where the spicule structures become bright, which is consistent with this idea. 

\subsection{Spicule Temperatures and \mml\ Brightness Temperatures}

Past work on spicule temperatures is somewhat equivocal. \cite{Beckers1972} concluded that spicule temperatures were $\approx 15\times 10^3$ K above a height of 2~Mm. \citet{Krall1976} report spicule temperatures of $13-16\times 10^3$ K for heights above 6~Mm. \citet{Matsuno1988} suggest that non-LTE modeling of these observations yields solutions consistent with a temperature of 7000~K and perform an analysis of H, Mg, and Ti lines during an eclipse to conclude that spicule temperature first decreases from 9000~K at a height of 2~Mm to 5000~K at 3.25~Mm and then increase to 7000~K at 4.7~Mm. A recent analysis of IRIS observations of spicules by \citet{Alissandrakis2018} in the Mg~II~h and k, C~II, and Si~IV lines reports that spicule temperatures increase from $\approx\!8000-9000$~K below 5~Mm to more than $20\times 10^3$~K at a height of 9~Mm. Finally, based on IRIS and SDO/AIA observations, several authors report significant heating of spicular material to transition region temperatures as they rise to heights of 15~Mm (e.g., \citealt{Tian2014, Pereira2014, Skogsrud2015, RouppevanderVoort2015, DePontieu2017}).

If spicules are optically thick and significantly warmer than chromospheric temperatures one would expect to see a increase in brightness at the limb.  \citet{Wannier1983} observed the limb of the Sun interferometrically at 2.6~mm with a (one-dimensional) angular resolution of $6"$.  They report no ``limb spike" in brightness associated with an increase in brightness temperature above the limb, reporting a brightness temperature $T_b=6100\pm 800$ K. Similarly, observations of the limb at 3~mm by \citet{Belkora1992} during a solar eclipse with an interferometer, yielding an effective one-dimensional resolution perpendicular to the limb of $1.6"$, showed no evidence for a limb spike in $T_b$. They report a good correspondence between the 3~mm limb and the height of H$\alpha$ spicules. \citet{White1994} also report no limb spike during eclipse observations at 3~mm. On the basis of these admittedly spotty reports, there is no previous evidence for brightness temperatures significantly greater than chromospheric temperatures emitting at \mml\ above the solar limb.

ALMA has now made two-dimensional, high-resolution observations of spicules in both the 3~mm and 1.25~mm bands. We find that while the brightness temperature near the limb is higher than the nominal quiet Sun brightness temperatures, there is no evidence for $T_b$ higher than $12-13\times 10^3$~K anywhere in the ALMA 3~mm maps or higher than $10^4$~K at 1.25~mm, consistent with quiet Sun brightness temperatures near the limb inferred by \citet{Alissandrakis2022}. We have measured the variation of $T_b$ from spicules with height above 4~Mm at 3~mm and above 2~Mm at 1.25~mm (Fig.~9, Fig.~10a) and find that it is $\lesssim\!10^4$ K in all cases; we find no evidence for optically thick spicule emission with $T_e\approx T_b$ much in excess of $10^4$ K.  That is not to say, however, that $T_e$ does not increase with height, as we discuss below.

\subsection{Spicule Densities}


We now turn to the question of spicular density. We have measured the variation in the brightness temperature with height, $T_b(h)$, of extended structures observed to rise and fall, or rise and fade, with time at both 3 and 1.25~mm. We identify these structures with spicules, recognizing that the angular resolution of ALMA is not sufficient to fully resolve discrete spicules or to disentangle multiple spicules along the line of sight (see \S4.6). We selected times that were sufficiently well spaced that a given map in a given wavelength is roughly independent from the others considered. For the 3~mm maps, we considered 3 times in each scan spaced by 200~s. For the 1.25~mm maps we considered ten mosaics, for which the time separation is also 200~s or more. Within each map, 2-5 discrete spicules were identified (Fig.~9) and their variation in brightness with height was measured interactively for $T_b>1000$~K. A total of 35 and 32 spicules were evaluated at 1.25 and 3~mm, respectively. The results are shown in Fig.~10a.  We note that \citet{Nindos2018} measured a brightness temperature of 2560~K in a spicule at 3~mm at height of 6.2~Mm, consistent with our results (although see \citealt{Shimojo2020}). From Fig.~10a, we see that the brightness temperature declines exponentially with height with similar scale heights. Fitting the observed variation of brightness temperature with height, assumed to be exponential, we find that the scale heights of $T_b$ at 3 and 1.25~mm are $1.94\pm0.5$~Mm and $1.89\pm0.4$~Mm, respectively. These are comparable the to intensity scale heights reported for spicules observed in H$\alpha$ \citep{Beckers1968} and in EUV \citep{Withbroe1983}. More recently, \citet{Alissandrakis2018} reported average intensity scale heights of 1.9-2.2~Mm for spicules observed in the Si~IV and the O~IV lines by IRIS. We note, however, that \citet{Pereira2012} report a scale height of $3.09\pm0.49$ Mm in a CH environment, as observed in the Ca~II~H line. 

If $T_b(h)$ is measured for a given wavelength, the optical depth can be inferred if $T_e(h)$ is known; i.e., with $T_b(h)=T_e(h)[1-\exp(-\tau_{ff}(h)]$,

\begin{equation}
\tau_{ff}(h)=-\ln[1-T_b(h)/T_e(h)]. 
\end{equation}

\noindent and from Eqn.~(5) we then have

\begin{equation}
\Phi(h)  \approx n_e^2(h) D = { {\tau_{ff}(h) \nu^2 T_e(h)^{3/2} }\over {C \xi[T_e(h),\nu]} }
\end{equation}

\noindent where $C=9.786 \times 10^{-3}$. We then infer $n_e(h)$ from 

\begin{equation}
n_e(h)\approx \biggl[{{\tau_{ff}(h) \nu^{2} T_e(h)^{3/2}}\over {D C \xi[T_e(h),\nu]}}\biggr]^{1/2}
\end{equation}

\noindent where $D$ is the effective thickness occupied by spicular material. If we take spicules to be cylindrical with a typical diameter $D_S=0.5$ Mm, then $D=N D_S$ where $N$ is the number of spicules along the line of sight. Note that \citet{Beckers1972} adopted $D_S=0.85$~Mm whereas a study by \citet{Pereira2012} of {\sl Hinode} SOT observations found $D_S\approx 0.3$ Mm.

It is important to note that for optically thin emission at a given frequency we have $\tau \approx T_b/T_e$ and so the dependence of $n_e$ on temperature $T_e$ is

\begin{equation}
n_e \propto \sqrt{{T_b T_e^{1/2}}\over{\xi}}\sim {{T_e^{1/4}}\over \xi^{1/2}}
\end{equation} 

\noindent For an observed value of $T_b$ along a given line of sight the inferred value of $n_e$ is  insensitive to assumptions regarding the temperature of the emitting material. For example, for a wavelength of 3~mm the number densities inferred for $T_e=10^4$~K and $T_e=10^5$~K for observed values of $T_b(h)$ differ by only $\approx\!43\%$!

For illustrative purposes we consider two schematic models for the variation of spicule temperature with height: i) the spicule is isothermal with $T_e=15\times 10^3$~K above a height of 2~Mm (e.g., \citealt{Beckers1972, Alissandrakis2019}); ii) the temperature of the spicules is assumed to increase linearly from $15\times 10^3$~K at a height of 2 Mm to a transition region value of $10^5$ K at a height of 15 Mm \citep{DePontieu2011, Pereira2014, Skogsrud2015}.  We refer to the first model as the ``isothermal model" (ISO) and the second as the ``temperature ramp (TR)" model. We acknowledge that neither model is realistic;  nor should they be construed as making any assumptions about spicular heating mechanisms. As we shall see, the two schematic models for the variation of spicule temperature with height show the insensitivity of our conclusions to the temperature and distribution of plasma within spicules and, by extension, any specific spicule heating model. 

To estimate spicular electron number density $n_e$ we assume that it decreases exponentially with height: $n_e(h)=n_\circ\exp(-h/L_{n_e})$, where $L_{n_e}$ is the effective scale height of the electron number density. We begin by assuming that all of the spicular material along the line of sight is confined to a single structure ($N=1$) with a thickness $D_S=0.5$~Mm and deduce $n_e(h)$ for the two temperature models, modulo the assumed value of $D_S$. The inferred densities should be decreased by factor $\sqrt{0.5/0.85}=0.77$ for $D_S=0.85$~Mm and increased by a factor 1.29 if $D_S=0.3$~Mm. 




\begin{figure}[ht!]
\begin{center}
\includegraphics[width=2.25in]{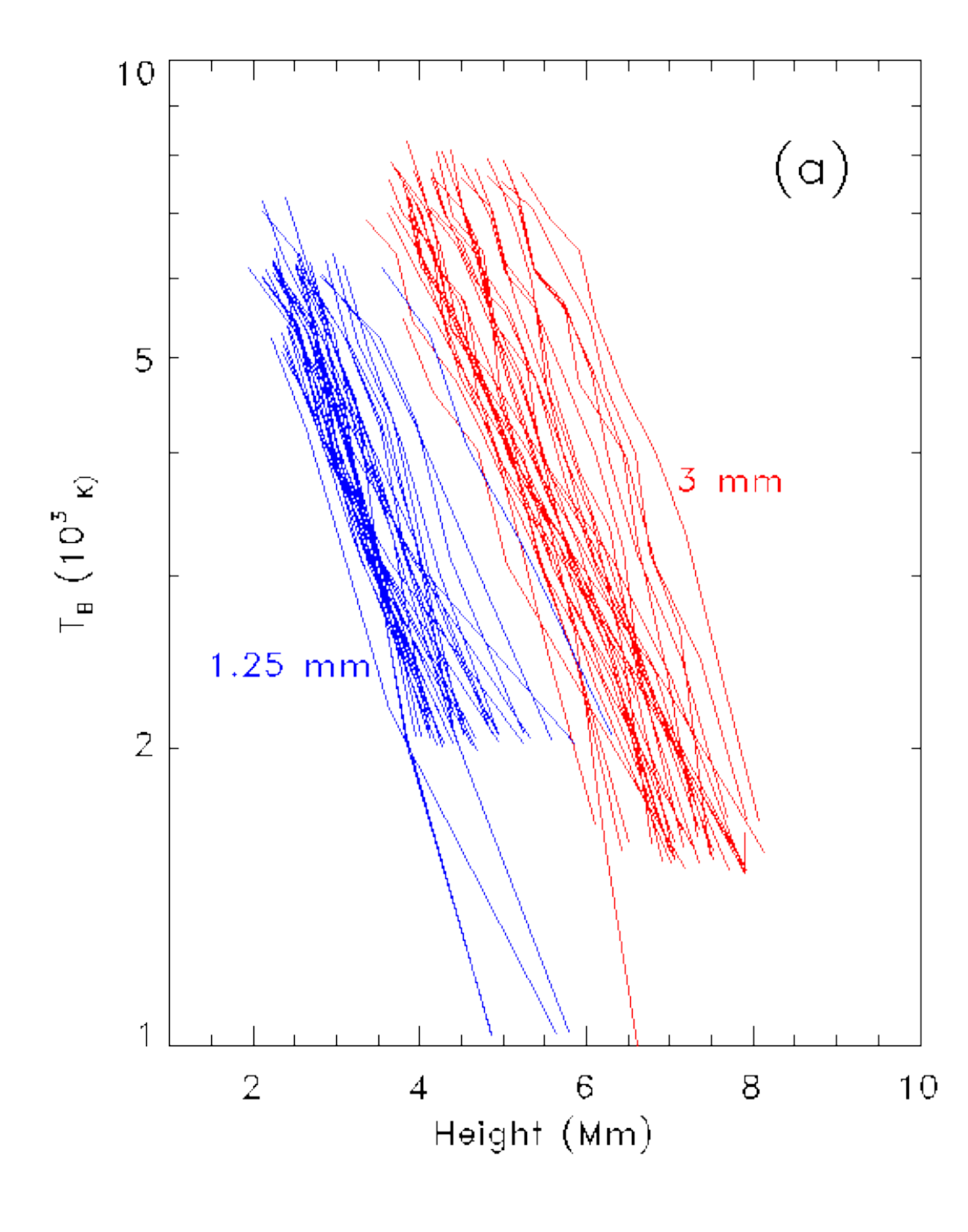}
\includegraphics[width=2.25in]{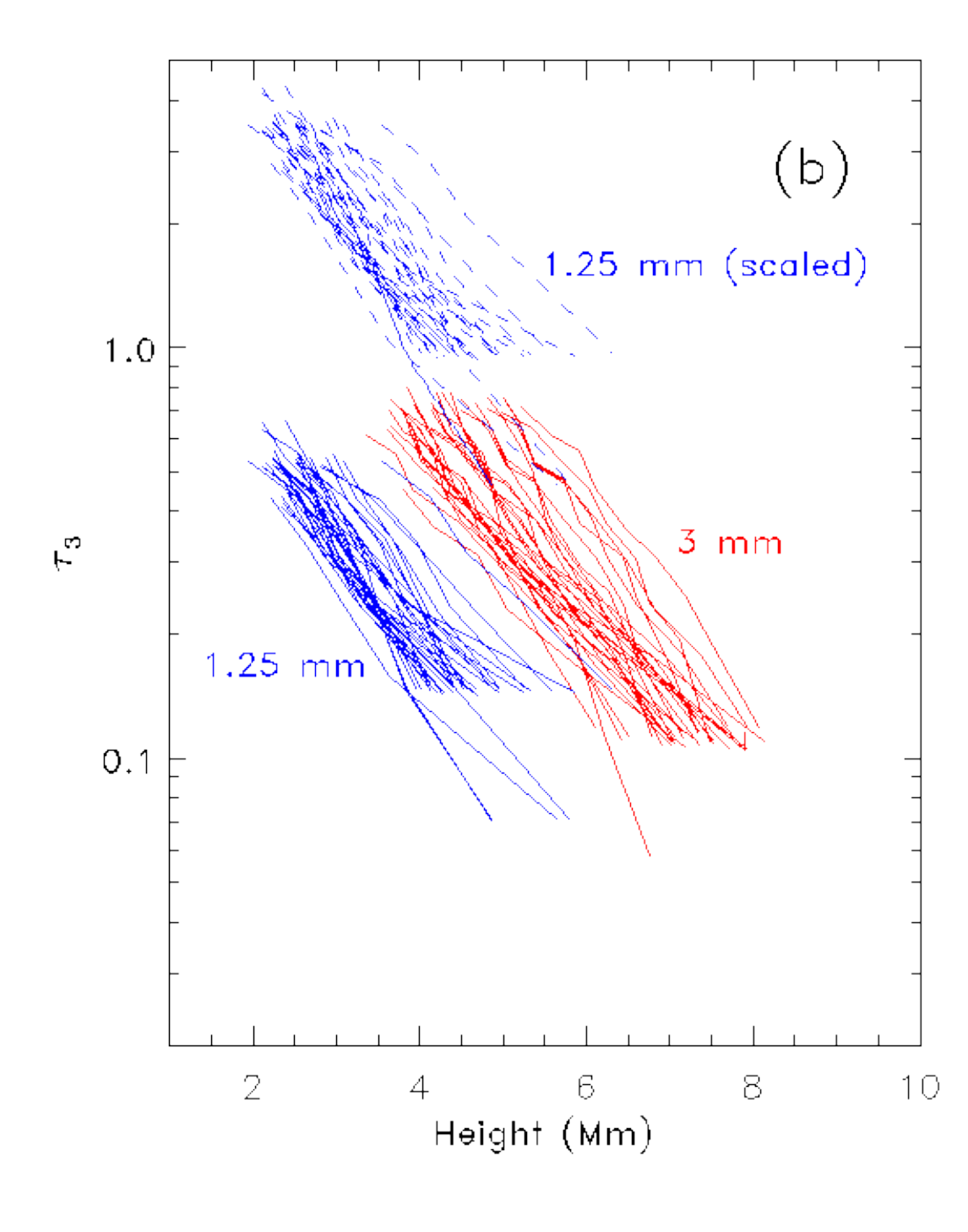}
\includegraphics[width=2.25in]{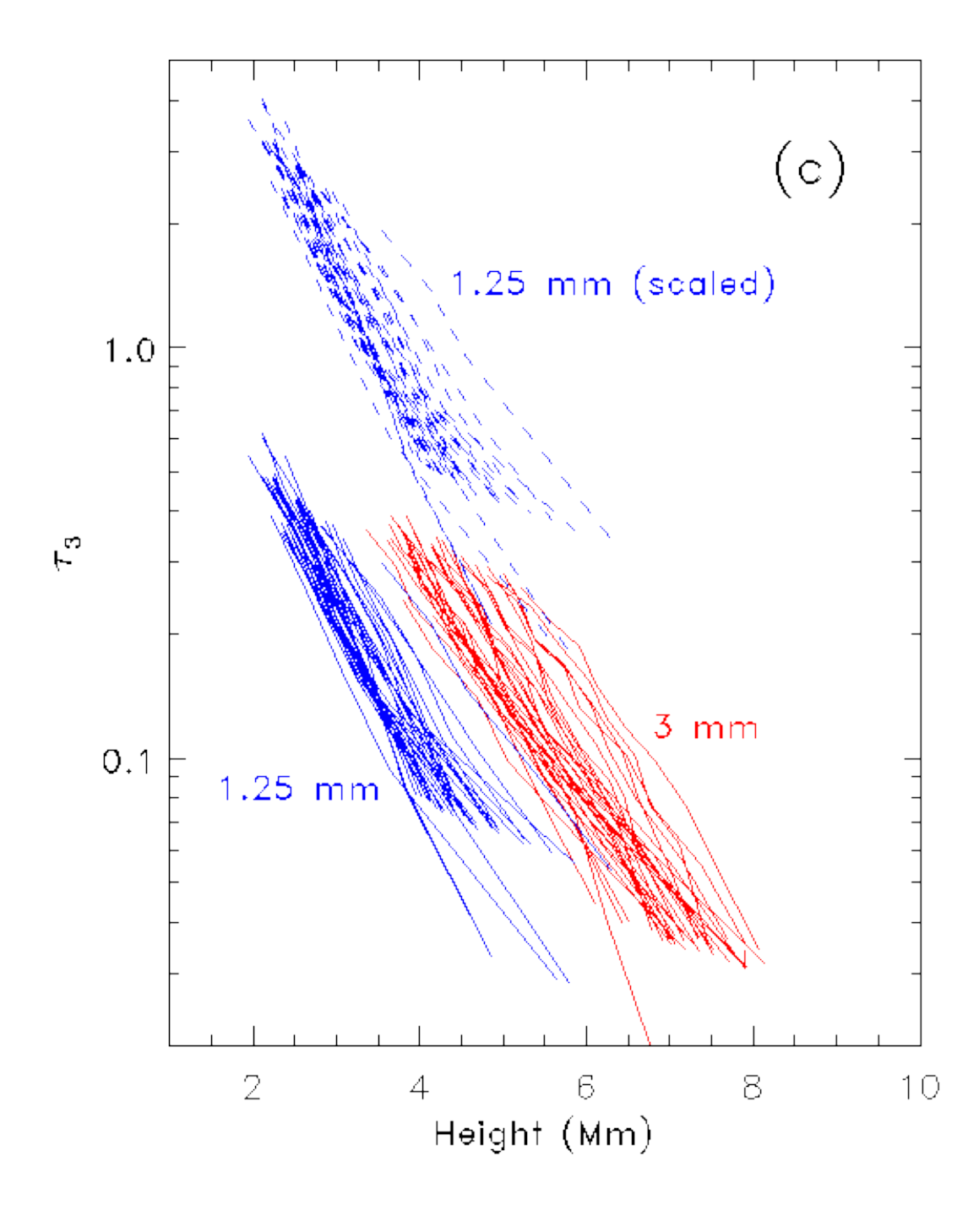}
\caption{a) Brightness temperature measurements as a function of height for an ensemble of spicules at 3~mm (red) and 1.25~mm (blue); b) the variation of optical depth with height for the ISO model for 3~mm (red). The dashed blue line are the corresponding optical depths at 1.25~mm. The solid blue lines are the optical depths at 3~mm obtained by scaling the 1.25~mm measurements according to Eqn.~(12); c) the same as (b) but for the TR model. Note the lack of continuity between the scaled 1.25~mm optical depths and those measured at 3~mm for both the ISO and TR models. }
\end{center}
\end{figure}

We fit the data using two approaches. First, there is an intrinsic variation by a factor $\sim\!2$ from spicule to spicule in the measured values of $T_b$ (Fig.~10a) and the inferred optical depths (Fig.~10b,c) at a given height. We therefore fit a density model $(n_\circ, L_{n_e}$) to each spicule individually at each wavelength and then computed the mean and standard deviation of the fitted parameters (Table~2). Specifically, we inferred $n_e(h)$ for a given spicule using Eqn.~9 assuming one of the two temperature models and then fit it to an exponential model. The standard deviation captures the intrinsic dispersion in spicule densities in this case. Second, we considered the spicule measurements in aggregate, fitting all measurements at 1.25~mm with a single fit, and all measurements at 3~mm with a single fit. Both the ``discrete" and``aggregate" approaches yield similar results, as do the separate fits to measurements in each wavelength. 
For the fits to the ISO model, we have $L_{n_e}\approx 3.5$~Mm for the discrete fits in both wavelengths. The scale height is somewhat higher for each wavelength when the data are fit in aggregate. For the TR model the scale height fit for the 3~mm data is larger than that found for the 1.25~mm data. This is not surprising as the temperature of the spicular material is cool low in the atmosphere but increases significantly with height.  Remarkably, the values of $n_\circ$ resulting from separate fits to the two wavelengths are quite similar, within the uncertainties. 

We were unable to observe spicules in the two wavelength bands simultaneously and, therefore, cannot fit single spicules using data from both wavelengths. We assume, however, that the measurements of spicule brightness as a function of height in the two wavelength bands are representative since the measurements in the two wavelength bands were of the same polar CH at a time separation of only $\sim\!3$ hrs. We can easily reference the 1.25~mm optical depth measurements to a wavelength of 3~mm using 

\begin{equation}
\tau_3=\tau_{1.25}{{\kappa_3}\over {\kappa_{1.25}}} = \tau_{1.25}{\xi_{1.25} \over \xi_3}\Bigl({3\over 1.25}\Bigr)^2
\end{equation}
\smallskip

\noindent and then fit the measurements in aggregate. However, as noted in \S4.3, in comparing the optical depths referenced to 3~mm as a function of height, a systematic difference was uncovered that we attribute to errors in the fits to the mean limb of each wavelength (Fig.~10b,c). That such systematic errors are present is perhaps not surprising given the relatively large uncertainties in the fits to the mean radius of the Sun. An increase in the difference in the 1.25~mm and 3~mm limb amounting to $\approx 1"$ is needed to reconcile the two sets of measurements. As discussed in \S4.3, we adjusted the mean radius by $-0.3"$ (1.25~mm) and by $+0.7"$ (3~mm). Fitting the aggregated data in both wavelengths to the two models we find that both yield similar values of $n_\circ$ while the scale height $L_{n_e}=3.54$~Mm for the ISO model and 3.92~Mm for the TR heating model (Table~2). 


Fig.~11 summarizes our results for fits to spicule densities (solid green lines) for the two temperature models and compares them to those of earlier workers, who inferred densities from non-LTE models of O/UV spectral lines or spectral line ratios for a single slab or cylinder of thickness $D_S$ to represent the spicule. We find that our values for the spicular electron number density and its decrease with height agree rather well with earlier values based on O/UV observations for both of the schematic models used for the variation of spicule temperature with height (ISO and TR). We emphasize that densities based on observations of  O/UV lines and line ratios are insensitive to temperatures much in excess of chromospheric whereas the \mml\ results are nearly equally sensitive to all temperatures above $10^4$~K. In fact, there is little difference in densities derived from the \mml\ observations for {\sl any} temperature model assumed for temperatures ranging from $\sim\!10^4-10^5$~K. 


\begin{deluxetable}{lccll}
\tablecaption{Spicule Density Parameters}
\tablehead{\colhead{Fit Type} &\colhead{Temperature Model} & \colhead{Wavelength (mm)} & \colhead{$n_\circ$ (cm$^{-3}$)} & \colhead{$L_{n_e}$ (Mm) }}
\startdata
Discrete &Isothermal& 1.25 & $2.74(0.61)\times 10^{11}$ &  $3.45(0.84)$\\
                 & & 3 &  $2.89(1.67)\times 10^{11}$ & $3.49(0.81)$ \\
&Temperature Ramp& 1.25 &  $2.50(0.52)\times 10^{11}$ &  $3.98(1.08)$\\
                 &  & 3 &  $2.26(1.13)\times 10^{11}$ & $4.23(1.03)$ \\
\tableline                 
Aggregate &Isothermal& 1.25 & $2.44(0.17)\times 10^{11}$ &  $3.61(0.27)$\\
                 && 3 &  $1.96(0.23)\times 10^{11}$ & $4.03(0.31)$ \\
                 && 1.25+3  & $2.44(0.11)\times 10^{11}$ & $3.54(0.10)$ \\
&Temperature Ramp& 1.25 &  $2.25(0.15)\times 10^{11}$ &  $4.13(0.34)$\\
                   && 3 &  $1.60(0.32)\times 10^{11}$ & $4.96(0.42)$ \\
                 && 1.25+3  & $2.27(0.10)\times 10^{11}$ & $3.92(0.12)$\\
\enddata
\end{deluxetable}

Since the intensity of the UV/EUV lines is $\propto n_e^2$, the density scale heights in Table 2 correspond to intensity scale heights of $\approx\!2$~Mm; consistent with \citet{Beckers1972}, \citet{Withbroe1983} and \citet{Alissandrakis2018}.  These are again consistent with most previous estimates implied by O/UV observations of spicules discussed at the beginning of this section, but are significantly smaller than the density scale height implied by \citet{Pereira2012}, a CH value of 6.18~Mm. 

The results derived for a single, cylindrical spicule represent upper limits. In order to estimate more typical spicular densities, we must take into account the spicule filling factor, as we now discuss.  

\begin{figure}[ht!]
\begin{center}
\includegraphics[width=2.25in]{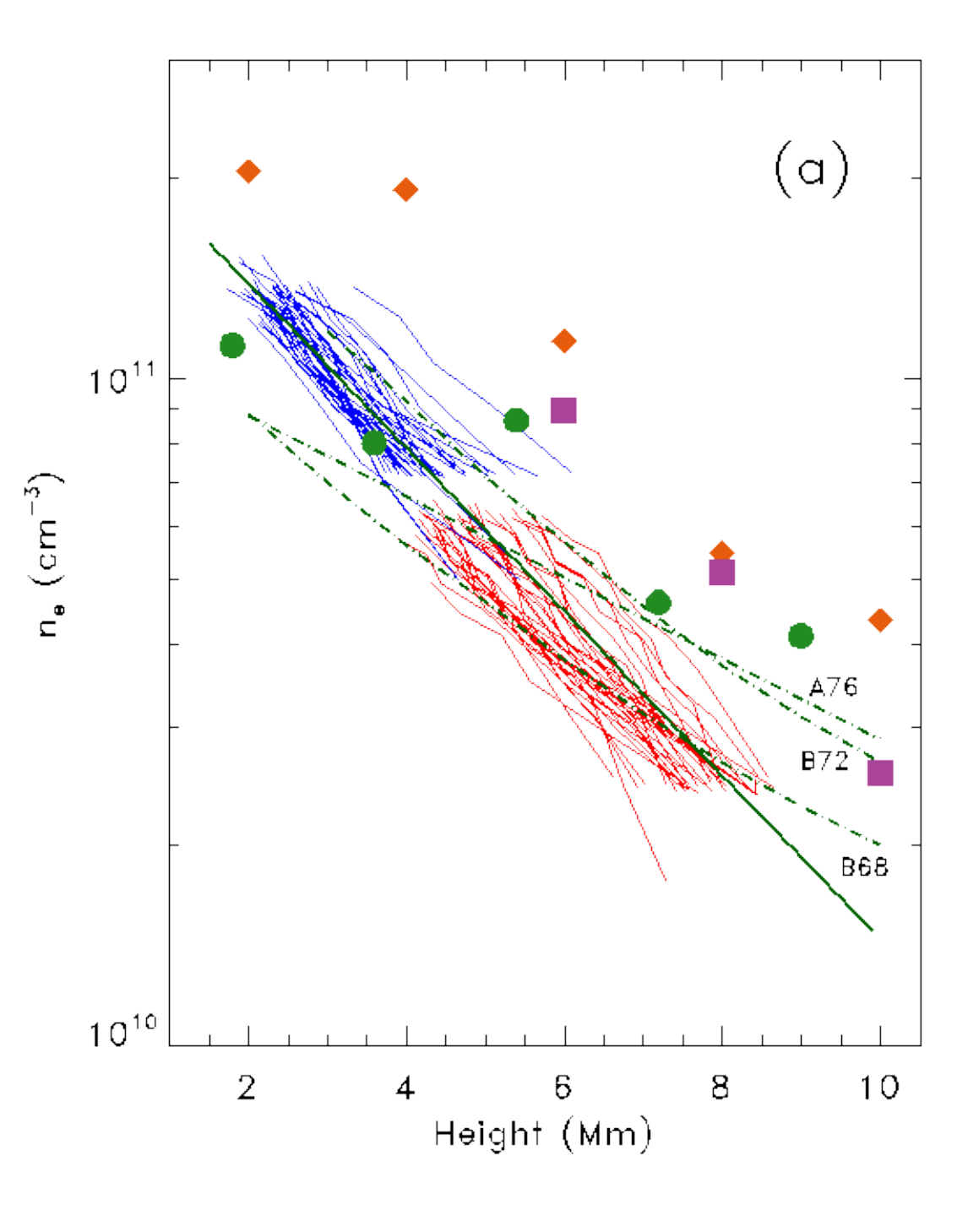}
\includegraphics[width=2.25in]{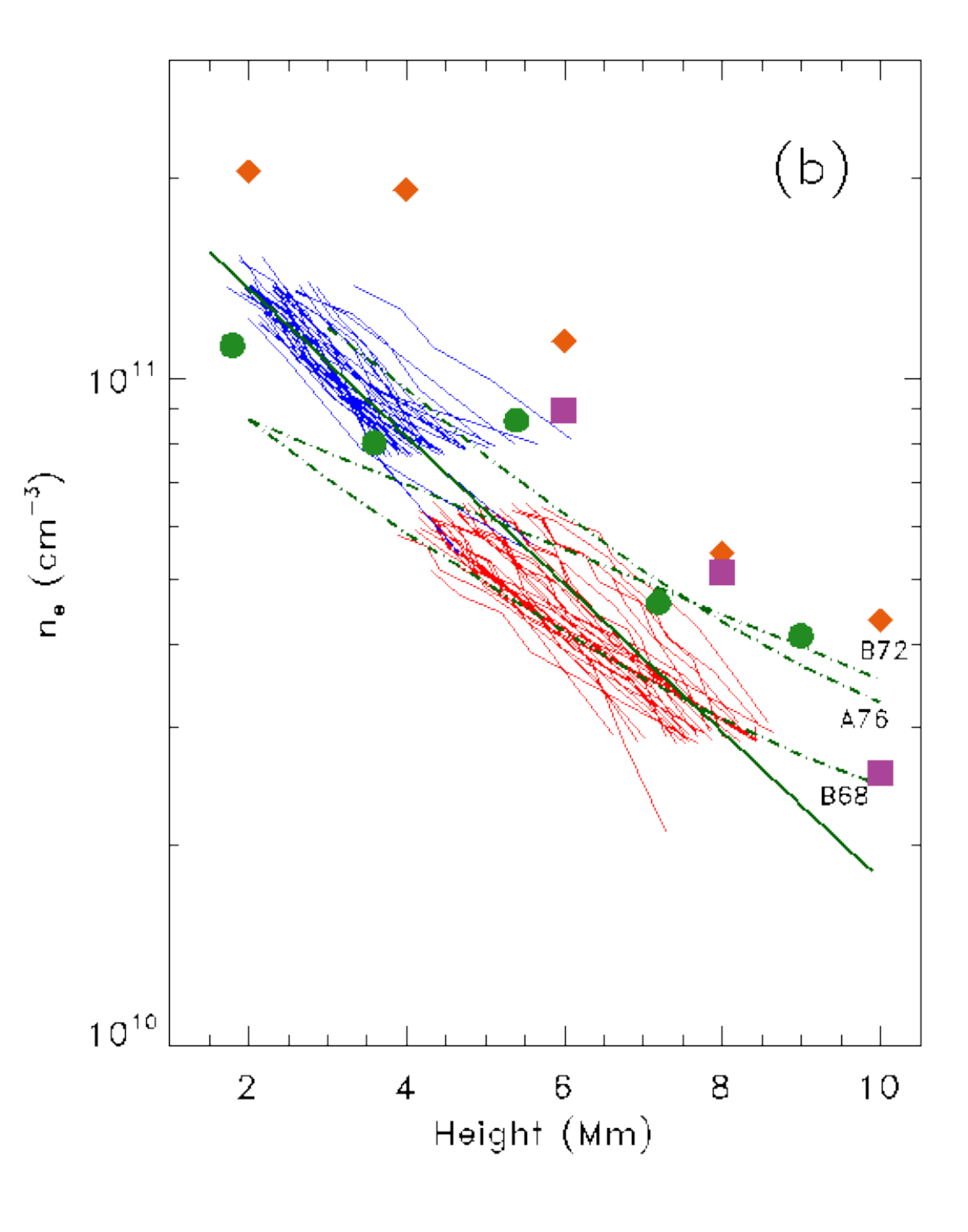}
\caption{Spicule densities inferred from the \mml\ observations. Densities based on 1.25~mm measurements are shown in blue and those based on 3~mm measurements are shown in red. In both panels, the solid green line represents the fit to the aggregated 1.25 and 3~mm data. The data points are from \citet{Alissandrakis2018} (filled green circles), \citet{Beckers1972} (purple squares), and from \citet{Krall1976} (orange diamonds). Note that a low-density solution has been dropped from the points from Alissandrakis {\sl et al.} for a height of 5.4~Mm.  a) Spicules are assumed to be isothermal with a temperature of $1.5\times 10^4$~K. The dash-dot lines labeled ``B68", ``B72", and ``A76" are the densities resulting from spicular filling factors based on the model of \citet{Beckers1968, Beckers1972}, and \citet{Athay1976}, respectively. b) Spicules are assumed to be heated from a temperature of $1.5\times 10^4$~K at a height of 2~Mm to $10^5$~K at a height of 15~Mm. The dash-dot lines again indicate densities corrected for filling factors derived from the authors indicated. }
\end{center}
\end{figure}

\subsection{Spicule Filling Factor}

Historically, the areal filling factor -- the fraction of the solar surface populated by spicules -- has received a good deal of attention. Summaries are given by \citet{Beckers1968, Beckers1972} and \citet{Athay1976} and are largely based on observations in H$\alpha$. More recently, discrete structures that had been considered ``spicules" have themselves been shown to display substructure.  \citet{Skogsrud2014} found that, when observed with very high angular resolution in H$\alpha$, individual ``spicules'' with transverse scales of order 0.5-1~Mm comprise multiple threads. These threads move in tandem as part of the overall spicule dynamics.  In addition, there are hints that threads may be of different temperatures \citep{Skogsrud2015, Depontieu2021}; that is, spicules may be multi-thermal, composed of threads with a range of temperatures. However, little is known about the distribution of plasma in threads as a function of temperature within the common, dynamical ``spicule'' structure. Given the insensitivity of electron number densities inferred from \mml\ observations to spicule temperature, we consider a ``spicule" to be the ensemble of substructures with common dynamics and consider observed values of $T_b$ to represent a useful mean. That is, while the spicule may comprise threads with temperatures ranging from $\sim 10^4$~K to $10^5$~K, the electron number density $n_e(h)$ in the common dynamical structure is insensitive to the temperature of the constituent threads or their distribution. We therefore do not consider the issue of multi-thermal threads and their filling factor within the spicule and instead focus on the areal filling factor of spicules $f_S$ as defined above and the corresponding line of sight filling factor ${f_S}^{1/2}$.

Surprisingly, spicule counts and filling factors have not been refined and updated by more recent studies to our knowledge.\citet{Beckers1968} presented a model of spicule temperature, density, and filling factor as a function of height, for which $f_S$ declines from $f_S=2.2\%$ at a height of 3 Mm, to 0.6\% at a height of 7~Mm (used as a fiducial by many authors), dropping below 0.1\% at 11~Mm, and rapidly declining at yet greater heights. Gathering the results of several previous workers, \citet{Beckers1972} concluded that the number of spicules seen in optical wavelengths declines exponentially with height with a scale height of 1.75~Mm. The decrease in the number of spicules as a function of height is equivalent to a decrease in areal filling factor with height. In this case $f_S(h)=\Pi(h)A_S/A_\odot$ where $\Pi(h)$ is the total number of spicules on the Sun above height $h$ \citep{Beckers1972}, $A_S=\pi D_S^2/4$ is the cross-sectional area of a spicule, $A_\odot$ is the surface area of the Sun. \citet{Athay1976} reported similar results. As indicated by Eqn.~(9), with $D=N D_S$, the estimate for the electron number density depends on the number of spicules along the line of sight as $N^{-1/2}$.  We have $N\sim D/D_S=f_S^{1/2} D_{LOS}/D_S$, where $D_{LOS}$ is the line of sight along some height $h$.  Following \citet{Athay1976} to estimate $D_{LOS}$ we find that, once corrected for the angular resolution of the ALMA observations, we expect no more than $N(h)\sim 10-20\ f_S(h)^{1/2}$ spicules along the line of sight on average. In practice, for a filling factor of 1\% we expect only 1-2 spicules along the line-of-sight on average with variation by a factor of a few. For example, spicules occur in network cell boundaries. The ALMA maps of the limb in each wavelength band contain 2-3 cells along the limb and along the line-of sight. If the line of sight aligns with one or more cell boundaries, the number of spicules may well be a factor of a few higher along that particular line of sight. However, since the estimated density $n_e(h)\propto f_S^{-1/4}$, the differences are minor in practice. To show this we have computed spicule densities corrected for the filling factor using values tabulated or inferred from \citet{Beckers1968, Beckers1972}, and \citet{Athay1976} for both the ISO and the TR models (Fig.~11ab). We see that the corrected densities do not depart by a large factor from the aggregate fit ($1.25+3$~mm) given in Table~2 (solid green line) and remain roughly consistent with the densities based on O/UV measurements. In fact, the model based on the filling factor given by \citet{Athay1976} is a better fit to the O/UV data than the simple aggregate fit. It is important to emphasize that spicule counts on which filling factors are based were derived from observations in H$\alpha$, corresponding to relatively cool material. 

It is striking that the densities inferred from the \mml\ observations for both the ISO and TR models, as well as data corrected for the spicule filling factor, are consistent with those derived from O/UV observations. While it may not initially be surprising that the ISO model ($T_e=1.5\times 10^4$~K) is in agreement, it is perhaps less obvious that the TR model would also be in agreement. For the TR model (Fig.~11b), we suppose the spicule temperature increases from $1.5\times 10^4$~K at a height of 2~Mm to a transition region value of $10^5$~K at a height of 15~Mm. If the model was even qualitatively correct, spicules would be observed to fade with height in cool lines like H$\alpha$, Ca~II~H, Mg~II~k and appear at successively greater heights in lines in C~II, and yet hotter lines in He~II and Si~IV. Indeed, this idea goes back to \citet{Pneuman1978} and, more recently, the fading of spicules in the Ca~II~H line and their appearance in hotter lines was initially interpreted within this framework; e.g., \citet{Pereira2014}. If it were correct, the O/UV densities which are based on inversion of cool lines, would presumably decline much more rapidly with height than those derived from the \mml\ observations and the TR model because the densities derived from \mml\ are insensitive to the change in temperature with height. That is, ALMA would continue to be sensitive to material that was no longer visible in H$\alpha$, Ca~II, or Mg~II. This is not observed. ALMA- and O/UV-derived densities decline with height at approximately the same rate.

On the other hand, the fact that spicules undergo some degree of heating is indisputable. But how much material is heated and how is it distributed within (threads) or around (sheaths, tips) spicules? What is the mean spicule differential emission measure (DEM) as a function of height? Consider the height-time diagrams shown for emission from spicules in lines spanning a wide range of temperatures (\citealt{Skogsrud2015}; their Fig.~5):  Emission at all temperatures is present along the length of each of the spicules displayed, from their base in the chromosphere up to their maximum extension. While spicules in Ca~II~H often fade with height, those in Mg~II do not \textcolor{blue}{(see \S1)}. In fact, with the exception of Ca~II, $\approx 2/3$ or more of the 54 spicules analyzed by \citet{Skogsrud2015} displayed nearly identical parabolic paths in height-time diagrams for all other lines and temperatures. We suggest that rather than fading from view in a given spectral line due to heating -- suggesting a change in filling factor as a function of height and temperature -- multi-thermal material is present in spicules at all heights. The change in filling factor does not appear to be temperature-dependent. We conclude that spicule heights seen in all temperatures do not extend much more than 15~Mm, as was also concluded by \citet{Withbroe1983} in his analysis of EUV emission from spicules. 

To summarize, our upper limits to spicule densities compare favorably with estimates based on O and UV estimates. When corrected for the line-of-sight filling factor, the density variation with height flattens somewhat and is arguably in better agreement with O/UV values.  Given the insensitivity of density estimates to spicule temperature, similar results are obtained for spicule densities for both the ISO and TR heating models. We suggest that, in fact, whatever the temperature of spicular material and its DEM distribution, the electron number density derived from the \mml\ observations captures material at all likely temperatures. We return to this point in \S5.

\subsection{Spicule Lifetimes and Kinematics at \mml}


\begin{figure}[ht!]
\begin{center}
\includegraphics[width=6in]{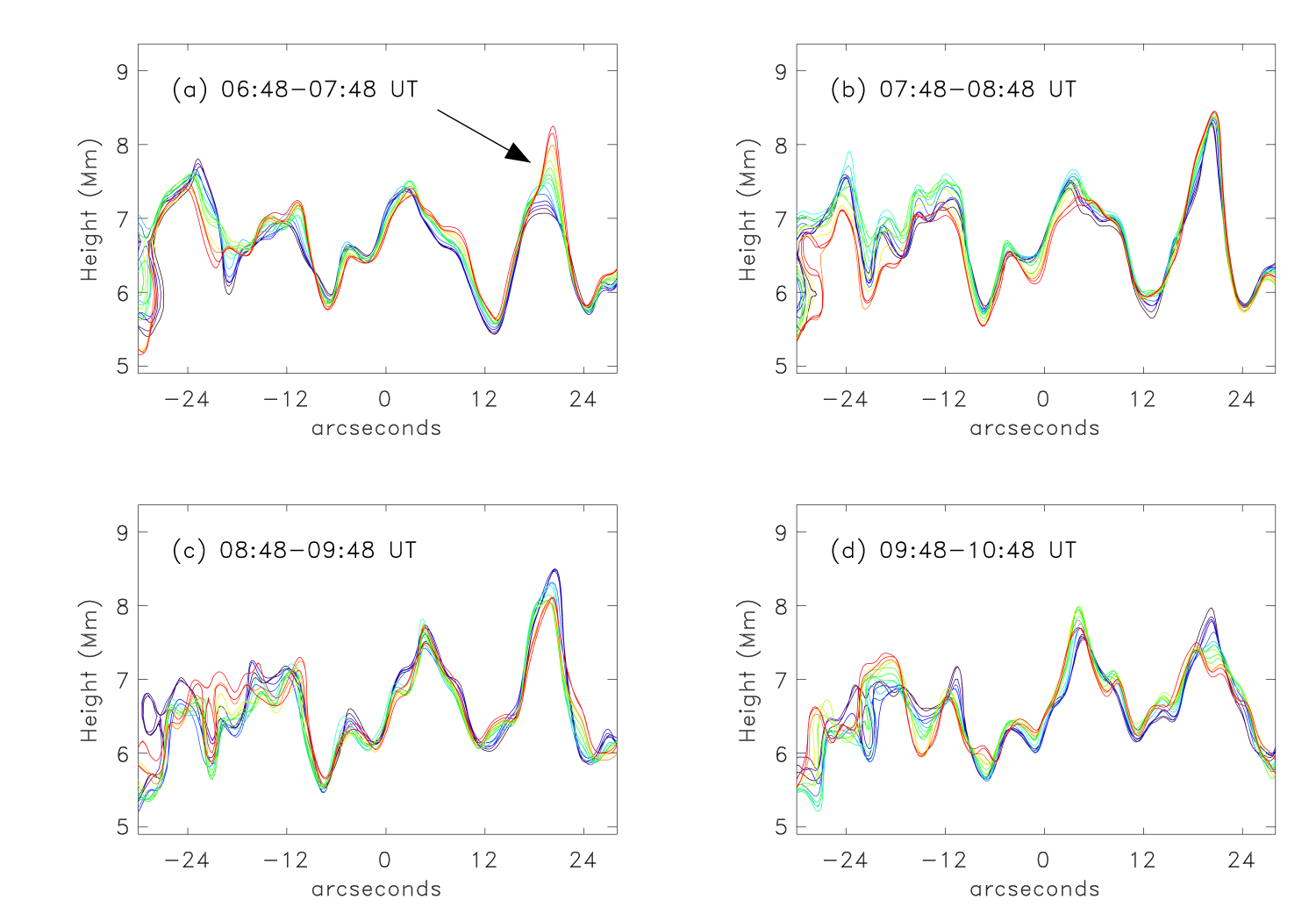}
\caption{Contours of a constant brightness temperature of 1500~K. Each contour represents a 4~s interval and each panel represents 1~min of time. Time ranges are given as mm:ss after 14:00 UT. The contours are color coded such that time advances from violet through the rainbow spectrum to red. a) the rise of a prominent spicule structure (right side) during the course of 1~min; b) the structure stalls; c,d) and then retracts.}
\end{center}
\end{figure}

In this section we briefly consider spicule lifetimes and motions. We refer primarily to 3~mm observations, for which we have several time series available, each of 10~min duration and with a cadence of 2~s. In order to enhance the visibility of spicule structures the data were background-subtracted and a radial-gradient filter was applied. In viewing a 10~min time series of filtered snapshot images one has the impression of constant motion, with radial, non-radial, and transverse motions. However, the angular resolution is marginal and the contrast between spicules is low. It is therefore difficult to identify discrete spicules in order to track their apparent motions. A detailed analysis of spicule lifetimes and kinematics is therefore outside the scope of the present work. We instead present a few examples of spicule motions seen in the 3~mm time series.

We examined the time series at 3~mm in two ways: stack plots and time-distance plots. Fig.~12 shows an example of stack plots in which the 1500~K contour is plotted every 4~s for four successive minutes in the 3~mm band. The contours advance in time from indigo to red in each panel. The spicule or spicule bundle appear as features that persist for the several minutes shown but the contours change significantly in detail. For example, the rightmost peak (arrow) shows the rise of a spicule or spicule bundle (12a); it then stalls (12b) and retracts (12c,d) suggesting a lifetime of $\approx\!4$~min. A review of the time sequence, however, shows that multiple structures are implicated, as also suggested in Fig.~12d. 

While stack plots provide an overall impression of spicules rising and falling, many motions are non-radial. Therefore, we have also formed several time-distance plots where the variation in $T_b$ on a cut along or across a spicule structure is plotted as a function of time (Fig.~13). Since the cuts are not radial, in general, the plot ordinate is the distance along the cut rather than height. Fig.~13a shows a striking example of a spicule oriented at $\approx 50^\circ$ to the radial direction; a cut is considered that is transverse to the spicule. The time-distance plot shows a transverse oscillation, with a lateral displacement of $\sim 1$ Mm in approximately 50~s. We note that \citet{dePontieu2007a} report an example of a transverse oscillation of a spicule with a similar lateral displacement, observed in the Ca II H line (3968 \AA) by the {\sl Hinode} SOT. More generally, \citet{dePontieu2007a} suggest that Alfv\'en waves with periods $100-500$~s are ubiquitous in the chromosphere. \citet{Okamoto2011} analyzed a sample of 89 spicules and found Alfv\'en waves with a median periods of 45~s, similar to that observed at 3~mm. Fig.~13b shows the time-distance plot of spicule shown in Fig.~12. The cut is nearly radial in this instance. The spicule rises at a speed of $\approx\!55$ km-s$^{-1}$, stalls for nearly a minute, and then retracts slowly although there are signs that more than one structure is present. Fig.~13c shows an example of a structure that changes from an angle of $\sim\!60^\circ$ to a more nearly radial orientation, traversing more than 2 Mm in 40~s, implying a transverse speed of more than 50 km-s$^{-1}$. The cut is again roughly transverse to the spicule motion.

The angular resolution is better at 1.25~mm but the image cadence is too long to reliably track the motions of discrete spicules, although there are examples where a given spicule appears to persist for at least two maps; i.e., for $\sim\!2-4$~min.  While these examples show that spicules show diverse motions at \mml, as they do in other wavelengths, more quantitative and comprehensive assessment of spicule kinematics at \mml\ awaits observations of spicules at \mml\ with both sub-arcsecond angular resolution and a time resolution of $\sim\!10$s. 

\begin{figure}[ht!]
\begin{center}
\includegraphics[width=3.75in]{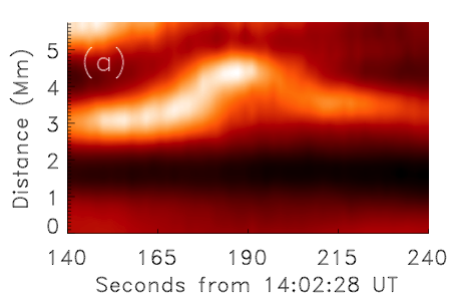}
\includegraphics[width=3.75in]{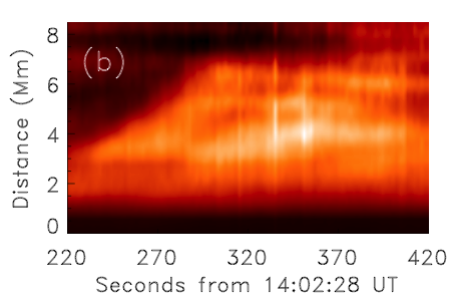}
\includegraphics[width=3.75in]{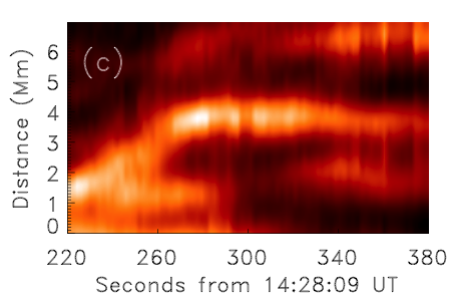}
\caption{Position-time plots of $T_b$ spicule-like structures: a) a structure that displays a transverse oscillation with a period of $\approx\!50$~s; b) a structure that rises with a speed of $\approx\!60$ km-s$^{-1}$ and then stalls and retracts slightly; c) a spicule that changes orientation from $\sim\!60^\circ$ to the radial direction to a stable radial orientation.}
\end{center}
\end{figure}


\section{Summary and Discussion}

We have presented ALMA observations of solar spicules in a coronal hole at \mml\ with arcsecond angular resolution and a time resolution of 2~s (3~mm) and 106~s (1.25~mm). In this paper, we have summarized some of the unique challenges associated with solar observations with ALMA near the limb, including trade-offs associated with mosaicking, the use of a heterogeneous array, corrections for ``seeing" using self-calibration techniques, and the recovery of flux on large spatial scales. We  have measured a number of properties of spicules in the north polar region of the Sun in a coronal hole in each wavelength band. We have qualitatively compared the data with observations made at optical, UV, and EUV wavelengths and have made quantitative measurements of the column emission measure and electron number density as a function of height above the solar limb. Our results may be summarized as follows:

\begin{enumerate}
\item Multi-wavelength comparison: We compared our maps with SJI images observed in the IRIS 2696 A bandpass, dominated by Mg II k line emission, and the IRIS 1400\AA\ bandpass, dominated by Si IV line emission at selected times. We found a good correspondence between spicules observed in the UV and those observed at \mml\ (Fig.~8). 
\item Limb extension at \mml: We established the mean solar limb at wavelengths of 3~mm and 1.25~mm by fitting the limb of the full disk total power maps. Small adjustments to the solar limb were needed to reconcile the opacity measurements in the two wavelength bands. We find that the mean limb is 5.4~Mm above the photospheric limb at 3~mm and 2~Mm above the limb at 1.25~mm. The minimum limb, which we take to be the height of the chromosphere, is at heights of 3~Mm and 1~Mm at wavelengths of 3~mm and 1.25~mm, respectively. 
\item EUV absorption: In agreement with \citet{Yokoyama2018} and \citet{Alissandrakis2019} we find that coronal EUV emission is absorbed by foreground spicule structures observed at \mml. We attribute the absorption to neutral hydrogen, neutral helium, and singly ionized helium. We find that the column depth of these species is sufficient to render the upper chromosphere and lower parts of spicules optically thick to coronal EUV emission but is insufficient to play a significant role in the 3~mm and 1.25~mm emission, which we attribute to thermal free-free emission. 
\item Spicule brightness temperature: We measured the brightness temperature $T_b$ of spicules as a function of height and find that  $T_b \lesssim 10^4$~K low in the chromosphere where they are optically thick. It then declines rapidly in both wavelengths above 2~Mm (1.25~mm) and 4~Mm (3~mm) with a scale height of $\approx\!2$~Mm in both wavelengths. The emission is optically thin over these heights. The effective temperature $T_e$ of spicules is not otherwise well constrained by the observations, however. 
\item Spicule densities: From the variation of $T_b(h)$ we infer the variation in optical depth $\tau(h)$ and column emission measure for two schematic models of $T_e(h)$. Assuming spicular material is confined to a cylindrical structure with a diameter $D_S=0.5$~Mm, we establish the variation in spicule density with height as a means of comparing our results to similar measurements made using O/UV data. The spicular electron number density declines exponentially with a scale height $L_{n_e}\approx 3.5$~Mm (ISO model) or $\approx\!4$~Mm (TR model). We emphasize that spicule densities inferred from \mml\ emission are quite insensitive to spicule temperature. We find that our upper limits are comparable to spicule densities inferred from O/UV observations tabulated by \citet{Alissandrakis2018}. Taking the line of sight filling factor into account, mean spicular densities decrease and the variation with height is somewhat flatter than is the variation when the filling factor is ignored.
\item Spicule lifetimes: Given the limited angular resolution and low contrast at 3~mm and the limited imaging cadence at 1.25~mm it is difficult to assess spicule lifetimes. We have nevertheless isolated several structures that persist for 100-200~s but the lifetimes are not well-constrained. 
\item Spicule kinematics at \mml: One has the impression of constant motion in time sequences of the 3~mm scans but with radial and non-radial motions. As is the case in establishing spicule lifetimes at \mml\ it is likewise difficult to identify discrete spicules and establish their speeds and trajectories. Time-distance plots were nevertheless constructed for a small number of spicules and we find that the inferred speeds are consistent of spicules $50-60$ km s$^{-1}$. We showed a vivid example of an oscillating spicule and one example for which a spicule appears to reorient from a highly oblique to a nearly radial position. 
\end{enumerate}

We now consider our results in context of the role spicules play in the mass and energy budget of the solar atmosphere. Many authors have suggested the possible importance of spicules as a means of supplying mass to the solar atmosphere sufficient to offset losses due to the solar wind. The mass loss rate of the solar wind is $\dot{M}\approx2-3\times 10^{-14}$ M$_\odot$~yr$^{-1}=1-2\times 10^{13}$ ions and electrons cm$^{-2}$~s$^{-1}$.  Assuming -- as have many authors previously -- a spicular density $n_e=6\times 10^{10}$~cm$^{-3}$, a speed $v_S=25$ km s$^{-1}$, and a filling factor $f=0.6\%$, referenced to a height of 7~Mm, spicules provide a mass flux $q_S=n_e v_S f$ to the low solar atmosphere that is two orders of magnitude greater than that needed to supply the solar wind mass flux; i.e, a rate of $\sim 10^{15}$ cm$^{-2}$ s$^{-1}$ (e.g., \citealt{Beckers1972, Pneuman1978, Athay1982, Withbroe1983, Sterling2000}, and many others). Clearly, most of the mass flux carried upward by spicules must return to the chromosphere. 

\citet{Beckers1972} pointed out, however, that the kinetic energy flux $f\rho v_S^3/2$ carried by solar spicules, ($\rho$ being the mass density) averaged over the solar surface, is only $\approx 5\times 10^3$ erg cm$^{-2}$ s$^{-1}$, far less than that needed to compensate for energy losses from the solar corona: $\approx\!5\times10^5$ erg cm$^{-2}$ s$^{-1}$ in a coronal hole \citep{Withbroe1988}.  Noting the transition region network downflows of a few km s$^{-1}$ discovered by instruments on board rockets, {\sl Skylab}, and OSO-8,  however, \citealt{Pneuman1978} pointed out that they carried significant energy flux in the form of enthalpy, greatly exceeding that supposed to be carried by conduction, leading them to suggest that the downflows may represent ``spicular material returning to the chromosphere after being heated to coronal temperatures."  If this is true, spicules could indeed play a significant role in the mass and energy budget of the solar atmosphere. Based on {\sl Hinode}/SOT and SOHO or SDO observations, \citet{DePontieu2009} and \citet{DePontieu2011} revived the idea that spicular plasma is heated to high temperatures and, while most of it falls back to the chromosphere, sufficient plasma is heated to coronal temperatures and launched to coronal heights to play a significant role in coronal mass supply and energetics. In a study of spicules in H$\alpha$ using the Goode Solar Telescope and SDO/AIA 171\AA\ emission, \citet{Samanta2019} also argue that spicules may provide hot plasma to the corona. As we discussed in \S4, the simple model suggested by \citet{Pneuman1978} and embodied by TR model is not now thought to be realistic; in fact, spicules are multi-thermal although the details of how plasma at various temperatures is distributed within a spicule (threads, sheathes) and the mean spicule DEM as a function of height are not well-constrained. However, the ALMA observations of spicules are largely immune to these unknown factors. The inferred electron number density captures material distributed over a wide range of temperature, whatever it may be.  


Fig.~14 shows $q_S$ carried upward by spicules based on fits to the ALMA data for the two temperature models and the filling factors discussed in \S4.6. We use a spicule speed of $v_S=50$ km s$^{-1}$, commensurate with modern observations. Note that the inferred mass flux is comparable to the oft-cited value of $\sim10^{15}$ cm$^{-2}$ s$^{-1}$ near the reference height of 7~Mm. We have also replotted the mass flux shown by \citet{Pneuman1978} in their Fig.~3. We see that in all cases the mass flux drops below $10^{13}$~cm${-2}$ s$^{-1}$ at a height of 11-14~Mm; i.e., below that value needed to sustain the solar corona against losses to the solar wind. For the \mml\ observations, as we argued in \S4.6, we cannot appeal to the idea that the material fades from view due to heating since ALMA is roughly equally sensitive to all temperatures from $10^4-10^5$ K. There is simply not enough spicular material, hot or otherwise, extending to significant heights. We suggest that the effective decrease in filling factor and, hence, mass flux simply reflects the reduction in the number of spicules reaching a given height, with few exceeding 15~Mm in height. 

\begin{figure}[ht!]
\begin{center}
\includegraphics[width=5in]{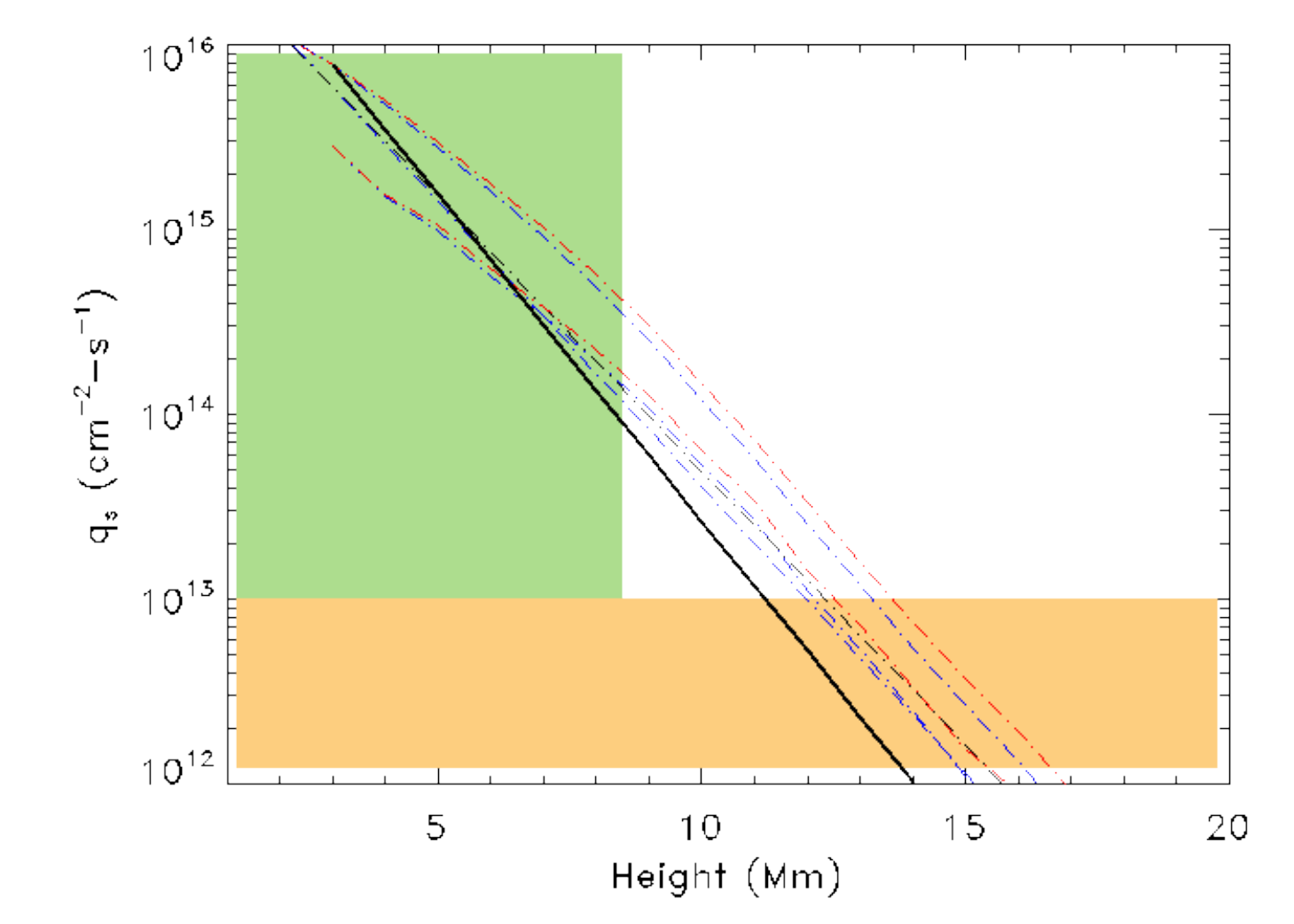}
\caption{Mass flux as a function of height for the two temperature models explored and the models of the spicule filling factor. The dot-dash blue lines represent the ISO model and the dot-dash red lines represent the TR model. The three models for filling factor (not labeled) described in \S5.6 all yield similar results. The solid black line is the mass flux estimated by \citet{Pneuman1978}. The pale green box indicates the range of heights for which \mml\ measurements are available whereas the pale orange region indicates the level below which the mass flux cannot sustain the corona against losses due to the solar wind.   }
\end{center}
\end{figure}

We acknowledge that our \mml\ measurements extend to heights of only 8.5~Mm and that caution is therefore warranted in extrapolating fits to greater heights. Nevertheless, our results are clearly in tension with claims that spicules play an important role in the mass budget of the solar atmosphere, suggesting that spicules do not carry sufficient mass into the solar corona to offset losses due to the solar wind outflow. Nor, as a result, do they carry substantial amounts of material heated to high temperatures into the corona. A number of other authors, have also questioned the importance of spicules in the heating the corona from a variety of perspectives \citep{Klimchuk2012, Judge2012, Patsourakos2014, SowMondal2022}. Does this mean that spicules are irrelevant, energetically? Perhaps not. {\sl Hinode} and IRIS observations provide line-of-sight velocities and are sensitive to the waves in spicules \citep{He2009, Tavabi_2011}. \citet{DePontieu2007} report observations of ubiquitous transverse velocity oscillations in spicules with amplitudes of 10-25 km s$^{-1}$, displacements of 0.5-1~Mm, on periods of 100-500 s. These are interpreted as Alfv\'en waves (or fast kink waves) propagating into the Sun’s corona carrying an energy flux of $10^5$ ergs cm$^{-2}$ s$^{-1}$, perhaps sufficient to play a role in heating the quiet corona and/or accelerating the solar wind. \citet{Suematsu2008} report both transverse motions and spicular rotation or torsional oscillation; see also \citet{DePontieu2012}. A detailed study of transverse spicule motions by \citet{Okamoto2011} found that the situation with respect to waves in spicules is complex, with upward propagating, downward propagating, and standing waves present. The wave periods in question are shorter than those presented by \citet{DePontieu2007} and, while they carry a significant energy flux upward their observations suggest that the bulk of it is reflected at the transition region. \citet{Zaqarashvili2009} reviewed oscillations and waves in spicules.  A review of chromospheric waves, including those in spicules, fibrils, and mottles by \citet{Jess2015} indicates that solar magnetoseismology of these waves offers rich diagnostic potential for understanding their role in the chromospheric and coronal heating. 

\section{Future Prospects}

Continuum observations of solar spicules at \mml\ with high angular resolution and high time resolution provide a new suite of spicule diagnostics that are complementary to observations in spectral lines at O/UV wavelengths. The fact that the source function is Planckian at \mml\ and that the emission is in the Rayleigh-Jeans regime greatly simplifies the radiative transfer and its interpretation. There remain limitations, however. While the time resolution at 3~mm is excellent, the angular resolution available in ALMA Cycle 6 in 2018 was not sufficient to fully resolve discrete spicules. More recently, a larger array configuration (C43-4) has been approved for solar observations in the 3~mm band that provides a nominal resolution of $0.92"$. While the angular resolution at 1.25~mm is perhaps sufficient to spatially resolve spicules as kinematically coherent structures, if not their threadlike substructure, the use of a mosaicking imaging strategy to enlarge the field of view limited the cadence of complete mosaics to no better than $\approx 2$~min, which is not sufficient to adequately resolve spicule kinematics at this wavelength. Both high angular resolution and time resolution are needed.

Looking forward, while it is desirable, it is not yet possible to perform simultaneous multi-band observations of the Sun with ALMA. Nevertheless, future observations should exploit the fact that ALMA can now observe in four wavelength bands: 0.86, 1.25, 1.51, and 3~mm. Additional bands may be commissioned for solar observing, including the recently-added ALMA Band 1 (7.5~mm). The 7.5~mm band would be extremely useful for observing spicules to greater heights than was possible here because the optical depth at a given height above the limb would be $\sim\!4$ greater at 7.5~mm than that at 3~mm, allowing the results presented here to be confirmed or refuted. However, multi-wavelength observing programs with ALMA must accept the reality that the observations must be executed over multiple sessions spanning hours, days, or even multiple antenna configurations. For example, workers may wish to statistically characterize spicules in a common environment -- e.g., coronal holes -- with matched angular resolution. To achieve nominal angular resolution of 0.8-1" in all four wavelength bands would require observing in antenna configurations C43-1 (0.86~mm), C43-2 (1.25~mm), C43-3 (1.51~mm), and C43-4 (3~mm), and once available, C43-6 (7.5 mm). Alternatively, given the importance of robustly restoring large angular scales at the limb, workers may wish to optimize sampling of the inner part of the {\sl uv} plane in each band to allow more effective exploitation of, for example, subtraction of the limb response from the data. In addition, the mapping strategy for wavelength bands shorter than the 3~mm wavelength band should carefully consider the trade-offs between the field of view and the mapping cadence. Three- or four-point mosaics, yielding a time cadence of 20-30~s, would resolve most spicule kinematics. Finally, joint \mml\, O, UV, and EUV observations should be fully exploited to jointly constrain spicule plasma parameters. Of particular importance is constraining the mean scale height of spicule emission in UV bands, and of determining the DEM of spicules. 


\acknowledgments
This paper makes use of the following ALMA data: ADS/JAO.ALMA\#2018.1.00199.S. ALMA is a partnership of ESO (representing its member states), NSF (USA) and NINS (Japan), together with NRC (Canada), MOST and ASIAA (Taiwan), and KASI (Republic of Korea), in cooperation with the Republic of Chile. The Joint ALMA Observatory is operated by ESO, AUI/NRAO and NAOJ. The National Radio Astronomy Observatory is a facility of the National Science Foundation operated under cooperative agreement by Associated Universities, Inc. AN acknowledges support by the ERC Synergy Grant (GAN: 810218) "The Whole Sun". SW acknowledges support for basic research from AFOSR through LRIR 23RVCOR003. Data were acquired by GONG instruments operated by NISP/NSO/AURA/NSF with contributions from NOAA. IRIS is a NASA small explorer mission developed and operated by LMSAL with mission operations executed at NASA Ames Research Center and major contributions to downlink communications funded by ESA and the Norwegian Space Centre. Data from the Solar Dynamics Observatory is used courtesy of NASA/SDO and the AIA, EVE, and HMI science teams. 

The authors thank Bart De Pontieu for his careful reading of a draft manuscript and for his perspective and critical input, which improved the paper. 

\vspace{5mm}
\facilities{{ALMA, GONG, IRIS, SDO}}
\software{python, CASA, AIPS, IDL}

\bibliography{Spicules.bib}{}

\begin{thebibliography}{}
\expandafter\ifx\csname natexlab\endcsname\relax\def\natexlab#1{#1}\fi
\providecommand{\url}[1]{\href{#1}{#1}}
\providecommand{\dodoi}[1]{doi:~\href{http://doi.org/#1}{\nolinkurl{#1}}}
\providecommand{\doeprint}[1]{\href{http://ascl.net/#1}{\nolinkurl{http://ascl.net/#1}}}
\providecommand{\doarXiv}[1]{\href{https://arxiv.org/abs/#1}{\nolinkurl{https://arxiv.org/abs/#1}}}

\bibitem[{Alissandrakis {et~al.}(2022{\natexlab{a}})Alissandrakis, Bastian, \&
  Braj{\v{s}}a}]{Alissandrakis2022a}
Alissandrakis, C.~E., Bastian, T.~S., \& Braj{\v{s}}a, R. 2022{\natexlab{a}},
  Frontiers in Astronomy and Space Sciences, 9, 981320,
  \dodoi{10.3389/fspas.2022.981320}

\bibitem[{Alissandrakis {et~al.}(2022{\natexlab{b}})Alissandrakis, Bastian, \&
  Nindos}]{Alissandrakis2022}
Alissandrakis, C.~E., Bastian, T.~S., \& Nindos, A. 2022{\natexlab{b}}, \aap,
  661, L4, \dodoi{10.1051/0004-6361/202243774}

\bibitem[{Alissandrakis {et~al.}(2020)Alissandrakis, Nindos, Bastian, \&
  Patsourakos}]{Alissandrakis2020}
Alissandrakis, C.~E., Nindos, A., Bastian, T.~S., \& Patsourakos, S. 2020,
  Astronomy {\&} Astrophysics, 640, A57, \dodoi{10.1051/0004-6361/202038461}

\bibitem[{Alissandrakis {et~al.}(2017)Alissandrakis, Patsourakos, Nindos, \&
  Bastian}]{Alissandrakis2017}
Alissandrakis, C.~E., Patsourakos, S., Nindos, A., \& Bastian, T.~S. 2017,
  \aap, 605, A78, \dodoi{10.1051/0004-6361/201730953}

\bibitem[{Alissandrakis \& Valentino(2019{\natexlab{a}})}]{Alissandrakis2019}
Alissandrakis, C.~E., \& Valentino, A. 2019{\natexlab{a}}, \solphys, 294, 96,
  \dodoi{10.1007/s11207-019-1486-7}

\bibitem[{Alissandrakis \& Valentino(2019{\natexlab{b}})}]{Alissandrakis2019a}
---. 2019{\natexlab{b}}, \solphys, 294, 146, \dodoi{10.1007/s11207-019-1545-0}

\bibitem[{Alissandrakis \& Vial(2023)}]{Alissandrakis2023}
Alissandrakis, C.~E., \& Vial, J.~C. 2023, \solphys, 298, 18,
  \dodoi{10.1007/s11207-023-02111-y}

\bibitem[{Alissandrakis {et~al.}(2018)Alissandrakis, Vial, Koukras, Buchlin, \&
  Chane-Yook}]{Alissandrakis2018}
Alissandrakis, C.~E., Vial, J.~C., Koukras, A., Buchlin, E., \& Chane-Yook, M.
  2018, \solphys, 293, 20, \dodoi{10.1007/s11207-018-1242-4}

\bibitem[{Antolin {et~al.}(2018)Antolin, Schmit, Pereira, De~Pontieu, \&
  De~Moortel}]{Antolin2018}
Antolin, P., Schmit, D., Pereira, T. M.~D., De~Pontieu, B., \& De~Moortel, I.
  2018, \apj, 856, 44, \dodoi{10.3847/1538-4357/aab34f}

\bibitem[{Anzer \& Heinzel(2005)}]{Anzer2005}
Anzer, U., \& Heinzel, P. 2005, \apj, 622, 714, \dodoi{10.1086/427817}

\bibitem[{Athay(1976)}]{Athay1976}
Athay, R.~G. 1976, The solar chromosphere and corona: Quiet sun, Vol.~53,
  \dodoi{10.1007/978-94-010-1715-2}

\bibitem[{{Athay} \& {Holzer}(1982)}]{Athay1982}
{Athay}, R.~G., \& {Holzer}, T.~E. 1982, \apj, 255, 743, \dodoi{10.1086/159873}

\bibitem[{Bastian {et~al.}(2022)Bastian, Shimojo, B{\'a}rta, White, \&
  Iwai}]{Bastian2022}
Bastian, T.~S., Shimojo, M., B{\'a}rta, M., White, S.~M., \& Iwai, K. 2022,
  Frontiers in Astronomy and Space Sciences, 9, 977368,
  \dodoi{10.3389/fspas.2022.977368}

\bibitem[{Beckers(1968)}]{Beckers1968}
Beckers, J.~M. 1968, \solphys, 3, 367, \dodoi{10.1007/BF00171614}

\bibitem[{Beckers(1972)}]{Beckers1972}
---. 1972, \araa, 10, 73, \dodoi{10.1146/annurev.aa.10.090172.000445}

\bibitem[{{Belkora} {et~al.}(1992){Belkora}, {Hurford}, {Gary}, \&
  {Woody}}]{Belkora1992}
{Belkora}, L., {Hurford}, G.~J., {Gary}, D.~E., \& {Woody}, D.~P. 1992, \apj,
  400, 692, \dodoi{10.1086/172031}

\bibitem[{Bose {et~al.}(2019)Bose, Henriques, Joshi, \& Rouppe van~der
  Voort}]{Bose2019}
Bose, S., Henriques, V. M.~J., Joshi, J., \& Rouppe van~der Voort, L. 2019,
  \aap, 631, L5, \dodoi{10.1051/0004-6361/201936617}

\bibitem[{Bose {et~al.}(2021)Bose, Joshi, Henriques, \& Rouppe van~der
  Voort}]{Bose2021}
Bose, S., Joshi, J., Henriques, V. M.~J., \& Rouppe van~der Voort, L. 2021,
  \aap, 647, A147, \dodoi{10.1051/0004-6361/202040014}

\bibitem[{{Carlsson} {et~al.}(2019){Carlsson}, {De Pontieu}, \&
  {Hansteen}}]{Carlsson2019}
{Carlsson}, M., {De Pontieu}, B., \& {Hansteen}, V.~H. 2019, \araa, 57, 189,
  \dodoi{10.1146/annurev-astro-081817-052044}

\bibitem[{Carlsson \& Stein(2002)}]{Carlsson2002}
Carlsson, M., \& Stein, R.~F. 2002, \apj, 572, 626, \dodoi{10.1086/340293}

\bibitem[{Chintzoglou {et~al.}(2018)Chintzoglou, De~Pontieu,
  Mart{\'\i}nez-Sykora, Pereira, Vourlidas, \& Tun~Beltran}]{Chintzoglou2018}
Chintzoglou, G., De~Pontieu, B., Mart{\'\i}nez-Sykora, J., {et~al.} 2018, \apj,
  857, 73, \dodoi{10.3847/1538-4357/aab607}

\bibitem[{Cornwell(1989{\natexlab{a}})}]{Cornwell1989b}
Cornwell, T. 1989{\natexlab{a}}, in Astronomical Society of the Pacific
  Conference Series, Vol.~6, Synthesis Imaging in Radio Astronomy, ed. R.~A.
  {Perley}, F.~R. {Schwab}, \& A.~H. {Bridle}, 277.
\newblock \url{https://ui.adsabs.harvard.edu/abs/1989ASPC....6..277C}

\bibitem[{Cornwell(1989{\natexlab{b}})}]{Cornwell1989a}
Cornwell, T. 1989{\natexlab{b}}, in Synthesis imaging in radio astronomy,

\bibitem[{Cornwell \& Fomalont(1989)}]{Cornwell1989}
Cornwell, T., \& Fomalont, E.~B. 1989, in Astronomical Society of the Pacific
  Conference Series, Vol.~6, Synthesis Imaging in Radio Astronomy, ed. R.~A.
  {Perley}, F.~R. {Schwab}, \& A.~H. {Bridle}, 185.
\newblock \url{https://ui.adsabs.harvard.edu/abs/1989ASPC....6..185C}

\bibitem[{Cornwell {et~al.}(1993)Cornwell, Holdaway, \& Uson}]{Cornwell1993}
Cornwell, T.~J., Holdaway, M.~A., \& Uson, J.~M. 1993, \aap, 271, 697.
\newblock \url{https://ui.adsabs.harvard.edu/abs/1993A&A...271..697C}

\bibitem[{Cotton(2015)}]{Cotton2015}
Cotton, W. 2015, Obit Memo 41: Image Combination by Feathering, Tech. rep.,
  National Radio Astronomy Observatory

\bibitem[{Daw {et~al.}(1995)Daw, Deluca, \& Golub}]{Daw1995}
Daw, A., Deluca, E.~E., \& Golub, L. 1995, \apj, 453, 929,
  \dodoi{10.1086/176453}

\bibitem[{{De Pontieu} {et~al.}(2012){De Pontieu}, {Carlsson}, {Rouppe van der
  Voort}, {Rutten}, {Hansteen}, \& {Watanabe}}]{DePontieu2012}
{De Pontieu}, B., {Carlsson}, M., {Rouppe van der Voort}, L.~H.~M., {et~al.}
  2012, \apjl, 752, L12, \dodoi{10.1088/2041-8205/752/1/L12}

\bibitem[{{De Pontieu} {et~al.}(2007){De Pontieu}, {Hansteen}, {Rouppe van der
  Voort}, {van Noort}, \& {Carlsson}}]{DePontieu2007}
{De Pontieu}, B., {Hansteen}, V.~H., {Rouppe van der Voort}, L., {van Noort},
  M., \& {Carlsson}, M. 2007, \apj, 655, 624, \dodoi{10.1086/509070}

\bibitem[{{De Pontieu} {et~al.}(2017){De Pontieu}, {Mart{\'\i}nez-Sykora}, \&
  {Chintzoglou}}]{DePontieu2017}
{De Pontieu}, B., {Mart{\'\i}nez-Sykora}, J., \& {Chintzoglou}, G. 2017, \apjl,
  849, L7, \dodoi{10.3847/2041-8213/aa9272}

\bibitem[{De~Pontieu {et~al.}(2009)De~Pontieu, McIntosh, Hansteen, \&
  Schrijver}]{DePontieu2009}
De~Pontieu, B., McIntosh, S.~W., Hansteen, V.~H., \& Schrijver, C.~J. 2009,
  \apjl, 701, L1, \dodoi{10.1088/0004-637X/701/1/L1}

\bibitem[{De~Pontieu {et~al.}(2007)De~Pontieu, McIntosh, Carlsson, Hansteen,
  Tarbell, Schrijver, Title, Shine, Tsuneta, Katsukawa, Ichimoto, Suematsu,
  Shimizu, \& Nagata}]{dePontieu2007a}
De~Pontieu, B., McIntosh, S.~W., Carlsson, M., {et~al.} 2007, Science, 318,
  1574, \dodoi{10.1126/science.1151747}

\bibitem[{De~Pontieu {et~al.}(2021)De~Pontieu, Polito, Hansteen, Testa, Reeves,
  Antolin, N{\'o}brega-Siverio, Kowalski, Martinez-Sykora, Carlsson, McIntosh,
  Liu, Daw, \& Kankelborg}]{Depontieu2021}
De~Pontieu, B., Polito, V., Hansteen, V., {et~al.} 2021, \solphys, 296, 84,
  \dodoi{10.1007/s11207-021-01826-0}

\bibitem[{Del~Zanna {et~al.}(2021)Del~Zanna, Dere, Young, \&
  Landi}]{DelZanna2021}
Del~Zanna, G., Dere, K.~P., Young, P.~R., \& Landi, E. 2021, \apj, 909, 38,
  \dodoi{10.3847/1538-4357/abd8ce}

\bibitem[{DePontieu {et~al.}(2007)DePontieu, McIntosh, Hansteen, Carlsson,
  Schrijver, Tarbell, Title, Shine, Suematsu, Tsuneta, Katsukawa, Ichimoto,
  Shimizu, \& Nagata}]{DePontieu_2007}
DePontieu, B., McIntosh, S., Hansteen, V.~H., {et~al.} 2007, Publications of
  the Astronomical Society of Japan, 59, S655, \dodoi{10.1093/pasj/59.sp3.s655}

\bibitem[{{DePontieu} {et~al.}(2011){DePontieu}, {McIntosh}, {Carlsson},
  {Hansteen}, {Tarbell}, {Boerner}, {Martinez-Sykora}, {Schrijver}, \&
  {Title}}]{DePontieu2011}
{DePontieu}, B., {McIntosh}, S.~W., {Carlsson}, M., {et~al.} 2011, Science,
  331, 55, \dodoi{10.1126/science.1197738}

\bibitem[{DePontieu {et~al.}(2014)DePontieu, Title, Lemen, Kushner, Akin,
  Allard, Berger, Boerner, Cheung, Chou, Drake, Duncan, Freeland, Heyman,
  Hoffman, Hurlburt, Lindgren, Mathur, Rehse, Sabolish, Seguin, Schrijver,
  Tarbell, Wülser, Wolfson, Yanari, Mudge, Nguyen-Phuc, Timmons, van Bezooijen,
  Weingrod, Brookner, Butcher, Dougherty, Eder, Knagenhjelm, Larsen, Mansir,
  Phan, Boyle, Cheimets, DeLuca, Golub, Gates, Hertz, McKillop, Park, Perry,
  Podgorski, Reeves, Saar, Testa, Tian, Weber, Dunn, Eccles, Jaeggli,
  Kankelborg, Mashburn, Pust, Springer, Carvalho, Kleint, Marmie, Mazmanian,
  Pereira, Sawyer, Strong, Worden, Carlsson, Hansteen, Leenaarts, Wiesmann,
  Aloise, Chu, Bush, Scherrer, Brekke, Martinez-Sykora, Lites, McIntosh,
  Uitenbroek, Okamoto, Gummin, Auker, Jerram, Pool, \&
  Waltham}]{DePontieu_2014}
DePontieu, B., Title, A.~M., Lemen, J.~R., {et~al.} 2014, Solar Physics, 289,
  2733, \dodoi{10.1007/s11207-014-0485-y}

\bibitem[{Dere {et~al.}(1997)Dere, Landi, Mason, Monsignori~Fossi, \&
  Young}]{Dere1997}
Dere, K.~P., Landi, E., Mason, H.~E., Monsignori~Fossi, B.~C., \& Young, P.~R.
  1997, \aaps, 125, 149, \dodoi{10.1051/aas:1997368}

\bibitem[{Ekers \& Rots(1979)}]{Ekers1979}
Ekers, R.~D., \& Rots, A.~H. 1979, in Astrophysics and Space Science Library,
  Vol.~76, IAU Colloq. 49: Image Formation from Coherence Functions in
  Astronomy, ed. C.~{van Schooneveld}, 61, \dodoi{10.1007/978-94-009-9449-2\_7}

\bibitem[{{Ewell} {et~al.}(1993){Ewell}, {Zirin}, {Jensen}, \&
  {Bastian}}]{Ewell1993}
{Ewell}, M.~W., J., {Zirin}, H., {Jensen}, J.~B., \& {Bastian}, T.~S. 1993,
  \apj, 403, 426, \dodoi{10.1086/172213}

\bibitem[{Fontenla {et~al.}(1990)Fontenla, Avrett, \& Loeser}]{Fontenla1990}
Fontenla, J.~M., Avrett, E.~H., \& Loeser, R. 1990, The Astrophysical Journal,
  355, 700, \dodoi{10.1086/168803}

\bibitem[{Golding {et~al.}(2014)Golding, Carlsson, \& Leenaarts}]{Golding2014}
Golding, T.~P., Carlsson, M., \& Leenaarts, J. 2014, \apj, 784, 30,
  \dodoi{10.1088/0004-637X/784/1/30}

\bibitem[{Greisen(2003)}]{Greisen2003}
Greisen, E.~W. 2003, in Information Handling in Astronomy - Historical Vistas
  (Springer Netherlands), 109--125, \dodoi{10.1007/0-306-48080-8_7}

\bibitem[{Grevesse {et~al.}(2007)Grevesse, Asplund, \& Sauval}]{Grevesse2007}
Grevesse, N., Asplund, M., \& Sauval, A.~J. 2007, \ssr, 130, 105,
  \dodoi{10.1007/s11214-007-9173-7}

\bibitem[{Harvey {et~al.}(1996)Harvey, Hill, Hubbard, Kennedy, Leibacher,
  Pintar, Gilman, Noyes, Title, Toomre, Ulrich, Bhatnagar, Kennewell,
  Marquette, Patron, Saa, \& Yasukawa}]{Harvey1996}
Harvey, J.~W., Hill, F., Hubbard, R.~P., {et~al.} 1996, Science, 272, 1284,
  \dodoi{10.1126/science.272.5266.1284}

\bibitem[{He {et~al.}(2009)He, Tu, Marsch, Guo, Yao, \& Tian}]{He2009}
He, J.-S., Tu, C.-Y., Marsch, E., {et~al.} 2009, Astronomy {\&} Astrophysics,
  497, 525, \dodoi{10.1051/0004-6361/200810777}

\bibitem[{{Horne} {et~al.}(1981){Horne}, {Hurford}, {Zirin}, \& {de
  Graauw}}]{Horne1981}
{Horne}, K., {Hurford}, G.~J., {Zirin}, H., \& {de Graauw}, T. 1981, \apj, 244,
  340, \dodoi{10.1086/158711}

\bibitem[{Iijima \& Yokoyama(2017)}]{Iijima2017}
Iijima, H., \& Yokoyama, T. 2017, \apj, 848, 38,
  \dodoi{10.3847/1538-4357/aa8ad1}

\bibitem[{Jess {et~al.}(2015)Jess, Morton, Verth, Fedun, Grant, \&
  Giagkiozis}]{Jess2015}
Jess, D.~B., Morton, R.~J., Verth, G., {et~al.} 2015, \ssr, 190, 103,
  \dodoi{10.1007/s11214-015-0141-3}

\bibitem[{{Judge} {et~al.}(2012){Judge}, {de Pontieu}, {McIntosh}, \&
  {Olluri}}]{Judge2012}
{Judge}, P.~G., {de Pontieu}, B., {McIntosh}, S.~W., \& {Olluri}, K. 2012,
  \apj, 746, 158, \dodoi{10.1088/0004-637X/746/2/158}

\bibitem[{Klimchuk(2012)}]{Klimchuk2012}
Klimchuk, J.~A. 2012, Journal of Geophysical Research: Space Physics, 117, n/a,
  \dodoi{10.1029/2012ja018170}

\bibitem[{Kosugi {et~al.}(2007)Kosugi, Matsuzaki, Sakao, Shimizu, Sone,
  Tachikawa, Hashimoto, Minesugi, Ohnishi, Yamada, Tsuneta, Hara, Ichimoto,
  Suematsu, Shimojo, Watanabe, Shimada, Davis, Hill, Owens, Title, Culhane,
  Harra, Doschek, \& Golub}]{Kosugi2007}
Kosugi, T., Matsuzaki, K., Sakao, T., {et~al.} 2007, Solar Physics, 243, 3,
  \dodoi{10.1007/s11207-007-9014-6}

\bibitem[{Krall {et~al.}(1976)Krall, Bessey, \& Beckers}]{Krall1976}
Krall, K.~R., Bessey, R.~J., \& Beckers, J.~M. 1976, \solphys, 46, 93,
  \dodoi{10.1007/BF00157556}

\bibitem[{Kundu(1971)}]{Kundu1971}
Kundu, M.~R. 1971, \solphys, 21, 130, \dodoi{10.1007/BF00155783}

\bibitem[{Lang(1999)}]{Lang1999}
Lang, K.~R. 1999, Astrophysical formulae.
\newblock \url{https://ui.adsabs.harvard.edu/abs/1999acfp.book.....L}

\bibitem[{Langangen {et~al.}(2008)Langangen, De~Pontieu, Carlsson, Hansteen,
  Cauzzi, \& Reardon}]{Langangen2008}
Langangen, {\O}., De~Pontieu, B., Carlsson, M., {et~al.} 2008, \apjl, 679,
  L167, \dodoi{10.1086/589442}

\bibitem[{Lemen {et~al.}(2011)Lemen, Title, Akin, Boerner, Chou, Drake, Duncan,
  Edwards, Friedlaender, Heyman, Hurlburt, Katz, Kushner, Levay, Lindgren,
  Mathur, McFeaters, Mitchell, Rehse, Schrijver, Springer, Stern, Tarbell,
  Wuelser, Wolfson, Yanari, Bookbinder, Cheimets, Caldwell, Deluca, Gates,
  Golub, Park, Podgorski, Bush, Scherrer, Gummin, Smith, Auker, Jerram, Pool,
  Soufli, Windt, Beardsley, Clapp, Lang, \& Waltham}]{Lemen2011}
Lemen, J.~R., Title, A.~M., Akin, D.~J., {et~al.} 2011, Solar Physics, 275, 17,
  \dodoi{10.1007/s11207-011-9776-8}

\bibitem[{{Lindsey} {et~al.}(1981){Lindsey}, {Hildebrand}, {Keene}, \&
  {Whitcomb}}]{Lindsey1981}
{Lindsey}, C., {Hildebrand}, R.~H., {Keene}, J., \& {Whitcomb}, S.~E. 1981,
  \apj, 248, 830, \dodoi{10.1086/159207}

\bibitem[{{Lindsey} {et~al.}(1992){Lindsey}, {Jefferies}, {Clark}, {Harrison},
  {Carter}, {Watt}, {Becklin}, {Roellig}, {Braun}, \& {Naylor}}]{Lindsey1992}
{Lindsey}, C., {Jefferies}, J.~T., {Clark}, T.~A., {et~al.} 1992, \nat, 358,
  308, \dodoi{10.1038/358308a0}

\bibitem[{Linsky(1973)}]{Linsky1973}
Linsky, J.~L. 1973, \solphys, 28, 409, \dodoi{10.1007/BF00152312}

\bibitem[{Loukitcheva {et~al.}(2004)Loukitcheva, Solanki, Carlsson, \&
  Stein}]{Loukitcheva2004}
Loukitcheva, M., Solanki, S.~K., Carlsson, M., \& Stein, R.~F. 2004, \aap, 419,
  747, \dodoi{10.1051/0004-6361:20034159}

\bibitem[{Macris(1957)}]{Macris1957}
Macris, C.~J. 1957, Annales d'Astrophysique, 20, 179.
\newblock \url{https://ui.adsabs.harvard.edu/abs/1957AnAp...20..179M}

\bibitem[{{Mart{\'\i}nez-Sykora} {et~al.}(2018){Mart{\'\i}nez-Sykora}, {De
  Pontieu}, {De Moortel}, {Hansteen}, \& {Carlsson}}]{MartinezSykora2018}
{Mart{\'\i}nez-Sykora}, J., {De Pontieu}, B., {De Moortel}, I., {Hansteen},
  V.~H., \& {Carlsson}, M. 2018, \apj, 860, 116,
  \dodoi{10.3847/1538-4357/aac2ca}

\bibitem[{Mart{\'\i}nez-Sykora {et~al.}(2017)Mart{\'\i}nez-Sykora, De~Pontieu,
  Hansteen, Rouppe van~der Voort, Carlsson, \& Pereira}]{MartinezSykora2017}
Mart{\'\i}nez-Sykora, J., De~Pontieu, B., Hansteen, V.~H., {et~al.} 2017,
  Science, 356, 1269, \dodoi{10.1126/science.aah5412}

\bibitem[{{Mart{\'\i}nez-Sykora} {et~al.}(2013){Mart{\'\i}nez-Sykora}, {De
  Pontieu}, {Leenaarts}, {Pereira}, {Carlsson}, {Hansteen}, {Stern}, {Tian},
  {McIntosh}, \& {Rouppe van der Voort}}]{MartinezSykora2013}
{Mart{\'\i}nez-Sykora}, J., {De Pontieu}, B., {Leenaarts}, J., {et~al.} 2013,
  \apj, 771, 66, \dodoi{10.1088/0004-637X/771/1/66}

\bibitem[{Matsuno \& Hirayama(1988)}]{Matsuno1988}
Matsuno, K., \& Hirayama, T. 1988, \solphys, 117, 21,
  \dodoi{10.1007/BF00148569}

\bibitem[{McMullin {et~al.}(2007)McMullin, Waters, Schiebel, Young, \&
  Golap}]{Mcmullin2007}
McMullin, J.~P., Waters, B., Schiebel, D., Young, W., \& Golap, K. 2007, in
  Astronomical Society of the Pacific Conference Series, Vol. 376, Astronomical
  Data Analysis Software and Systems XVI, ed. R.~A. {Shaw}, F.~{Hill}, \& D.~J.
  {Bell}, 127.
\newblock \url{https://ui.adsabs.harvard.edu/abs/2007ASPC..376..127M}

\bibitem[{Nikolic {et~al.}(2013)Nikolic, Bolton, Graves, Hills, \&
  Richer}]{Nikolic2013}
Nikolic, B., Bolton, R.~C., Graves, S.~F., Hills, R.~E., \& Richer, J.~S. 2013,
  Astronomy {\&} Astrophysics, 552, A104, \dodoi{10.1051/0004-6361/201220987}

\bibitem[{{Nindos} {et~al.}(2018){Nindos}, {Alissandrakis}, {Bastian},
  {Patsourakos}, {De Pontieu}, {Warren}, {Ayres}, {Hudson}, {Shimizu}, {Vial},
  {Wedemeyer}, \& {Yurchyshyn}}]{Nindos2018}
{Nindos}, A., {Alissandrakis}, C.~E., {Bastian}, T.~S., {et~al.} 2018, \aap,
  619, L6, \dodoi{10.1051/0004-6361/201834113}

\bibitem[{Okamoto \& De~Pontieu(2011)}]{Okamoto2011}
Okamoto, T.~J., \& De~Pontieu, B. 2011, \apjl, 736, L24,
  \dodoi{10.1088/2041-8205/736/2/L24}

\bibitem[{Patsourakos {et~al.}(2014)Patsourakos, Klimchuk, \&
  Young}]{Patsourakos2014}
Patsourakos, S., Klimchuk, J.~A., \& Young, P.~R. 2014, \apj, 781, 58,
  \dodoi{10.1088/0004-637X/781/2/58}

\bibitem[{{Pereira} {et~al.}(2012){Pereira}, {De Pontieu}, \&
  {Carlsson}}]{Pereira2012}
{Pereira}, T. M.~D., {De Pontieu}, B., \& {Carlsson}, M. 2012, \apj, 759, 18,
  \dodoi{10.1088/0004-637X/759/1/18}

\bibitem[{{Pereira} {et~al.}(2013){Pereira}, {De Pontieu}, \&
  {Carlsson}}]{Pereira2013}
---. 2013, \apj, 764, 69, \dodoi{10.1088/0004-637X/764/1/69}

\bibitem[{Pereira {et~al.}(2014)Pereira, De~Pontieu, Carlsson, Hansteen,
  Tarbell, Lemen, Title, Boerner, Hurlburt, W{\"u}lser, Mart{\'\i}nez-Sykora,
  Kleint, Golub, McKillop, Reeves, Saar, Testa, Tian, Jaeggli, \&
  Kankelborg}]{Pereira2014}
Pereira, T. M.~D., De~Pontieu, B., Carlsson, M., {et~al.} 2014, \apjl, 792,
  L15, \dodoi{10.1088/2041-8205/792/1/L15}

\bibitem[{Pesnell {et~al.}(2011)Pesnell, Thompson, \& Chamberlin}]{Pesnell2011}
Pesnell, W.~D., Thompson, B.~J., \& Chamberlin, P.~C. 2011, Solar Physics, 275,
  3, \dodoi{10.1007/s11207-011-9841-3}

\bibitem[{Phillips {et~al.}(2015)Phillips, Hills, Bastian, Hudson, Marson, \&
  Wedemeyer}]{Phillips2015}
Phillips, N., Hills, R., Bastian, T., {et~al.} 2015, in Astronomical Society of
  the Pacific Conference Series, Vol. 499, Revolution in Astronomy with ALMA:
  The Third Year, ed. D.~{Iono}, K.~{Tatematsu}, A.~{Wootten}, \& L.~{Testi},
  347.
\newblock \doarXiv{1502.06122}

\bibitem[{Pneuman \& Kopp(1978)}]{Pneuman1978}
Pneuman, G.~W., \& Kopp, R.~A. 1978, \solphys, 57, 49,
  \dodoi{10.1007/BF00152043}

\bibitem[{{Roellig} {et~al.}(1991){Roellig}, {Becklin}, {Jefferies}, {Kopp},
  {Lindsey}, {Orrall}, \& {Werner}}]{Roellig1991}
{Roellig}, T.~L., {Becklin}, E.~E., {Jefferies}, J.~T., {et~al.} 1991, \apj,
  381, 288, \dodoi{10.1086/170650}

\bibitem[{Rouppe van~der Voort {et~al.}(2015)Rouppe van~der Voort, De~Pontieu,
  Pereira, Carlsson, \& Hansteen}]{RouppevanderVoort2015}
Rouppe van~der Voort, L., De~Pontieu, B., Pereira, T. M.~D., Carlsson, M., \&
  Hansteen, V. 2015, \apjl, 799, L3, \dodoi{10.1088/2041-8205/799/1/L3}

\bibitem[{Rouppe van~der Voort {et~al.}(2009)Rouppe van~der Voort, Leenaarts,
  de~Pontieu, Carlsson, \& Vissers}]{RouppevanderVoort2009}
Rouppe van~der Voort, L., Leenaarts, J., de~Pontieu, B., Carlsson, M., \&
  Vissers, G. 2009, \apj, 705, 272, \dodoi{10.1088/0004-637X/705/1/272}

\bibitem[{Rutten(2017)}]{Rutten2017}
Rutten, R.~J. 2017, \aap, 598, A89, \dodoi{10.1051/0004-6361/201629238}

\bibitem[{Samanta {et~al.}(2019)Samanta, Tian, Yurchyshyn, Peter, Cao,
  Sterling, Erd{\'e}lyi, Ahn, Feng, Utz, Banerjee, \& Chen}]{Samanta2019}
Samanta, T., Tian, H., Yurchyshyn, V., {et~al.} 2019, Science, 366, 890,
  \dodoi{10.1126/science.aaw2796}

\bibitem[{Sault {et~al.}(1996)Sault, Stavley-Smith, \& Brouw}]{Sault1996}
Sault, R.~J., Stavley-Smith, L., \& Brouw, W.~N. 1996, Astronomy and
  Astrophysics Supplement Series, 120, 375

\bibitem[{{Sekse} {et~al.}(2013){Sekse}, {Rouppe van der Voort}, {De Pontieu},
  \& {Scullion}}]{Sekse2013}
{Sekse}, D.~H., {Rouppe van der Voort}, L., {De Pontieu}, B., \& {Scullion}, E.
  2013, \apj, 769, 44, \dodoi{10.1088/0004-637X/769/1/44}

\bibitem[{Shimojo {et~al.}(2017)Shimojo, Bastian, Hales, White, Iwai, Hills,
  Hirota, Phillips, Sawada, Yagoubov, Siringo, Asayama, Sugimoto, Braj{\v{s}}a,
  Skoki{\'c}, B{\'a}rta, Kim, de~Gregorio-Monsalvo, Corder, Hudson, Wedemeyer,
  Gary, De~Pontieu, Loukitcheva, Fleishman, Chen, Kobelski, \&
  Yan}]{Shimojo2017}
Shimojo, M., Bastian, T.~S., Hales, A.~S., {et~al.} 2017, \solphys, 292, 87,
  \dodoi{10.1007/s11207-017-1095-2}

\bibitem[{{Shimojo} {et~al.}(2020){Shimojo}, {Kawate}, {Okamoto}, {Yokoyama},
  {Narukage}, {Sakao}, {Iwai}, {Fleishman}, \& {Shibata}}]{Shimojo2020}
{Shimojo}, M., {Kawate}, T., {Okamoto}, T.~J., {et~al.} 2020, \apjl, 888, L28,
  \dodoi{10.3847/2041-8213/ab62a5}

\bibitem[{{Skogsrud} {et~al.}(2014){Skogsrud}, {Rouppe van der Voort}, \& {De
  Pontieu}}]{Skogsrud2014}
{Skogsrud}, H., {Rouppe van der Voort}, L., \& {De Pontieu}, B. 2014, \apjl,
  795, L23, \dodoi{10.1088/2041-8205/795/1/L23}

\bibitem[{Skogsrud {et~al.}(2015)Skogsrud, Rouppe van~der Voort, De~Pontieu, \&
  Pereira}]{Skogsrud2015}
Skogsrud, H., Rouppe van~der Voort, L., De~Pontieu, B., \& Pereira, T. M.~D.
  2015, \apj, 806, 170, \dodoi{10.1088/0004-637X/806/2/170}

\bibitem[{Sow~Mondal {et~al.}(2022)Sow~Mondal, Klimchuk, \&
  Sarkar}]{SowMondal2022}
Sow~Mondal, S., Klimchuk, J.~A., \& Sarkar, A. 2022, \apj, 937, 71,
  \dodoi{10.3847/1538-4357/ac879b}

\bibitem[{Sterling(2000)}]{Sterling2000}
Sterling, A.~C. 2000, \solphys, 196, 79, \dodoi{10.1023/A:1005213923962}

\bibitem[{Sterling(2021)}]{Sterling2021}
Sterling, A.~C. 2021, in Solar Physics and Solar Wind, ed. N.~E. {Raouafi} \&
  A.~{Vourlidas}, Vol.~1, 221, \dodoi{10.1002/9781119815600.ch6}

\bibitem[{Suematsu {et~al.}(2008)Suematsu, Ichimoto, Katsukawa, Shimizu,
  Okamoto, Tsuneta, Tarbell, \& Shine}]{Suematsu2008}
Suematsu, Y., Ichimoto, K., Katsukawa, Y., {et~al.} 2008, in Astronomical
  Society of the Pacific Conference Series, Vol. 397, First Results From
  Hinode, ed. S.~A. {Matthews}, J.~M. {Davis}, \& L.~K. {Harra}, 27.
\newblock \url{https://ui.adsabs.harvard.edu/abs/2008ASPC..397...27S}

\bibitem[{Tavabi {et~al.}(2011)Tavabi, Koutchmy, \&
  Ajabshirizadeh}]{Tavabi_2011}
Tavabi, E., Koutchmy, S., \& Ajabshirizadeh, A. 2011, New Astronomy, 16, 296,
  \dodoi{10.1016/j.newast.2010.11.005}

\bibitem[{Team {et~al.}(2022)Team, Bean, Bhatnagar, Castro, Donovan~Meyer,
  Emonts, Garcia, Garwood, Golap, Gonzalez~Villalba, Harris, Hayashi, Hoskins,
  Hsieh, Jagannathan, Kawasaki, Keimpema, Kettenis, Lopez, Marvil, Masters,
  McNichols, Mehringer, Miel, Moellenbrock, Montesino, Nakazato, Ott, Petry,
  Pokorny, Raba, Rau, Schiebel, Schweighart, Sekhar, Shimada, Small, Steeb,
  Sugimoto, Suoranta, Tsutsumi, van Bemmel, Verkouter, Wells, Xiong, Szomoru,
  Griffith, Glendenning, \& Kern}]{Team2022}
Team, C., Bean, B., Bhatnagar, S., {et~al.} 2022, \pasp, 134, 114501,
  \dodoi{10.1088/1538-3873/ac9642}

\bibitem[{Thompson {et~al.}(2017)Thompson, Moran, \& Swenson}]{Thompson2017}
Thompson, A.~R., Moran, J.~M., \& Swenson, G.~W. 2017, Interferometry and
  Synthesis in Radio Astronomy (Springer International Publishing),
  \dodoi{10.1007/978-3-319-44431-4}

\bibitem[{Tian {et~al.}(2014)Tian, DeLuca, Cranmer, De~Pontieu, Peter,
  Mart{\'\i}nez-Sykora, Golub, McKillop, Reeves, Miralles, McCauley, Saar,
  Testa, Weber, Murphy, Lemen, Title, Boerner, Hurlburt, Tarbell, Wuelser,
  Kleint, Kankelborg, Jaeggli, Carlsson, Hansteen, \& McIntosh}]{Tian2014}
Tian, H., DeLuca, E.~E., Cranmer, S.~R., {et~al.} 2014, Science, 346, 1255711,
  \dodoi{10.1126/science.1255711}

\bibitem[{Tsiropoula {et~al.}(2012)Tsiropoula, Tziotziou, Kontogiannis,
  Madjarska, Doyle, \& Suematsu}]{Tsiropoula2012}
Tsiropoula, G., Tziotziou, K., Kontogiannis, I., {et~al.} 2012, \ssr, 169, 181,
  \dodoi{10.1007/s11214-012-9920-2}

\bibitem[{Tsuneta {et~al.}(2008)Tsuneta, Ichimoto, Katsukawa, Nagata, Otsubo,
  Shimizu, Suematsu, Nakagiri, Noguchi, Tarbell, Title, Shine, Rosenberg,
  Hoffmann, Jurcevich, Kushner, Levay, Lites, Elmore, Matsushita, Kawaguchi,
  Saito, Mikami, Hill, \& Owens}]{Tsuneta2008}
Tsuneta, S., Ichimoto, K., Katsukawa, Y., {et~al.} 2008, \solphys, 249, 167,
  \dodoi{10.1007/s11207-008-9174-z}

\bibitem[{Vernazza {et~al.}(1976)Vernazza, Avrett, \& Loeser}]{Vernazza1976}
Vernazza, J.~E., Avrett, E.~H., \& Loeser, R. 1976, \apjs, 30, 1,
  \dodoi{10.1086/190356}

\bibitem[{Vernazza {et~al.}(1981)Vernazza, Avrett, \& Loeser}]{Vernazza1981}
---. 1981, \apjs, 45, 635, \dodoi{10.1086/190731}

\bibitem[{Verner {et~al.}(1996)Verner, Ferland, Korista, \&
  Yakovlev}]{Verner1996}
Verner, D.~A., Ferland, G.~J., Korista, K.~T., \& Yakovlev, D.~G. 1996, \apj,
  465, 487, \dodoi{10.1086/177435}

\bibitem[{{Wannier} {et~al.}(1983){Wannier}, {Hurford}, \&
  {Seielstad}}]{Wannier1983}
{Wannier}, P.~G., {Hurford}, G.~J., \& {Seielstad}, G.~A. 1983, \apj, 264, 660,
  \dodoi{10.1086/160639}

\bibitem[{White \& Kundu(1994)}]{White1994}
White, S.~M., \& Kundu, M.~R. 1994, in Infrared Solar Physics, ed. D.~M.
  {Rabin}, J.~T. {Jefferies}, \& C.~{Lindsey}, Vol. 154, 167.
\newblock \url{https://ui.adsabs.harvard.edu/abs/1994IAUS..154..167W}

\bibitem[{White {et~al.}(2017)White, Iwai, Phillips, Hills, Hirota, Yagoubov,
  Siringo, Shimojo, Bastian, Hales, Sawada, Asayama, Sugimoto, Marson,
  Kawasaki, Muller, Nakazato, Sugimoto, Braj{\v{s}}a, Skoki{\'c}, B{\'a}rta,
  Kim, Remijan, de~Gregorio, Corder, Hudson, Loukitcheva, Chen, De~Pontieu,
  Fleishmann, Gary, Kobelski, Wedemeyer, \& Yan}]{White2017}
White, S.~M., Iwai, K., Phillips, N.~M., {et~al.} 2017, \solphys, 292, 88,
  \dodoi{10.1007/s11207-017-1123-2}

\bibitem[{Withbroe(1983)}]{Withbroe1983}
Withbroe, G.~L. 1983, \apj, 267, 825, \dodoi{10.1086/160917}

\bibitem[{Withbroe(1988)}]{Withbroe1988}
---. 1988, \apj, 325, 442, \dodoi{10.1086/166015}

\bibitem[{Wootten \& Thompson(2009)}]{Wootten2009}
Wootten, A., \& Thompson, A. 2009, Proceedings of the IEEE, 97, 1463,
  \dodoi{10.1109/jproc.2009.2020572}

\bibitem[{Yokoyama {et~al.}(2018)Yokoyama, Shimojo, Okamoto, \&
  Iijima}]{Yokoyama2018}
Yokoyama, T., Shimojo, M., Okamoto, T.~J., \& Iijima, H. 2018, \apj, 863, 96,
  \dodoi{10.3847/1538-4357/aad27e}

\bibitem[{Zaqarashvili \& Erd{\'e}lyi(2009)}]{Zaqarashvili2009}
Zaqarashvili, T.~V., \& Erd{\'e}lyi, R. 2009, \ssr, 149, 355,
  \dodoi{10.1007/s11214-009-9549-y}

\bibitem[{Zhang {et~al.}(1998)Zhang, White, \& Kundu}]{Zhang1998}
Zhang, J., White, S.~M., \& Kundu, M.~R. 1998, \apjl, 504, L127,
  \dodoi{10.1086/311587}

\bibitem[{Zhang {et~al.}(2012)Zhang, Shibata, Wang, Mao, Matsumoto, Liu, \&
  Su}]{Zhang2012}
Zhang, Y.~Z., Shibata, K., Wang, J.~X., {et~al.} 2012, \apj, 750, 16,
  \dodoi{10.1088/0004-637X/750/1/16}

\end{thebibliography}
\bibliographystyle{aasjournal}

\appendix
\section{ALMA Solar Data Reduction Challenges}

Interferometric and total power observations of the Sun with ALMA have been described by \citet{Shimojo2017, White2017}, and \citet{Bastian2022}. As the community has gained experience with ALMA data, a number of challenges have become apparent. We briefly discuss several of these challenges here and their implications for the reduction and analysis of the spicule observations in the two ALMA wavelength bands.

\subsection{Self-calibration}

Self-calibration is a mature technique employed in radio synthesis imaging (e.g., \citealt{Cornwell1989}) that is analogous to adaptive optics used with optical telescopes. Water vapor in the atmosphere introduces variations in the electrical path length of the incident \mml\ radiation, resulting in time-variable phase changes at each antenna --  in effect corrupting the antenna gain. This can lead to a baseline-dependent loss of coherence, typically increasing with baseline length, and a reduction in the effective angular resolution. However, if the coherence scale is comparable to or greater than the extent of the antenna array, ``seeing'' is dominated by image wander. Non-solar observations have recourse to measurements of phase variations using water vapor radiometers (WVRs; \citet{Nikolic2013}). ALMA uses WVRs to monitor sky brightness temperature variations using the 183~GHz water line. From these, changes in the electrical path length over each antenna toward the source can be inferred and used to correct variations in phase, at least in part. Such measurements are not available to solar observations because the WVRs saturate when pointing at the Sun. Instead, self-calibration techniques must be exploited to correct antenna-based phase variations due to the sky. A well-known drawback to self-calibration is that absolute position information is lost unless the data are referenced to a position standard. We use the limb of the Sun as the position reference. The specifics of the self-calibration strategy and the use of the solar limb as the position reference for each of the two wavelength bands are described in \S3.1 and \S3.2.

\begin{figure}[ht!]
\begin{center}
\includegraphics[angle=0, width=6in]{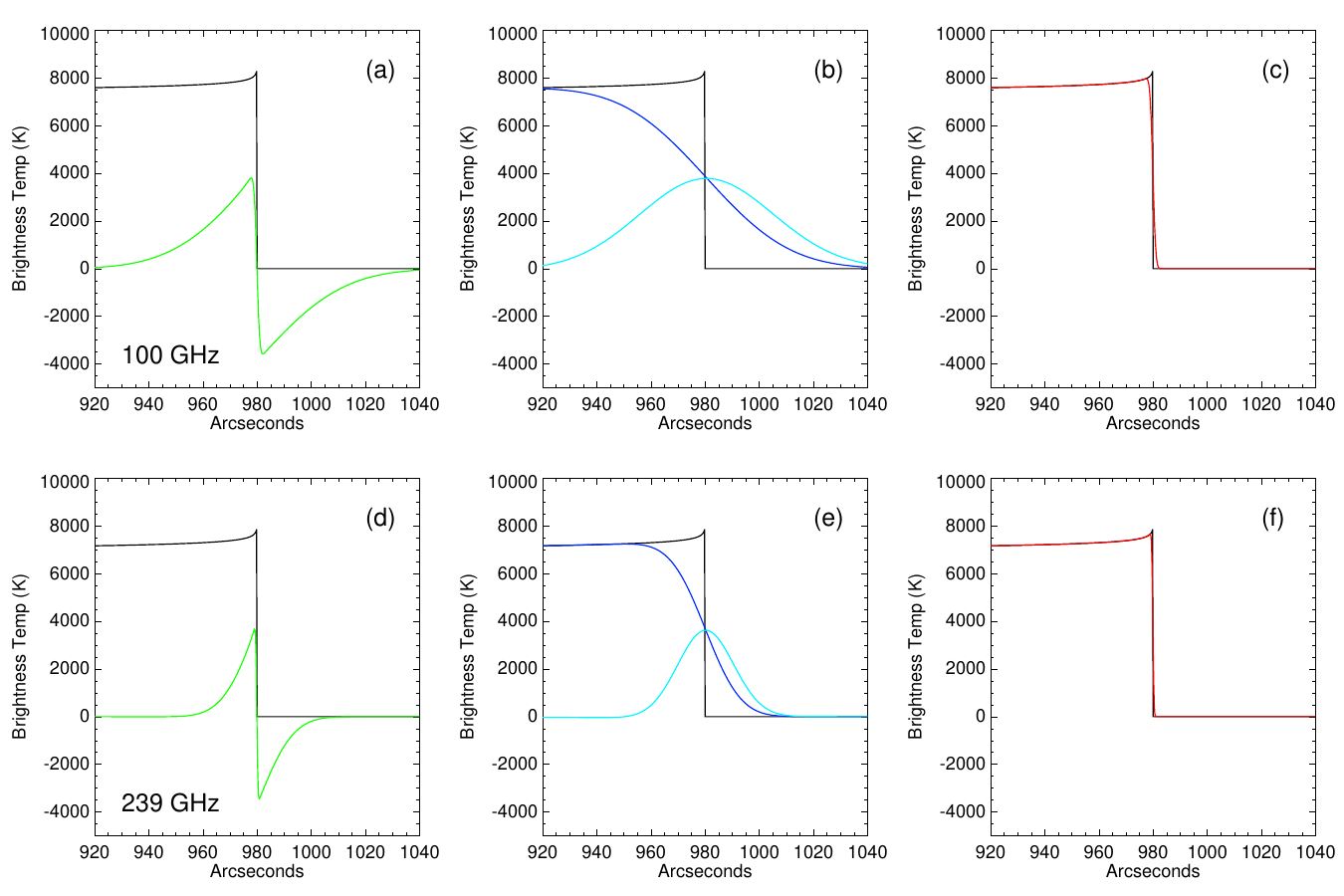}
\caption{Recovery of large angular scales. Panels a-c show the steps schematically for 3~mm observations while panels d-f show the same for 1.25~mm observations. All panels show the assumed mean brightness distribution near the limb (black line). Panels (a) and (d) show the idealized brightness profile (green line) resulting from observations with the ALMA array. Since the shortest baselines and, hence, the largest spatial frequencies, are not measured there is strong ``interferometric overshoot" at the limb. Panels (b) and (e) show the brightness profile resulting from TP observations normally made by a single 12~m antenna (dark blue line). Also shown is the radial gradient of the TP brightness profile (cyan line), discussed in \S4.3. Panels (c) and (f) show the combination of the interferometric data and the single dish measurements results in a restored profile (red line) that represents ``true" brightness profile convolved with a Gaussian representing the nominal resolution of the array. }
\end{center}
\end{figure}


\subsection{Missing Spatial Frequencies}

An interferometric array like ALMA acts as a high-pass filter, measuring only those angular frequencies corresponding to the range of antenna baselines available. Since there is a minimum antenna spacing in the array, there is a maximum angular scale measured. For observations of quiet Sun sources on the solar disk, this is not necessarily a significant problem, at least for single-pointing observations. To first order, the interferometric image represents the distribution of brightness variations relative to the mean. That is, the mean flux has been resolved out and the average brightness across the image is zero. For interferometric maps of fields on the solar disk, simply adding the mean brightness (corresponding to the ``zero-spacing" flux density) to the map largely corrects for the missing short-spacing flux although it is preferable to restore the missing angular scales explicitly. The situation is more complicated at solar limb, however, where any interferometer displays a strong response to the fact that the Sun's brightness drops precipitously to nearly zero. Since the average brightness of the interferometric image remains zero, the response manifests itself as a bright limb and a strong negative-brightness feature above the limb that we refer to as interferometric ``overshoot". Overshoot has a significant impact on any emission features extending above the limb (e.g., spicules) that cannot be corrected using deconvolution alone. It also leads to an incorrect brightness temperature being inferred for brightness near the limb. We illustrate the phenomenon schematically in Fig.~15. The effect of the high-pass filtering by the interferometric array is to reduce the brightness temperature at the limb by a factor $\sim 2$. Inclusion of the brightness distribution on angular scales that are not measured by the array -- using separate observations or a model distribution -- is therefore critical. 

The ALMA solar observing mode addresses the general problem of missing short spacings by providing low-resolution, full disk TP maps of the Sun in spectral windows identical to those used for interferometric imaging, as discussed in \S2.2.2. A TP map provides the brightness distribution on angular scales not measured by the interferometric array, allowing an observer to, in effect, fill in the missing short baselines using techniques such as ``feathering" \citep{Cotton2015}. For reasons given in the main text, we elected to use models of the full disk TP data.

\subsection{Heterogeneous Array}

A second way ALMA solar observations addresses the problem of missing short baselines is to use the ACA array of up to twelve 7m antennas together with the 12m array to increase the number of short baselines measured \citep{Shimojo2017}; i.e., to use a heterogeneous array. Representative sampling of the aperture, also referred to as {\sl uv} coverage, for a single pointing, spectral window, and integration time in each of the two wavelength bands is shown in Fig.~16 for the inner part of the array. The $u$ and $v$ axes, given in wavelength units, are the orthogonal coordinates of the samples in the Fourier domain. The grey disk shown in the middle of each plot shows the effective {\sl uv} coverage filled in by a TP map. The blue points indicate the location of visibility samples in the {\sl uv} plane and their  size indicates whether the visibility was measured by a 7m $\times$ 7m, 7m $\times$ 12m, or a 12m $\times$ 12m antenna pair. For both bands, the additional coverage provided by the TP observations neatly fills in hole in the {\sl uv} coverage provided by heterogeneous array comprising the C43-3 12-m array and the ACA. There is little overlap between the TP and INT coverage in the {\sl uv} plane for the C43-3 configuration, however. It is also notable that the sampling density of the inner {\sl uv} plane is much sparser at 1.25~mm than it is at 3~mm.  The paucity of points measured on the shortest antenna baselines has implications for the maximum recoverable scale $\theta_{MRS}$. As discussed in the {\sl ALMA Technical Handbook}\footnote{https://almascience.nrao.edu/proposing/technical-handbook}, $\theta_{MRS}$ is the largest angular scale that can be robustly recovered from the data. It does not simply correspond to $\lambda/L_{min}$ where $L_{min}$ is the minimum antenna baseline. Rather, based on extensive simulations, $\theta_{MRS}\approx 0.983\lambda/L_5$ radians ($\lambda,\ L_5$ in m), where $L_5$ is the radius within which $5\%$ of antennas baselines are located. For the 12m array configuration C43-3, we have $\theta_{MRS}\approx 16.2"$ and 7" for the 3~mm and 1.25~mm bands, corresponding to {\sl uv} radii of 12.3~k$\lambda$ and 30~k$\lambda$, respectively. While inclusion of the ACA can improve on this significantly for full synthesis imaging -- i.e., for observations of duration $\sim\!2$ hrs to ensure dense sampling of short baselines -- inspection of Fig.~16 shows that just a handful of ACA visibilities overlap with the angular coverage provided by TP maps for snapshot imaging and there is consequently very little overlap between the TP maps and ACA/$L_5$ baselines, an additional challenge to robust recovery of the largest angular scales. 

Fig. 16 shows the sampling function for 3 mm (left panel) for the inner part of the {\sl uv plane}. The minimum baseline sampled
is 4~k$\lambda$ and so angular scales larger than $58.3"$ are not sampled by the 12m array. The number of visibilities
contributed by baselines involving 7m antennas is limited for the domain shown: just 28 7m$\times$7m and 18 7m$\times$12m
baselines compared with 145 12m$\times$12m baselines (note that a given baseline samples a visibility $V (u,v)$ and its complex
conjugate at $V^\ast(-u,-v)$. The use of the heterogeneous array does not greatly improve sampling of short baselines
and, as we discuss in the next subsection, visibilities involving correlations between 7m$\times$7m antennas and 7m$\times$12m antennas are in any case strongly down-weighted relative to those in the 12m array and therefore do not yield significant improvements to imaging fidelity. We therefore used only 12m array visibilities for imaging time series of ALMA 3 mm data.

In the case of the 1.25~mm data, the inner part of the {\sl uv} plane is much more sparsely sampled than is the case for the 3~mm observations. Out of a total of 1225 baselines, only 45 have baselines less than 20~k$\lambda$, corresponding to angular scales $\gtrsim\!10"$. The number of different baseline types also differs (Fig.~16, right panel). Of the 45 baselines shown, 20 are 7m$\times$7m baselines, 2 are 7m$\times$12m baselines, and 23 are 12m$\times$12m baselines; there are comparable numbers of 7m$\times$7m and 12m$\times$12m baselined. We elected to use all available visibilities -- i.e., the heterogeneous array comprised of the 12~m array and the ACA --  to improve sampling of the largest spatial scales.  Even so, the relative contributions of the 7m$\times$7m and 7m$\times$12m baselines are down-weighted relative to those of the 12m array. Mitigating this to some degree is the mosaicking strategy (see below) used to image an angular domain comparable to that of the 3~mm images, improving the effective coverage of the inner part of the {\sl uv} domain (\citealt{Ekers1979}; Fig.~16b).


\begin{figure}
\begin{center}
\includegraphics[angle=90,clip,width=6.5in]{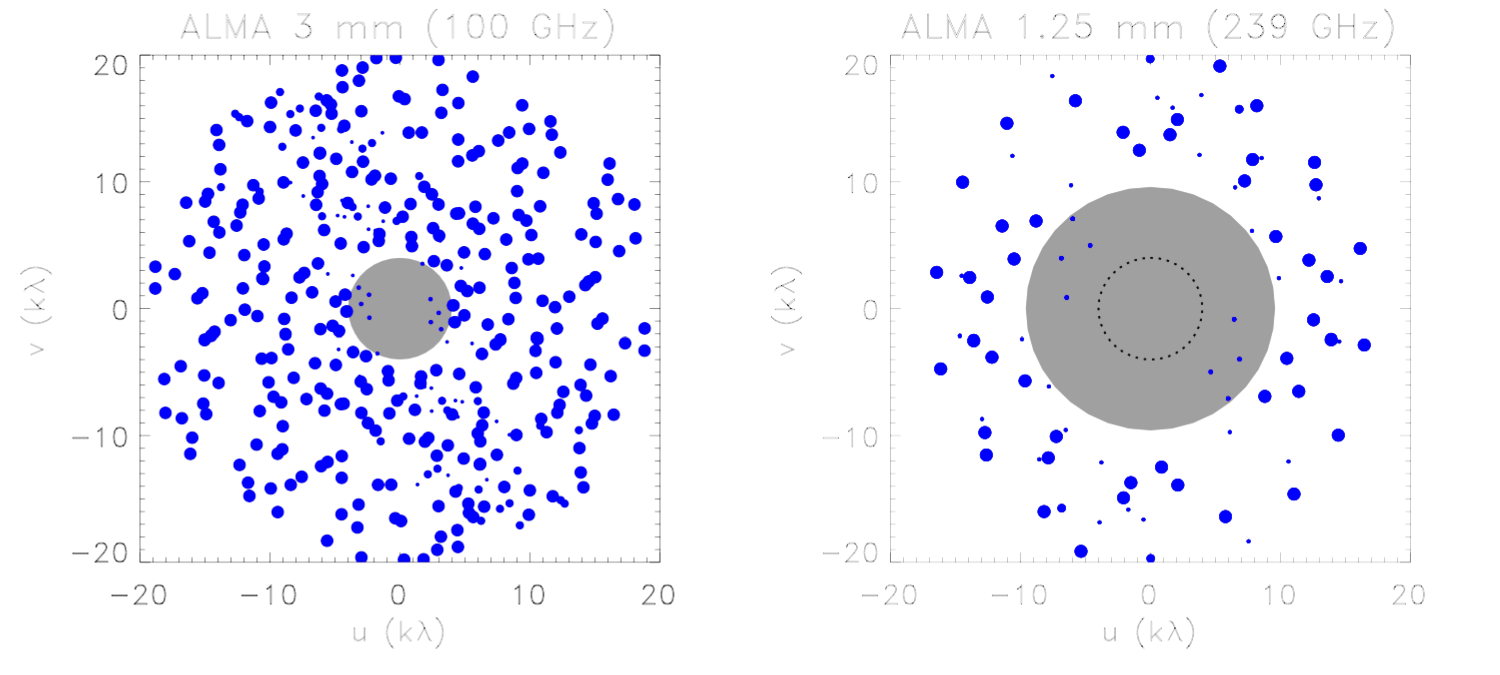}
\caption{The ALMA {\sl uv} coverage in the inner part of the array. The blue points represent baseline locations. Their size schematically indicates their relative weight: the largest size corresponds to antennas in the 12-m array, those of intermediate size corresponded to a 7m$\times$12m baseline, and those of the smallest size correspond to a 7m$\times$7m baseline. The grey disk shows the effective {\sl uv} coverage provided by the single dish total power maps. (a) the 3~mm data; (b) the 1.25~mm data. Note that because the baselines scale inversely with wavelength, the {\sl uv} coverage is correspondingly sparser for the 1.25~mm observations relative to those at 3~mm in the inner part of the {\sl uv} plane.  The dashed line indicates the effective increase in coverage of angular scales in the inner part of the {\sl uv} plane through mosaicking.}
\end{center}
\end{figure}


\subsection{Mosaicking}

A third way in which larger angular scales can be recovered is through use of the mosaicking technique \citep{Ekers1979, Cornwell1989b}.  The FOV of a single pointing at 1.25~mm is small ($24.3"$ for the 12~m array) and one must therefore use mosaicking techniques \citep{Cornwell1989a, Cornwell1993, Sault1996} to map a larger field of view. Mosaicking entails a sequential observation of a user-specified grid of pointings. In order to recover the spatial scales over the range sampled by the mosaic pattern, the pointings must sample the angular domain at the Nyquist frequency or better.  In so doing, angular scales up to the size of the domain sampled are recovered. In the case of the 1.25~mm observations described here, fourteen-point mosaics were constructed, sampling a domain of $\sim\!1'$, comparable in size to the 3~mm FOV (Fig.~1). While this technique works well for static sources, solar spicules are fundamentally dynamic in nature. Hence, observers must consider the trade-offs between imaging an adequate field of view using multi-point mosaics, the time required to perform such a mosaic, and the characteristic time over which the phenomenon of interest varies -- in this case, spicules. As noted, we use all available baselines in the heterogeneous array and mosaicing techniques to improve our coverage of the largest angular scale. We discuss our 1.25~mm mosaicking approach further in \S3.6.


\subsection{Data Weights}

The observations on 2018 December 25 were performed using 41 antennas in the 12-m array and 9 antennas in the ACA, yielding a total of 1225 antenna baselines: 820 12m$\times$12m baselines, 369 7m$\times$12m baselines, and 36 7m$\times$7m baselines. All antenna pairs were correlated in the baseline correlator. The initial stages of imaging proceed by gridding the weighted visibility data and then performing inverse Fourier transformation. The CASA task {\sl tclean} was used to perform the 1.25~mm mosaicking and to properly handle the two antenna sizes during data gridding ({\sl gridder=mosaicft}).  As noted by \citet{Bastian2022} the weight assigned to a visibility measurement by antenna $i$ and antenna $j$ is $1/\sigma_{ij}^2$ where $\sigma_{ij}^2$ is the statistical variance of the visibility. The expression for $\sigma_{ij}^2$  is

\begin{equation}
\sigma_{ij}^2 = {{2 k_B}\over{A_{e,i} A_{e,j}}} { {(T_{sys,i}+T_{ant,i})(T_{sys,j}+T_{ant,j})}\over{2\Delta\nu\Delta t}}
\end{equation}

\noindent Here, $A_{e,i}$ is the effective area of antenna $i$ (which may be different from that of antenna $j$), $k_B$ is Boltzmann's constant, $\Delta t$ is the integration time, and $T_{sys,i}$ and $T_{ant,i}$ are the system temperature and antenna temperature of antenna $i$. We have neglected efficiency factors of order unity in this expression. $T_{ant,i}=S A_{e,i}/2 k_B$, where $S$ is the flux density of the source. For most sidereal sources, $T_{ant} <<T_{sys}$ and it can neglected. However, when observing the Sun, $T_{ant} > T_{sys}$ and it can no longer be neglected in flux calibration (Shimojo et al. 2017) or in computing visibility weights. All other things being equal; i.e., $T_{sys}$, $T_{ant}$, frequency bandwidth, and integration time, the relative visibility weights scale as $(7/12)^4:(7/12)^2:1=0.116:0.340:1$ for antenna baselines involving 7m$\times$7m, 7m$\times$12m, and 12m$\times$12m antennas, respectively.  Coupled with the number of baselines of each type, the contribution of ACA baselines is strongly down-weighted relative to the 12m array and mixed 7m$\times$12m baselines are themselves down-weighted relative to the 12m array. 


\end{document}